\newif\ifdraft
\drafttrue
\documentclass[10pt,twocolumn,lettersize]{article}

\usepackage[margin=0.7in]{geometry}

  \usepackage{cite}

\usepackage{hyperref}

\usepackage[cmex10]{amsmath}

  \usepackage[caption=false,font=footnotesize]{subfig}

\usepackage{listings}
\usepackage{qvt}
\usepackage{rotating}
\usepackage{xcolor}
\usepackage{amssymb,amsmath}
\usepackage{pifont}
\usepackage{comment}

\newtheorem{definition}{Definition}

\newcommand{\chk}{\textsc{Check}}
\newcommand{\rep}{\textsc{Repair}}
\newcommand{\dom}{\boldsymbol{M}}
\newcommand{\inc}{\mathcal{I}}
\newcommand{\cst}{\mathcal{C}}
\newcommand{\upd}{\mathcal{U}}
\newcommand{\rps}{\mathcal{R}}

\newcommand{\Badger}[0]{\emph{Badger}}
\newcommand{\Echo}[0]{\emph{Echo}}
\newcommand{\ModelAnalyzer}[0]{\emph{Model/Analyzer}}

\begin{document}

\title{A Feature-based Classification of Model Repair Approaches\\(First Draft)}

\author{Nuno~Macedo,
        Tiago~Jorge,
        and~Alcino~Cunha% <-this % stops a space
        }% <-this % stops an unwanted space

\ifdraft
\newcommand{\nunoHide}[1]{}
\newcommand{\tiagoHide}[1]{}
\newcommand{\nuno}[1]{{\color{blue}[#1]}}
\newcommand{\tiago}[1]{{\color{orange}[#1]}}
\newcommand{\alcino}[1]{{\color{red}[#1]}}
\else
\newcommand{\tiagoHide}[1]{}
\newcommand{\nuno}[1]{}
\newcommand{\tiago}[1]{}
\newcommand{\alcino}[1]{}
\fi

\twocolumn[
  \begin{@twocolumnfalse}
      
      \maketitle
      
\begin{abstract}
Consistency management, the ability to detect, diagnose and handle inconsistencies, is crucial during the development process in Model-driven Engineering (MDE). As the popularity and application scenarios of MDE expanded, a variety of different techniques were proposed to address these tasks in specific contexts. 
Of the various stages of consistency management, this work focuses on inconsistency fixing in MDE, where such task is embodied by model repair techniques.
This paper proposes a feature-based classification system for model repair techniques, based on an systematic review of previously proposed approaches. We expect this work to assist both the developers of novel techniques and the MDE practitioners looking for suitable solutions.
\end{abstract}

\end{@twocolumnfalse}
]

\section{Introduction}\label{sec:intro}

Model-driven Engineering (MDE) is a family of development processes that focus on models as the primary development artifact. As models are modified by different stakeholders, in a possibly distributed and heterogeneous environment, the consistency of the overall MDE environment must be constantly monitored. Therefore, \emph{consistency management}~\cite{NuseibehER:00,SpanoudakisZ:01}---which involves various techniques concerned with the detection, diagnosis, fixing and tracking of inconsistencies---is essential to MDE. In fact, besides being fundamental to preserve consistency as the models naturally evolve, the necessity for consistency management arises during various others MDE activities, like meta-model and constraint evolution~\cite{DemuthLE:13}, model refactoring~\cite{StraetenD:06}, variability modeling~\cite{BenavidesSC:10} or version merging~\cite{DamRE:14}.

Although inconsistencies may occur due to mistakes or imprudent decisions, the overall impact of the changes applied by the developers may not be immediately perceptible, especially considering the complexity of the MDE development environment. Moreover, inconsistencies may also reflect conflicting or alternative interpretations of the requirements or uncertainty and partial knowledge~\cite{EasterbrookN:96}. Thus, developing frameworks should not forbid the introduction of inconsistencies altogether, but tolerate them while still providing support for their detection~\cite{Balzer:91}. Notwithstanding, as the development progresses and conflicting interpretations converge, so are the models expected to evolve to a consistent version, and thus inconsistencies must eventually be \emph{fixed} (or \emph{resolved})~\cite{SpanoudakisZ:01}. To be manageable, these tasks must be supported by automated techniques that help the user decide how to repair the models so that the consistency of the environment is restored. In MDE this amounts to \emph{model repair} techniques that attempt to ameliorate the consistency level of the MDE environment by proposing updates to the models.

One of the main challenges of model repair is that for any given set of inconsistencies, there (possibly) exists an overwhelming number of repair updates that restore the consistency. Yet, since the selection of the most suitable repair is ultimately a choice of the developer, approaches to model repair must balance the automation level of the technique and the need for user guidance in the generation of the repairs. 
%\alcino{The next sentence misses something}
Some authors~\cite{PuissantSM:13} advocate the use of heuristics to tackle the presence of a large search space, the need for algorithms with a low computational complexity, and the absence of known optimal solutions. Others~\cite{RederE:12} advocate against fully automatic approaches that replace the role of the human designer in repairing models. According to the latter authors, repairing models should be an activity that goes hand in hand with the creative process of modeling. 
To render these tasks more manageable, a variety of techniques has been developed that assume a more controlled environment with more concrete goals, including change propagation~\cite{DamWP:06}, model synchronization~\cite{AntkiewiczC:07}, bidirectional model transformation~\cite{Stevens:07} or incremental model transformation~\cite{EtzlstorferKKLRSSW:13}.

%Indeed, methods on how to detect and repair inconsistencies vary widely. Moreover, they are implemented in different programming languages, on different operating systems, use different modelling languages, and different input and output formats.

Motivated by this diversity of approaches, this paper explores the landscape of model repair techniques and proposes a taxonomy for their classification. While such surveys have been performed in other areas of MDE, a systematic analysis of the design space for model repair techniques is still lacking. Following other successful classifications of MDE techniques (e.g.,~\cite{CzarneckiH:06} for model transformation), we present our classification axes as \emph{feature models}~\cite{KangCHNP:90}, diagrams developed with the goal of modeling alternative configurations in software product lines. We hope this will aid the MDE practitioner in need of model repair techniques---or the developer interested in developing novel solutions---in the selection of the technique he deems best-suited for his particular application domain. As proof of concept, we classify and compare some modern approaches to model repair under our classification system.
Although essential to model repair, we do not focus on the problem of detecting the inconsistencies and their causes, which is sufficiently rich to require a dedicated survey itself. Also left out of this survey are techniques that circumvent the problem by just forbidding the introduction of inconsistencies.

This paper is structured as follows. Section~\ref{sec:setting} starts by presenting and formalizing the model repair problem. Section~\ref{sec:features} then defines the feature-based taxonomy under which model repair techniques can be classified. This taxonomy is used in Section~\ref{sec:classes} to classify and compare three recent techniques for model repair. Lastly, Section~\ref{sec:related} discusses some related work, while Section~\ref{sec:conclusion} draws conclusions and final remarks. %\alcino{We must decide if we use ``inconsistency fixing'' always or both terms simultaneously, and in the latter case just make a footnote somewhere.} 

%\nuno{coisas que tem que ser ditas que ficaram de fora: TGGs não incrementais, sincronização em que a spec é uma transformação}

\section{Model Repair}
\label{sec:setting}

This section presents and formalizes the problem of consistency management, focusing particularly on model repair, the target of this work. 

\subsection{Overview}
\label{sec:overview}
In this section we introduce a couple of examples, inspired by modern model repair techniques~\cite{RederE:12,PuissantSM:13,MacedoC:14}, that will provide an overview of the model repair problem and illustrate the vastness of features that consistency management techniques may implement.

\begin{figure}
    \centering
    \includegraphics[scale=0.5]{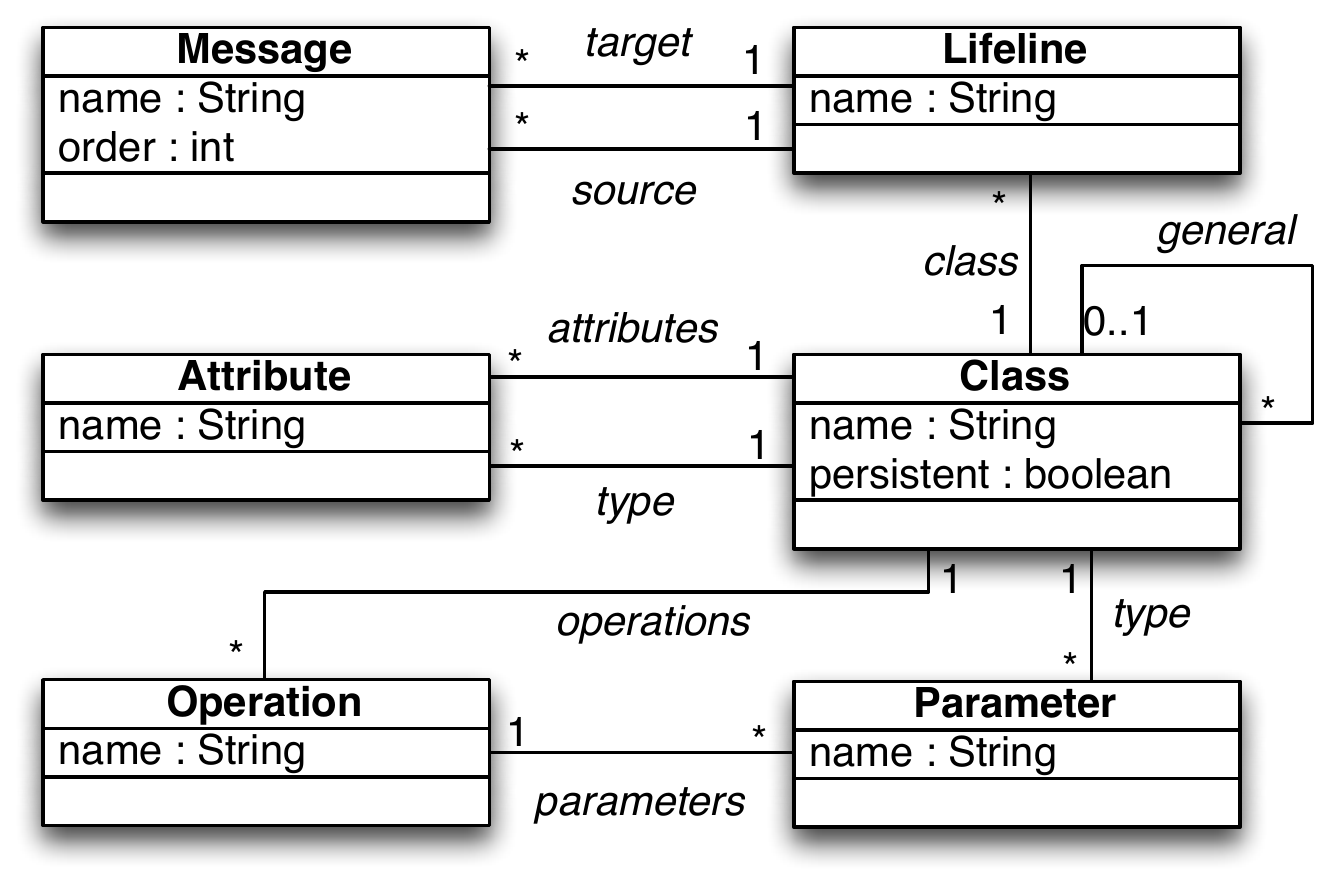}
    \caption{Simplified meta-model for class and sequence diagrams.}
    \label{fig:mm}
\end{figure}

While many approaches to consistency management are focused on particular kinds of models (e.g. UML diagrams), most recent approaches to model repair are meta-model independent: they allow the user to specify both the meta-models using some of the available meta-modeling languages like OMG's \emph{Model-driven architecture} (MDA) or the \emph{Eclipse Modeling Framework} (EMF). Figure~\ref{fig:mm} depicts one such meta-model, for representing very simplified class and sequence diagrams.

Although a meta-model defines which model instances are considered well-formed, there are a number of structural and behavioral properties that cannot be captured by meta-models alone. Thus, meta-models are usually annotated with additional \emph{intra}- and \emph{inter}-model constraints that restrict the internal structure of individual models and their relationship with other models, respectively. Ideally, the user should be allowed to define such constraints, typically using MDA's OCL~\cite{OCL:12} or some similar constraint language. 
One such constraint over class diagrams is that class generalization links must be acyclic. In OCL, this can be defined as follows for the meta-model in Fig.~\ref{fig:mm}:
%\tiago{acho que e melhor chamar atributos as propriedades, pk posso considerar que propriedade da classe e o seu nome ou se e persistente e confunde, atributo percebe-se logo}
\begin{lstlisting}[language=QVT]
context Class acyclic_generalization: 
    not self.closure(general)->includes(self)
\end{lstlisting}

\begin{figure}
    \centering
    \subfloat[][Initial state.]{\label{fig:ex_ay}\includegraphics[scale=0.5]{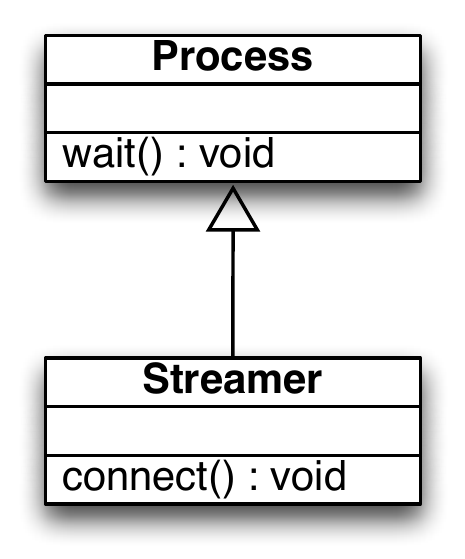}} \qquad
    \subfloat[][Updated state.]{\label{fig:ex_cy}\includegraphics[scale=0.5]{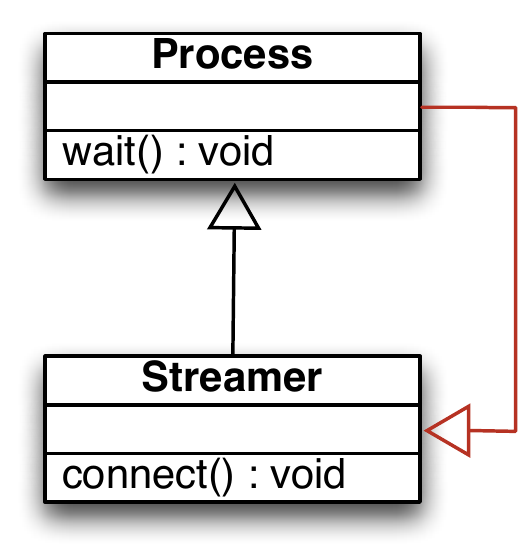}}
    \caption{Inconsistency in a class diagram.}
\end{figure}

Consider, as an example, the class diagram from Fig.~\ref{fig:ex_ay} conforming to Fig.~\ref{fig:mm}, depicting a tentative first version of the structure of a \emph{video on demand} (VOD) system (inspired by~\cite{PuissantSM:13}), consistent under \texttt{acyclic\_generalization}. Then, assume that at some point one of the developers, maybe oblivious of the whole inheritance tree or maybe disagreeing with previous design decisions, updated the model to the version depicted in Fig.~\ref{fig:ex_cy} by introducing a new generalization link (colored red), giving rise to a violation that breaks the \texttt{acyclic\_generalization} constraint. Since such updates can evidence conflicting interpretations of the requirements, the introduction of inconsistencies should not be forbidden but rather detected, diagnosed and resolved when deemed necessary.

Regarding the resolution of inconsistencies, the focus of our study, a variety of fixes can be applied depending on the stakeholders intentions. However, the available choices are also limited by the support provided by the modeling framework. For instance, a modeling tool working in an online setting could have detected the user operation that led to the violation (the introduction of the new generalization link), and allow the user to either preserve it or undo it. In contrast, offline tools, operating in a state-based setting, would have no such information available, and would need to either make an arbitrary choice or present every alternative to the user. Other design choices regard how the repairs should be presented to the user: should the procedure just return the repaired model, or abstract instructions to guide the user in the repair process?

\begin{figure}
    \centering
    \subfloat[][Initial state.]{\label{fig:ex_cdsd0}\includegraphics[scale=0.5]{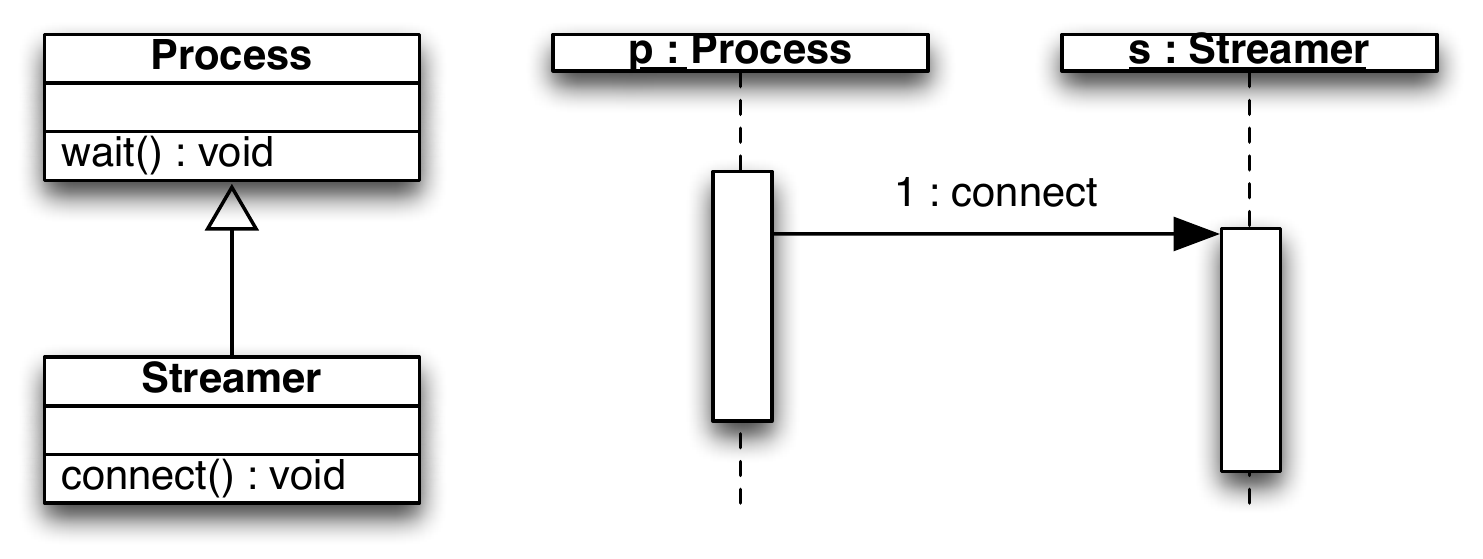}} \qquad
    \subfloat[][Updated state.]{\label{fig:ex_cdsd}\includegraphics[scale=0.5]{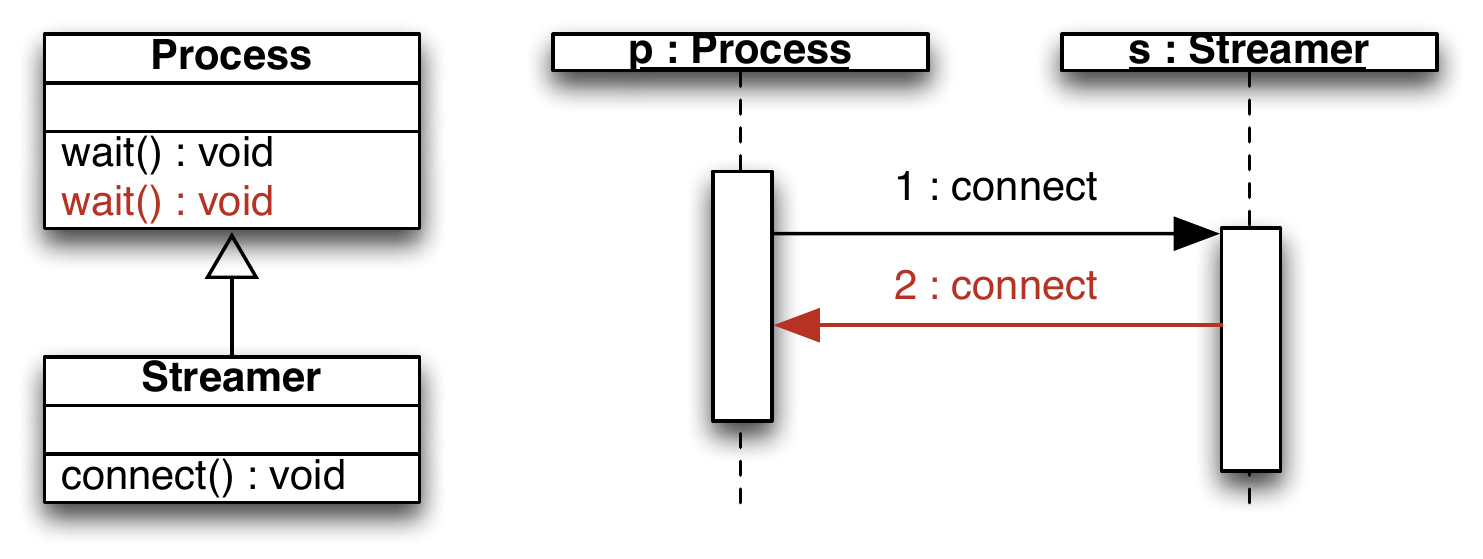}}
    \caption{Inconsistency between the diagrams.}
    \label{fig:ex_}
\end{figure}

The problem becomes more complex when various constraints coexist, which is the common scenario. Consider the coexistence of class and sequence diagrams. Besides internal consistency of the diagrams, consistency between them must also be maintained because some data of the two diagrams overlaps: messages refer to operations that must be available in the target lifeline's class. Since we have assumed that both kinds of diagrams share the same meta-model (much like UML diagrams), these kind of properties can still be defined as regular OCL constraints. This one in particular would take the shape: % \alcino{Being the same meta-model can we really call them inter-model?}
\begin{lstlisting}[language=QVT]
context Message message_operation: 
    self.target.class.operations->
        exists(o | o.name = self.name)
\end{lstlisting}
%\alcino{It is better to move this before the example, so that the user understands why is a new operation with the same name being introduced.} 
These constraints must coexist with those over the individual diagrams. For instance, another constraint that must hold in class diagrams is that the operations defined within a class must have unique names:
\begin{lstlisting}[language=QVT]
context Class unique_operations: 
    self.operations->
        forall(x,y | x.name = y.name => x = y)
\end{lstlisting}

The class and sequence diagrams from Fig.~\ref{fig:ex_cdsd0} are consistent under the constraints that have been defined. However, after two simultaneous updates over these models were performed---the introduction of a new operation and a new message (colored red), as depicted in Fig.~\ref{fig:ex_cdsd}---violations were introduced for both \texttt{message\_operation} and \texttt{unique\_operations}. 

\begin{figure*}
    \centering
    \subfloat[][Operation \emph{wait} renamed.]{\label{fig:ex_cd1}\includegraphics[scale=0.5]{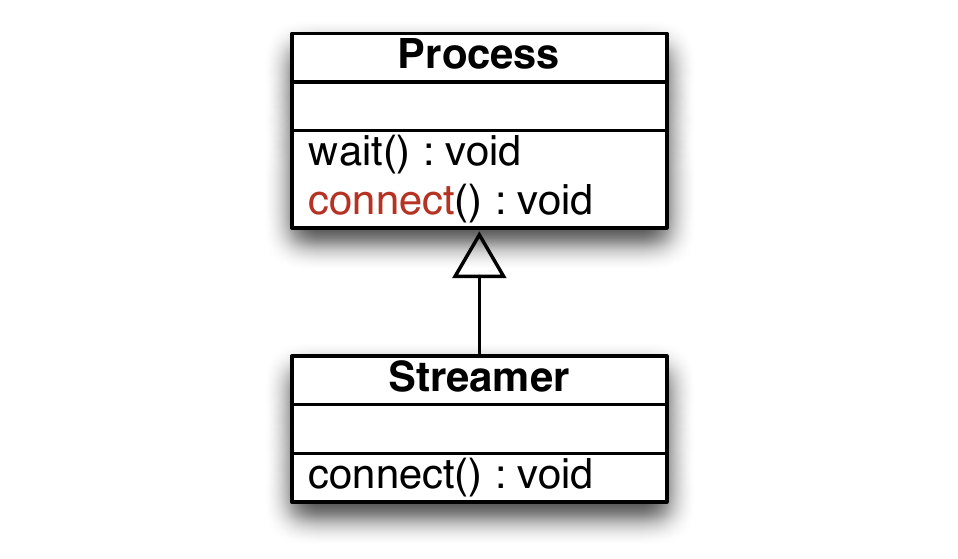}} \qquad
    \subfloat[][Operation \emph{connect} created.]{\label{fig:ex_cd2}\includegraphics[scale=0.5]{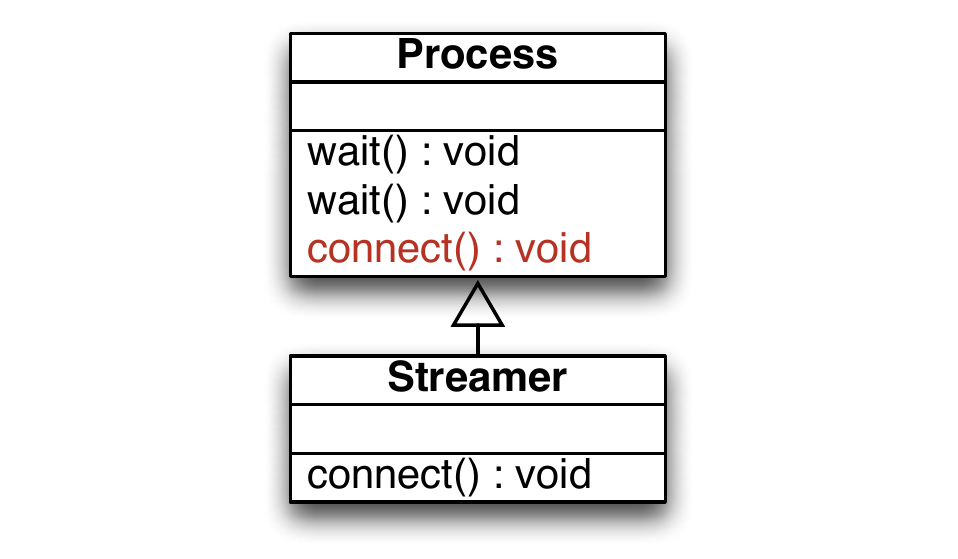}} \qquad
    \subfloat[][Generalization link created.]{\label{fig:ex_cd3}\includegraphics[scale=0.5]{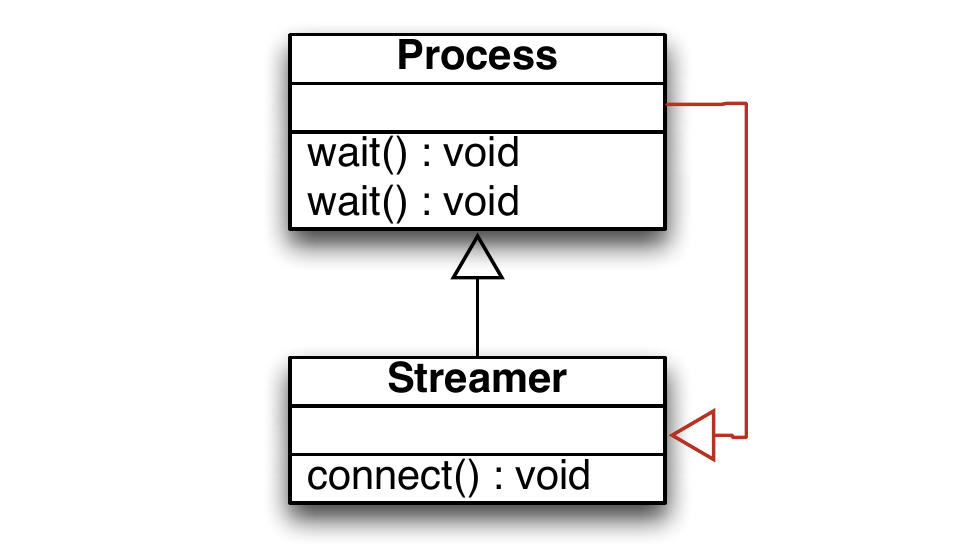}} \\
    \subfloat[][Message \emph{2 : connect} removed.]{\label{fig:ex_sd1}\includegraphics[scale=0.5]{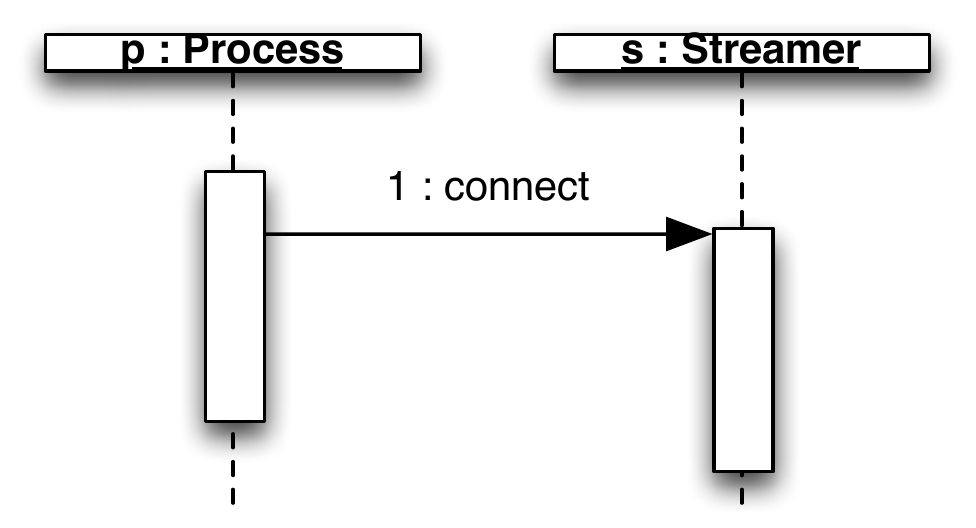}} \qquad
    \subfloat[][Message \emph{2 : connect} renamed.]{\label{fig:ex_sd2}\includegraphics[scale=0.5]{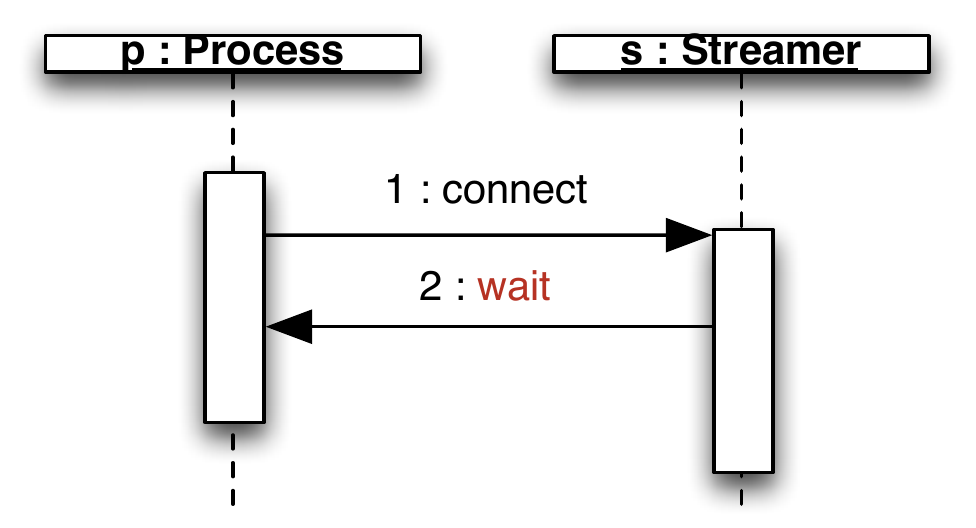}} \qquad
    \subfloat[][\smash{Target of message \emph{2 : connect} moved.}]{\label{fig:ex_sd3}\includegraphics[scale=0.5]{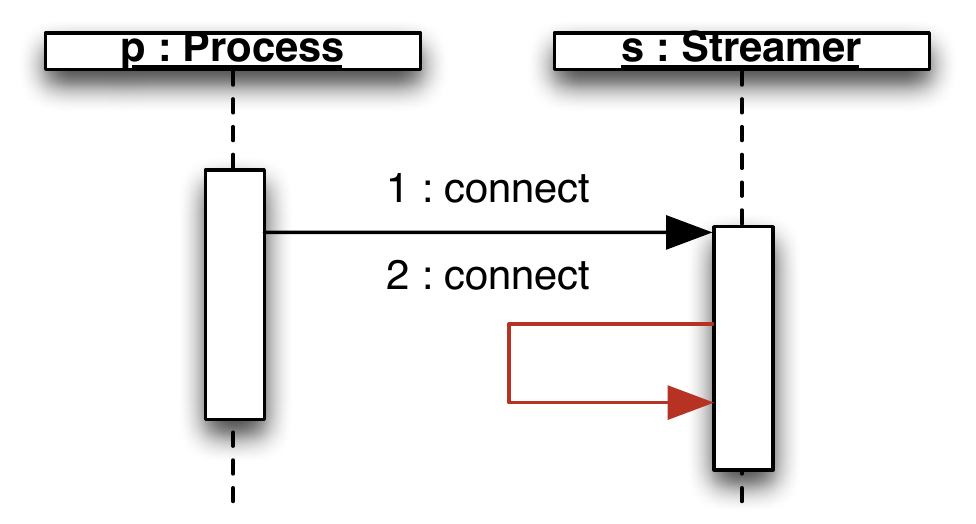}} 
    \caption{Possible repairs for the inconsistency between the diagrams.}
    \label{fix:ex_cdsd123}
\end{figure*}

When attempting to remove the violation of the \texttt{message\_operation} constraint, the developer should be aware of the impact that each of the acceptable repairs has on the other constraints. Figure~\ref{fix:ex_cdsd123} depicts a number of possible repairs that can be applied to the class diagram or to the sequence diagram that remove the \texttt{message\_operation} violation. %\alcino{Better captions to the sub-figures, please! For example, describe the repair: ``delete message x'', ``add operation x'', etc.}
However, some of these updates have (possibly undesirable) \emph{side-effects}: the repair applied in Fig.~\ref{fig:ex_cd1} also solves the violation caused by the \texttt{unique\_operations}---a positive side-effect---while the repair applied in Fig.~\ref{fig:ex_cd3} introduces a new violation by breaking \texttt{acyclic\_generalization}---a negative side-effect. Either way, it is important that the user is aware of these side-effects when choosing the fix to be applied, and thus model repair procedures should somehow consider all inconsistencies when generating the repairs. In this example it is also manifest that the number of valid repairs can quickly become too large for the user to handle. Thus, a variety of techniques have been proposed that try to balance the automation provided by the repair procedures and the required user input that reduces the number of generated repairs. This input includes, for instance, requiring the definition of repair hints for each constraint, assigning different priorities to constraints or model elements, or even disabling some edit operations.

As techniques were developed to handle more complex application domains, more specialized mechanisms to manage their consistency emerged. Such is the case of techniques designed to manage the consistency of models spread across heterogeneous modeling frameworks. A classical example of such scenario is the object-relational mapping, concerned with keeping class diagrams consistent with relational database schemas, so that data conforming to the former can be persisted in databases conforming to the latter. In such cases, unlike the UML sequence and class diagrams of the previous example, overlapping information can not be directly detected, and thus dedicated mechanisms to define inter-model consistency are required, like defining \emph{traceability links} or \emph{consistency relations}, as advocated in MDA's QVT Relations~\cite{QVT:11}. Dedicated to manage inter-model consistency, such techniques often disregard intra-model constraints altogether.

% and would take the following shape: \alcino{I'm not sure about including this example...} \nuno{yes, I was just trying to show that in certain inter-model scenarios simple OCL constraints may not suffice...}\alcino{Still, its going to be difficult to understand without seeing meta-models, etc. Maybe just refer QVT-R without the example.}
%\begin{lstlisting}[language=QVT]
%transformation cd2ds (cd:CD,ds:DS) {
%  top relation ClassToTable { 
%    n:String;
%    domain CD c:Class { 
%      persistent = true, 
%      name       = n };
%    domain DS t:Table { 
%      name       = n };
%    where { A2C(c,t); } 
%  }
%  relation AttributeToColumn { 
%    n:String; a:Attribute; g:Class;
%    domain CD c:Class {} { 
%      (c->closure(general)->includes(g) or g = c) and
%      g.attributes->includes(a) and a.name = n };
%    domain DS t:Table { 
%      column = l:Column { name = n } }; 
%  }    
%}
%\end{lstlisting}

It is easy to envision the complexity of model repair procedures over a considerable number of models and inter-related constraints, giving rise to multiple violations and an overwhelming number of acceptable repairs. Should all violations be removed, a subset of them, or only a certain class? Will the repairs introduce or remove other violations? Will the presentation of the repair alternatives be intuitive and manageable by the user? How can the user guide the generation of repairs so that they prove useful to him? A myriad of solutions has been proposed to address these and other problems. The remainder of this paper will attempt to shed a light on the design landscape of such techniques. 

\subsection{Formalization}
\label{sec:formal}
In order to properly classify model repair techniques, one must first formally define the artifacts which are to be analyzed. This section defines such artifacts in an abstract way, which are instantiated to particular shapes in the following section.

In this study, the MDE environment is considered to consist of a set of $k$ models $m_1,\dots,m_k$ that conform to a set of meta-models $M_1,...,M_k$, a fact denoted by $(m_1, \dots, m_k) \in M_1 \times \dots \times M_k$. In practice this product of meta-models can be seen as a single composed meta-model $\dom$, to which $(m_1,\dots,m_k)$ (usually abbreviated as $\boldsymbol{m}$) conforms. While $\dom$ defines the structural consistency of model instances, semantic properties must be defined by external constraints. Such constraints defined over the meta-models entail the notion of consistent environment state. We denote the universe of constraints supported by a model repair technique by $\cst$, from which the set of constraints $\{c_1,\dots,c_l\} \subseteq \cst$ can be drawn (usually abbreviated as $\boldsymbol{c}$). 
%\tiago{Segundo esta descricao estes design spaces parecem corresponder as gramaticas para exprimir meta-models/modelos/restricoes. Pela seccao 4, no entanto, estes domains abrangem mais coisas certo?} \nuno{hmm, sim, mas nunca dissemos que a sec. 4 apenas ia definir estes dominos, ha coisas que sao externas} 
%\alcino{Not very clear to me what are design spaces. Both $\dom$ and $\cst$ are denoted design spaces, which is confusing. Why are meta-models part of the state? Why are constraints defined over $\dom$? I found this paragraph / formalization quite confusing.} \nuno{revised, hopefully better now}\alcino{Just to be clear, $\dom$ is the set of meta-models in a particular MDE environment, while $\cst$ is the set of supported constraints, not the set of constraints in a particular MDE environment, right? It's better then before, but this non-uniformity still annoys me a bit.} \nuno{yes. but i'm open to suggestions :)}\alcino{Mudei o tipo de letra do M. O mathcal fica para tipos e o bold para tuplos / conjuntos de cenas.}

Prior to being removed, inconsistencies must be detected and diagnosed. Since inconsistencies are introduced by the different stakeholders as the models evolve, information regarding the performed updates may help the checking and repair procedures execute quicker and produce more accurate results. We denote such updates by $u \in \upd$, which, in general, contain at least information about the updated post-state $\boldsymbol{m}' \in \dom$, which can be retrieved by $\mathsf{post}(u)$. \nunoHide{when updates are simply the edit sequence, the post-state may not be defined} For instance, in frameworks that record the user's edit operations, updates may take the shape of a pair $(\boldsymbol{m},s)$, where $\boldsymbol{m}$ is the state of the environment prior to the update and $s$ denotes the applied edit operations. In such cases, the post-state is retrieved by applying $s$ to $\boldsymbol{m}$, i.e., $\mathsf{post}(\boldsymbol{m},s) = s(\boldsymbol{m})$. % \alcino{Since this is not for RAMICS, I think it would be better to use more clear functions like $\mathsf{post}$ or $\mathsf{pre}$}
If available, we denote the operation that retrieves the state of the environment prior to an update $u$ by $\mathsf{pre}(u) \in \dom$. Since inconsistency is expected to be tolerated during development, a pre-state $\boldsymbol{m}$ is not assumed to be fully consistent. %Updates $u,u' \in \upd$ can also be sequentially composed as $u' \circ u \in \upd$. %\alcino{Porque nao usar composicao?} 
%In state-based updates, this simply amounts to the post-state of $u'$, while in operation-based updates this consists of the concatenation of the edit operations from both updates. \tiago{nao percebi bem a ultima frase, o $u'$ cai um bocado do ceu}

Given a user update, a checking procedure will test whether the resulting state is consistent. Such consistency tests are not necessarily boolean, but may return more elaborate reports, like which constraints are being broken or the elements involved in the violations. Such checking reports can be compared for their ``inconsistency level'', e.g., when some violations are removed, the environment becomes ``more consistent'' but may still not be ``fully consistent''. Following the approach proposed by Stevens~\cite{Stevens:14}, we assume these inconsistency levels to form a partially ordered set $(\inc,\sqsubseteq)$. In general, but not necessarily, this partially ordered set has a least element denoting the highest level of consistency for the environment, which will be denoted by $\bot_\inc$. 

\begin{definition}[consistency checking]
    A consistency checking procedure $\chk : \mathbb{P} \cst \rightarrow \upd \rightarrow \inc$ calculates the inconsistency level $i \in \inc$ for an update $u \in \upd$ under constraints $\boldsymbol{c} \subseteq \cst$.
\end{definition}
%
%These techniques are essential to the development of model repair techniques, but their underlying mechanics are outside the scope of this paper.
%\alcino{Acho que o $\cst$ tem que ser $\mathbb{P} \cst$, pois o checker recebe um conjunto de constraints. Idem na definicao seguinte.}

At certain points during the development process, the stakeholders may wish to ameliorate the inconsistency level of the environment, by removing some of the detected violations. This is precisely the role of model repair procedures, whose goal is to, at least, decrease the level of inconsistency of the environment. Again, these techniques may produce repairs in a variety of shapes, whose universe we denote by $\rps$. Although the shape of the repairs $\rps$ is not necessarily the same as the updates $\upd$, it is assumed that from a repair $r \in \rps$ and the user update $u \in \upd$ that led to the current state, a repair update $u' \in \upd$ can be derived that applies $r$ to $u$: otherwise, the consistency checking procedure could not be executed after the application of repairs. For instance, if $u$ is simply represented by the post-state of the environment after user updates, and $r$ is a set of edit operations, the repaired $u'$ can be retrieved by applying the $r$ operations to the $u$ state. We denote this operation by $r(u) \in \upd$. As expected, if $\upd$ contains the pre-state of the update, then $\mathsf{pre}(r(u)) = \mathsf{post}(u)$.
%\alcino{Should the pre-state of $r(u)$ should be the same as the post-state of $u$? If so, clearly state that.}

\begin{figure}[t]
    \centering
    \includegraphics[width=0.4\textwidth]{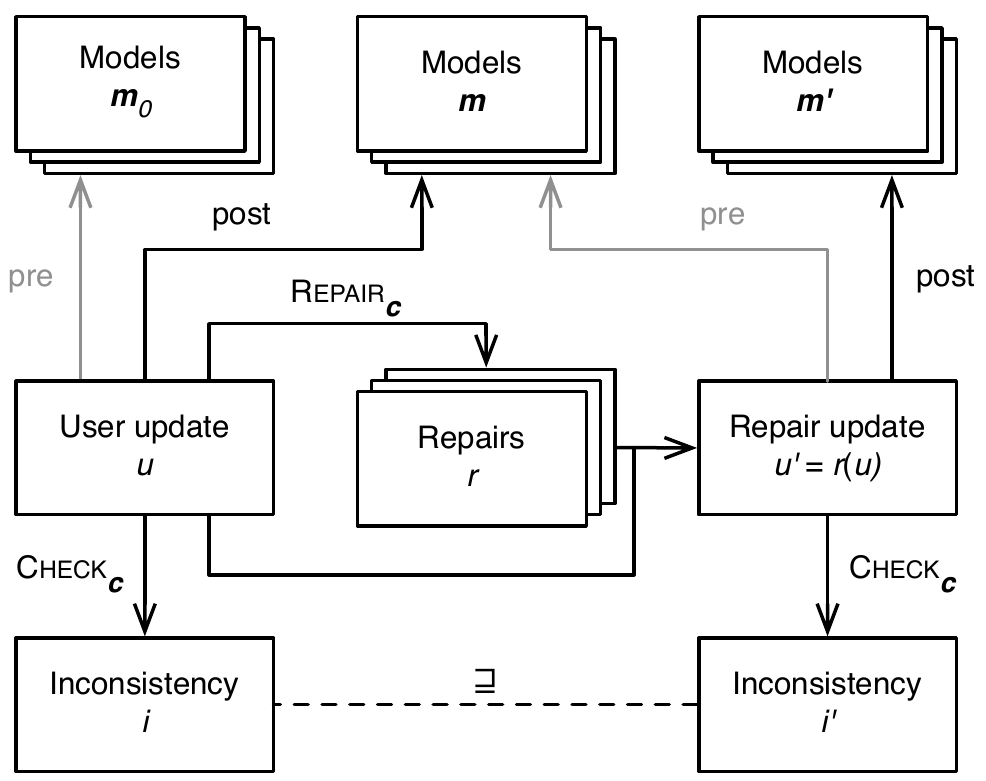}
    \caption{Consistency maintenance scheme.}
    \label{fig:overview}
\end{figure}

\begin{definition}[model repair] \label{def:rep}
    A model repair procedure $\rep : \mathbb{P} \cst \rightarrow \upd \rightarrow \mathbb{P} \rps$ calculates repairs for an update $u \in \upd$ under constraints $\boldsymbol{c} \subseteq \cst$. 
\end{definition}
We assumed that the repair procedure is able to access the checking procedure, and retrieve the inconsistency levels $\inc$ of the states.
The generated repairs do not necessarily recover full consistency, although they are expected to ameliorate the inconsistency level of the environment. %\nuno{warning: ao definir a correctness a partir daquilo que o checker consegue detetar ($\inc$) nao podemos e.g. dizer que o QVT nao e' correto, pq de facto ele nao consegue detetar as inconsistencias intra-modelo} \alcino{I don't think that is a problem $\boldsymbol{c}$ might contain QVT-R constraints together with intra-model OCL constraints.}
%\tiago{talvez seja melhor usar $u'$ no diagrama} 
Figure~\ref{fig:overview} presents our overview scheme for consistency maintenance. User updates $u$ are applied to an existing state $\boldsymbol{m}_0$, updates from which the modified state $\boldsymbol{m}$ can be obtained, and to which the checking procedure assigns an inconsistency level $i$. %\tiago{(ao update, para ser consistente com definicao)}
Given an update, and with access to the checking procedure, the repair procedure generates a set of possible repairs $r$, which, when applied to $u$, result in a repair update $u'$ from which the repaired state $\boldsymbol{m}'$ can be obtained, and whose inconsistency level $i'$ is expected to be at least the same as the one of $u$. (The $\mathsf{pre}$ operations are greyed out because the updates may not keep that information.)
%\alcino{Acho que o $\sqsubseteq$ na figura esta ao contrario.}

%\alcino{Maybe explain a bit the figure.} 
%\alcino{Also, why mention only the post-state of the repair? Maybe you should draw the post-state of the user update (same as pre-state of the repair update, I think) in between the pre-state of the user update and the post-state of the repair update.}
%\alcino{Also the lte arrow bellow is of a different nature of the others which are functions. Maybe draw it using a dashed line. Actually the user arrow is also weird - it should be reversed and labeled with the pre-state operation. And maybe label the update boxes as ``User update'' and ``Repair update'', and replace $u'$ by $r(u')$.}

\section{Feature-based Classification} 
\label{sec:features}

This section presents the collected classification features for model repair approaches, that instantiate the abstract artifacts defined in Section~\ref{sec:formal},  the mechanisms available to the user to customize them, and the behavior of the checking and repair procedures. The features are organized under the following axes:
\begin{description}
    \item[Domain] the domain space $\dom$ and mechanisms to customize it;
    \item[Constraint] the language of constraints $\cst$ and mechanisms to define them; 
    \item[Update] the shape of updates $\upd$; 
    \item[Check] the shape of inconsistency levels $\inc$ and how they are reported by $\chk$;
    \item[Repair] the shape of repairs $\rps$, the behavior of $\rep$ and mechanisms to control it.
%    \item[Deployment] the effective deployment of the of repair technique, if any. \nuno{potentially to remove}
\end{description}

Classification axes are organized as \emph{feature models}, hierarchical models that define the set of valid \emph{configurations} of features that a system may implement. Feature models are typically represented diagrammatically, as defined in Table~\ref{tab:features}. A child feature may only be selected if its parent is also selected. Children features may either be \emph{mandatory} (if the parent feature is selected, so must be the child), \emph{optional} (if the parent feature is selected, the child may or not be selected) or arranged in \emph{or groups} (if the parent is selected, at least one feature of the group must also be selected) or \emph{xor groups} (if the parent feature is selected, exactly one feature of the group must also be selected). Every feature model has a \emph{root} feature  that is always present in every configuration, and may comprise \emph{reference} features which point to other models. Finally, feature models may also be annotated with \emph{requires} and \emph{excludes} constraints that allow cross-tree implications.

\begin{table}[t]
    \centering
    \footnotesize
    \begin{tabular}{c|c|c|c}
        \includegraphics[height=1.1cm]{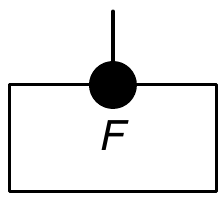} & Mandatory feature & \includegraphics[height=0.7cm]{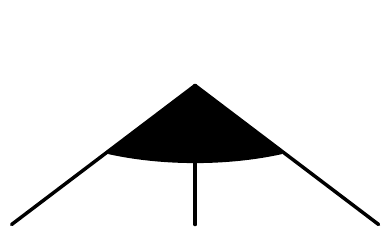} & Or group \\ \hline        
        \includegraphics[height=1.1cm]{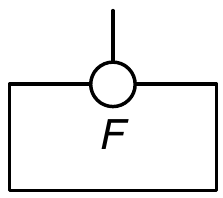} & Optional feature & \includegraphics[height=0.7cm]{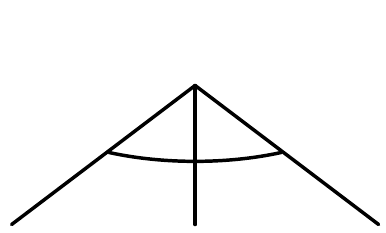} & Xor group \\ \hline
        \includegraphics[height=1.1cm]{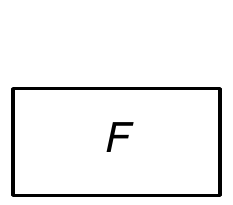} & Root feature &  \includegraphics[height=.9cm]{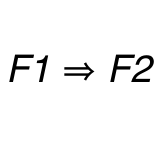} & Requires constraint \\ \hline
        \includegraphics[height=1.1cm]{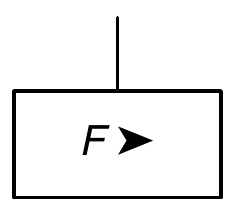} & Reference feature & \includegraphics[height=.9cm]{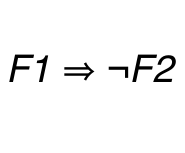} & Excludes constraint
    \end{tabular}
    \caption{Feature definition.}
    \label{tab:features}
\end{table}

\begin{figure}[t]
    \centering
    \includegraphics[scale=0.43]{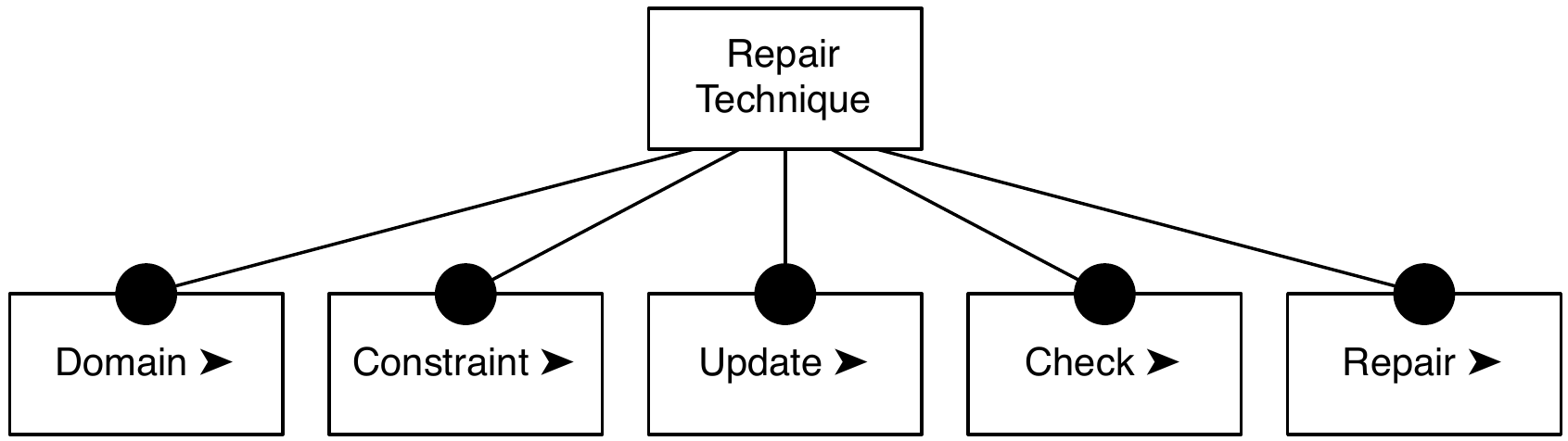}
    \caption{Model repair features.}
    \label{fig:fm_top}
\end{figure}

Our feature model that classifies model repair approaches is depicted in Fig.~\ref{fig:fm_top}, with \emph{Repair Technique} as its root, and a mandatory child feature for every main classification axis, referencing a separate and detailed feature model. These are explored in the succeeding sections.
% Note that the \emph{Deployment} feature is optional, because the approach may not have reached the that level of development.

\subsection{Domain} 
\label{sec:fm_domain}
\begin{figure*}[t]
    \centering
    \includegraphics[scale=0.43]{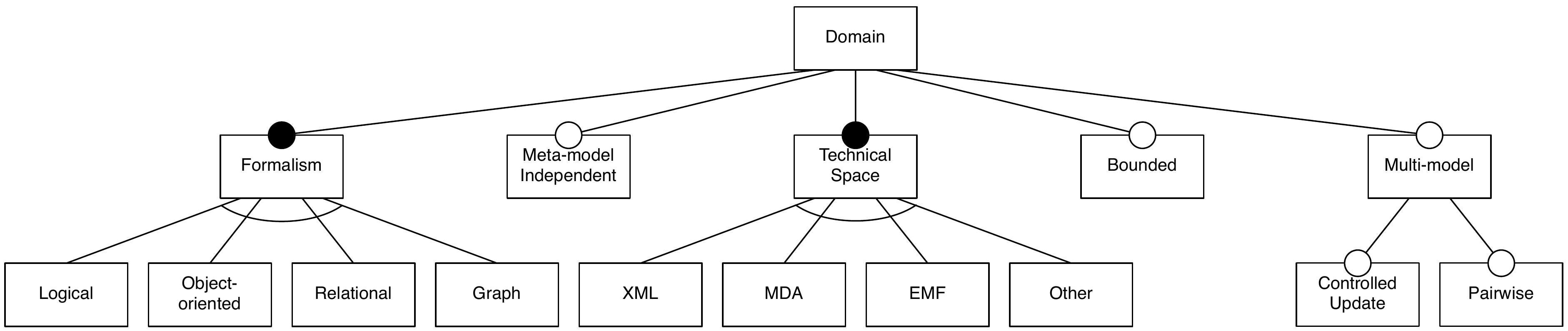}
    \caption{Domain features.}
    \label{fig:fm_domain}
\end{figure*}

The definition of the model space has great implications on the applicability of the technique, since it defines which model artifacts it is able to handle. Moreover, the mechanisms provided to the user to define such space affect the overall flexibility of the technique. The alternatives are explored below and depicted in Fig.~\ref{fig:fm_domain}.

\subsubsection{Formalism} \label{sec:fm_domain_formal} Apart from early human-centered approaches, that do not propose automated systems to manage consistency and consider informally defined artifacts~\cite{Easterbrook:91}, procedures $\chk$ and $\rep$ are designed to handle model instances $\boldsymbol{m}$ from $\dom$ represented using particular \emph{formalisms}. Typical formalisms include \emph{logical} representations in some abstract formal specification language~\cite{FinkelsteinGHKN:94,EasterbrookN:96,NuseibehR:99,StraetenMSJ:03,StraetenD:06,EramoPRV:08,SilvaMBB:10,PuissantMS:10,PuissantSM:13,SchoenboeckKEKSWW:14}, \emph{object-oriented} specifications~\cite{SpanoudakisF:97,LiuEM:02,EgyedLF:08,RederE:12,KolovosPP:08,ChechikLNDESS:09,DamW:10} or the support for \emph{relational} data structures~\cite{StraetenPM:11,MacedoGC:13,CunhaMG:14,MacedoC:14,MacedoCP:14} or \emph{graphs}~\cite{HausmannHS:02,EndersHGTT:02,WagnerGN:03,IvkovicK:04,KonigsS:06,MensSD:06,BeckerHLW:07,AmelunxenLSS:07,GieseW:09,KleinerFA:10,Kortgen:10,HegedusHRBV:11}. The chosen formalism is tightly connected with the kind of properties that the technique is able to check. For instance, reachability properties are more easily handled in relational or graph data structures. Note that, although related, this feature does not directly restrict the technical space on which model artifacts are designed (Section~\ref{sec:fm_domain_model}), which can be internally  converted to the underlying formalism. %\alcino{technical space? talvez ao referir o formato apresentar alguns exemplos (XML, ECore, etc) pois pode ser que o leitor esteja à espera de ver isso discutido nesta seccao - eu estava - e convem ficar claro que nao esta esquecido e que sera abordado no deployment} 

\nunoHide{some techniques focus on particular kinds of models, like agent-oriented design models~\cite{DamWP:06,DamW:11}, enterprise architectures~\cite{DamLG:10} or business process models~\cite{MafaziMS:14}. Can we fit these under object-oriented data structure?} 

\subsubsection{Meta-model Independent}
\label{sec:fm_domain_dependent}
Model repair approaches may aim to be independent of the application domain. Such \emph{meta-model independent} techniques provide the users with mechanisms to define the well-formedness rules of the model instances. This task may be delegated to different agents of the MDE process. For instance, in the ViewPoints framework~\cite{FinkelsteinGHKN:94,EasterbrookN:96} there are two well-defined roles: the designer of the viewpoint, that defines the meta-model, the constraints and the repair plans, and the owner of the viewpoint, that manages the view according to the designer's rules. Meta-model independent techniques~\cite{IvkovicK:04,EramoPRV:08,KolovosPP:08,GieseW:09,XiongHZSTM:09,KleinerFA:10,Kortgen:10,PuissantSM:13,MacedoGC:13,MacedoC:14,SchoenboeckKEKSWW:14} are more customizable and have wider applicability than those whose meta-model is fixed. Techniques with fixed meta-models are designed to act on specific domains, like those proposed to manage the consistency of UML diagrams specifically~\cite{LiuEM:02,StraetenMSJ:03,StraetenD:06,MensSD:06,EgyedLF:08,DamW:10,PuissantMS:10,RederE:12}. While with more limited applicability, knowing the shape of the model artifacts \emph{a priori} may allow the technique to have improved effectiveness and efficiency.

\subsubsection{Technical Space}
\label{sec:fm_domain_model}
This feature defines the \emph{technical space} in which the user is expected to specify the various relevant artifacts. These may be built around standard languages/architectures like \emph{XML}, \emph{MDA} or \emph{EMF}, or \emph{other} specific to the technique. This technical space defines the concrete model syntax that the technique is able to process, like XML~\cite{NentwichEF:03}, XMI~\cite{KleinerFA:10,MacedoGC:13,MacedoC:14}, UML~\cite{LiuEM:02,HausmannHS:02,WagnerGN:03,StraetenD:06,MensSD:06,EgyedLF:08,DamW:10,RederE:12}, or a technique-specific language~\cite{StraetenMSJ:03,PuissantMS:10}. These concrete artifacts are translated by the technique into their representation in the underlying formalism (Section~\ref{sec:fm_domain_formal}).

For meta-model independent techniques, this feature also specifies the meta-modeling language through which the user should specify the meta-models. Under MDA, these are expected to follow the MOF~\cite{MOF:14} standard~\cite{IvkovicK:04,KonigsS:06}, and those under EMF, Ecore\footnote{\url{http://eclipse.org/modeling/emf/}}~\cite{KolovosPP:08,KleinerFA:10,MacedoGC:13,MacedoC:14,SchoenboeckKEKSWW:14}. Again, techniques may not support standard meta-modeling languages, and require the user to define them through technique-specific mechanisms~\cite{PuissantSM:13}.

If the user is allowed to define or customize constraints (Section~\ref{sec:fm_constraint_spec}), this feature defines the language in which he is able to do so. Typically this amounts to some version of MDA's OCL~\cite{IvkovicK:04,KleinerFA:10,DamW:10,DamW:11,MacedoGC:13,SchoenboeckKEKSWW:14}, that is also prescribed in EMF, or it can be designed specifically for the technique~\cite{XiongHZSTM:09}. In techniques with support for inter-model constraints (Section~\ref{sec:fm_constraint_kind}), standard languages include MDA's QVT~\cite{QVT:11} standard~\cite{MacedoC:14,MacedoCP:14}.

\subsubsection{Bounded}  
Techniques may assume a \emph{bounded} universe of model elements, so that the repair procedure can be more manageable\footnote{Note that the domain being classified is effectively the search space available to the repair procedure.}. Such is the case of techniques that do not allow the creation of new elements, and thus are inherently bounded by the elements present in the current inconsistent state~\cite{NuseibehR:99,EgyedLF:08,StraetenPM:11}. Some techniques rely on bounded solvers but guarantee that this is opaque to the user, by iteratively introducing new model elements in the universe~\cite{MacedoC:14,CunhaMG:14}.
\tiagoHide{como o repair proc e que e bounded e nao o domain, fica estranho aqui, percebo que se esta a referir ao domain do repair mas still estranho, ou talvez aproveitar para sublinhar que este domain corresponde ao repair space?} \nunoHide{eu sei, ja esteve do lado do repair, depois eventualmente passou para aqui, mas se calhar gostava mais do outro lado; claro que este eixo refere-se sempre aquilo que os procedimentos conseguem processar, mas podes reforcar; nesta fase nao me apetece mudar :)}%\tiago{ footnote added}

\subsubsection{Multi-model} 
\label{sec:fm_domain_multi}
Model repair techniques may be designed with particular concerns about inter-model consistency and provide dedicated support for \emph{multi-model} scenarios, in which case each state $\boldsymbol{m} \in \dom$ is comprised by a set of multiple model instances. Such is the case of techniques that were developed to manage consistency in development environments with multiple views~\cite{Easterbrook:91,FinkelsteinGHKN:94,EasterbrookN:96,SpanoudakisF:97,GrundyHM:98,OlssonG:02,EndersHGTT:02,EramoPRV:08,MafaziMS:14}, for model synchronization~\cite{IvkovicK:04,KonigsS:06,BeckerHLW:07,KolovosPP:08,GieseW:09,Kortgen:10} and bidirectional~\cite{CicchettiREP:10,MacedoC:14} and multidirectional transformations~\cite{MacedoCP:14}.
By focusing in inter-model consistency, such techniques often disregard the internal consistency of the individual models, leading to overall inconsistent states.

In contrast, techniques may be defined to manage the internal consistency of a single model, in which case a state $\boldsymbol{m}$ consists of a single model instance $m$.
Without dedicated support, these techniques may still handle coexisting models, by merging the various models (and associated meta-models) into a ``dummy'' model conforming to a single meta-model, and expressing their seemingly inter-model constraints with that model's intra-model rules (Section~\ref{sec:fm_constraint_kind}). %\tiago{in the true sense of the word, here ``inter-model" refers to models conforming to distinct meta-models \nuno{nao me parece o melhor sitio para dizer isto}}\tiago{mas penso que convinha clarificar desde inicio} %\tiago{mudei ultima frase, aprovar...} \nuno{nunca vi holistic aplicado neste contexto?}.
Such is the case of techniques that manage the consistency between different UML diagrams, since they share the same meta-model~\cite{LiuEM:02,StraetenMSJ:03,MensSD:06,DamW:10,RederE:12}, as in the example from Section~\ref{sec:overview}. In this classification system, such environments are not considered multi-model (nor their constraints inter-model). 
However, depending on the context, specifying inter-model consistency as an internal constraint may prove to be more cumbersome. Moreover, techniques with native multi-model support may provide a finer control on how these models are repaired (e.g., bidirectional transformations assume that repairs are only applied to a single model). Such behavior can be simulated through distinguished constraints (Section~\ref{sec:fm_constraint_dynamic})---by temporarily introducing a constraint that restricts the valuation of one of the models~\cite{Macedo:14}---or through area selections (Section~\ref{sec:fm_domain_parametrizable})---by focusing the repair on a particular model~\cite{PuissantSM:13}.

%\alcino{Acho que as duas subsubseccoes seguintes deviam ainda ser mais aninhadas para reflectir a natureza hirerarquica do feature model. Talvez usar paragraphs? Idem no restante do artigo.} \nuno{So nao fiz isso antes pq os paragrafos neste template sao a coisa mais horrorosa... mudei para experimentar}
\paragraph{Controlled Update} Approaches with support for multiple models may only allow the user to update the state in a \emph{controlled} manner, typically only allowing updates over a single model~\cite{IvkovicK:04,GieseW:09,ChechikLNDESS:09} so that the propagation to the others is more easily managed. These is common in bidirectional transformation techniques, where the update is propagated from one of the models to the other: allowing concurrent updates could lead to conflicts that could not be resolved.
%\alcino{Acho que isto e uma feature que tem a ver com o domain, debaixo do multi-model (e ja agora nao gosto da desingacao ``single model'' que parece o oposto de multi-model). Acho que o feature group update deveria ser apenas a representation.}

\paragraph{Pairwise}
Techniques may focus on \emph{pairwise} consistency management~\cite{HausmannHS:02,IvkovicK:04,ChechikLNDESS:09,CicchettiREP:10,MacedoC:14}, since managing the consistency between only two models is more manageable. Such is the case of techniques built over \emph{triple graph grammars} (TGGs)~\cite{BeckerHLW:07,GieseW:09,Kortgen:10}. Pairwise consistency management can also be used to render the consistency management of multiple models more manageable~\cite{EndersHGTT:02,FinkelsteinGHKN:94}. However, not every constraint between multiple models can be decomposed into a set of binary constraints~\cite{MacedoCP:14}.

%\alcino{Acho que esta e o tecnhical space nao sao bem features, mas reference features, debaixo das quais deviam estar as varias opccoes, certo? Uma feature nao devia ser algo concreto uma tecnica suporta ou nao?} \nuno{tambem me faz confusao, mas a verdade e' que nao vamos conseguir ser exaustivos nestes casos. na realidade como estava na duvida segui o survey do Czarnecki, onde por exemplo na `Domain Language' e' uma feature mandatory normal.} \tiago{de qql forma tem de ser mandatory se e meta indep.}

\subsection{Constraint}
\begin{figure*}[t]
    \centering
    \includegraphics[scale=0.43]{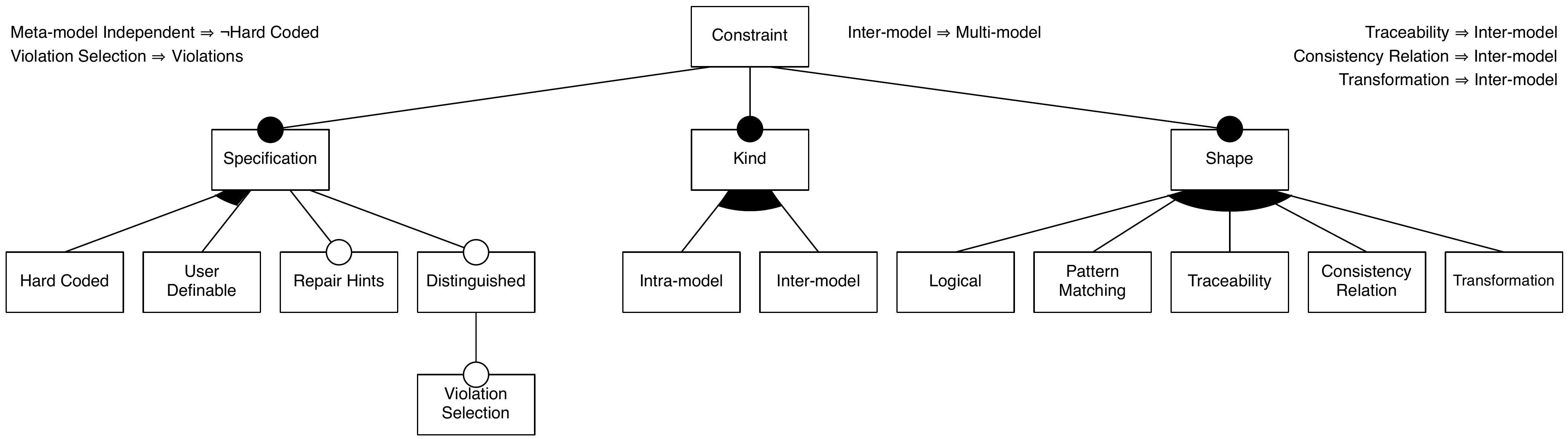}
    \caption{Constraint features.}
    \label{fig:fm_constraint}
\end{figure*}

The expressiveness of the constraint design space entails the class of problems that may be addressed by the technique, while the ability of the user to customize them impacts its general applicability (i.e., the shape of constraints $\boldsymbol{c} \subseteq \cst$ and how they are specified in the framework). These design choices are explored below and depicted in Fig.~\ref{fig:fm_constraint}. For techniques with external checking procedures (Section~\ref{sec:fm_repair_checking}), these features are assumed to regard the design choices of the associated checker, if identified by the authors. %\tiago{those design choices? mas tb tem sobre repair} \nuno{como assim?}

\subsubsection{Specification} \label{sec:fm_constraint_spec}
 Similar to the meta-model (Section~\ref{sec:fm_domain_dependent}), techniques may either have the constraints \emph{hard-coded}~\cite{OlssonG:02,StraetenMSJ:03,StraetenD:06,DamWP:06,AmelunxenLSS:07,EgyedLF:08,SilvaMBB:10,PuissantMS:10,HegedusHRBV:11} or provide the \emph{user} with mechanisms to define or customize them~\cite{IvkovicK:04,PuissantSM:13,KonigsS:06,EramoPRV:08,KolovosPP:08,XiongHZSTM:09,KleinerFA:10,RederE:12,MacedoGC:13,MacedoC:14}. Techniques may even provide a set of pre-defined constraints but allow the user to extend them~\cite{EasterbrookN:96,SchoenboeckKEKSWW:14} or restrict them~\cite{DamLG:10,DamW:10,DamW:11}. Many frameworks delegate such tasks to a repair administrator, rendering the process opaque to the software designer~\cite{FinkelsteinGHKN:94,EasterbrookN:96}.
Techniques that do not allow the user to define new rules are typically paired with fixed meta-model techniques, where both the meta-model and the constraints are fixed \emph{a priori} (techniques for managing consistency of UML diagrams being the classical example). Nonetheless, some techniques with fixed meta-model still allow the user to define additional constraints~\cite{RederE:12}.

\paragraph{Repair Hints}
\label{sec:fm_constraint_hints}
When defining the constraint, the user may be required to define hints on how to repair the model when such constraint is broken~\cite{IvkovicK:04,XiongHZSTM:09}. This contrasts with techniques where the repair procedures are automatically derived from the constraints. The extreme case occurs in rule-based approaches (Section~\ref{sec:fm_repair_core}) where the user is expected to specify the actual resolution rules for the inconsistencies. Although a laborious and error-prone activity that does not provide totality or correctness guarantees, repair hints are the more direct way to allow the user to control the behavior of the repair procedure, one that is tightly coupled with the definition of the constraint.

\paragraph{Distinguished}  \label{sec:fm_constraint_dynamic}
%\tiago{acho que o adjetivo sera distinct e nao distinguished = ilustre} \nuno{era no sentido de forma do verbo distinguish = recognize or treat as different; mas podes mudar e eu depois mudo na imagem}
Techniques may support the definition of \emph{distinguished} constraints that instruct the repair procedure to focus on certain constraints in relation to others. This could allow the user, for instance, to focus on intra-model constraints and instruct the repair procedure to temporarily disregard the inter-model consistency.
%Techniques may allow the user to define temporary constraints on every check and repair execution without actually modifying the set of constraints defined in the environment. Such \emph{focused} executions allow the user to impose constraints over the resulting updates. For instance, bidirectional transformations can also be simulated by defining a constraint that forces the valuation of one of the models to remain unchanged.
%
%Incremental techniques (Section~\ref{sec:fm_repair_incremental}) are typically not able to support such functionality.
%
%\alcino{A explicacao anterior nao e muito clara. E mais uma vez as seccoes nao aparecem pela ordem do diagrama.}
%
Distinguished constraints usually give rise to composite inconsistency levels (Section~\ref{sec:fm_check_composite}), that report the consistency of the environment regarding the different constraints. %Thus, for environment constraints $\boldsymbol{c} \in \cst$ and a distinguished constraint $c$, the checking procedure is typically typed $\chk_{(c,\boldsymbol{c})} : \upd \rightarrow \inc_1 \times \inc_2$, where $\inc_2$ is the shape of the report for the overall technique and $\inc_1$ is the shape of the report for the distinguished constraint.

The most common occurrence of distinguished constraints arises in techniques that allow the user to \emph{select} a specific \emph{violation} to be fixed~\cite{NentwichEF:03,DamW:10,DamLG:10,DamW:11,RederE:12}. Such approaches may be more scalable than resolving all inconsistencies at once by following a spirit of toleration. Rule-based approaches typically handle a single violation at a time, since the resolution hints are defined per constraint~\cite{EasterbrookN:96,LiuEM:02,EndersHGTT:02,HausmannHS:02,WagnerGN:03,StraetenMSJ:03,StraetenD:06,MensSD:06,KusterR:07,EgyedLF:08}.
Violation selection is only available in techniques whose checking procedure returns at least the set of found violations (Section~\ref{sec:fm_check_report}). In such cases, the composite report typically assesses whether the selected violation was effectively removed, and the impact of that repair over the other constraints of the environment. Since typical constraint languages like OCL do not allow the specification of constraints at the model level, violation selection is performed through mechanisms internal to the technique.

% Let $\boldsymbol{c} \in \cst$ be the set of constraints defined in the environment and $\boldsymbol{v} \in \inc_2$ the set of detected violations. Then, for each violation $v \in \boldsymbol{v}$ selected by the user, the checking procedure $\chk_{(v,\boldsymbol{v})} : \upd \rightarrow \inc_1 \times \inc_2$ usually tests whether the violation was fixed and the state of the remainder violations.

\subsubsection{Kind} 
\label{sec:fm_constraint_kind}
General-purpose model repair techniques act on \emph{intra-model} constraints, interpreting the environment as a single model restricted by internal constraints. Nonetheless, techniques that focus on multi-model domains (Section~\ref{sec:fm_domain_multi}) typically support the definition of \emph{inter-model} constraints that relate two or more models. While some of these focus on inter-model consistency and disregard the intra-model constraints, some approaches do consider both kinds of constraints~\cite{LiuEM:02,StraetenMSJ:03,EndersHGTT:02,MensSD:06,DamWP:06,DamW:10,DamW:11,MacedoGC:13}. In such cases, the shape (presented below) of the different classes of constraint may or not be identical.

%The technique may provide dedicated mechanisms to handle such constraints or simply merge the model domains into a single model and convert the inter-model constraint into an intra-model one \tiago{convertem mesmo? alterar para estar de acordo com o que esta atras}. Techniques may rely on the same mechanism \tiago{o de merge? ou queria se dizer common? e aqui sera inter-model no verdadeiro sentido da palavra certo? convem clarificar} to handle both kinds of inconsistencies. \nuno{ok, esta um pouco confuso; a unica coisa que queria dizer e q muitas tecnicas, tipo a nossa, suportam os dois tipos pq por baixo convertem tudo para o mesmo formalismo; mas se calhar nem e' relevante dizer isso; tem que ser reescrita esta seccao}

\subsubsection{Shape} 
The most common means to define intra-model consistency is through the definition of \emph{logical} constraints~\cite{NuseibehR:99,NentwichEF:03,StraetenMSJ:03,StraetenD:06,DamWP:06,KusterR:07,EgyedLF:08,KolovosPP:08,EramoPRV:08,XiongHZSTM:09,KleinerFA:10,SilvaMBB:10,PuissantMS:10,DamW:10,DamLG:10,DamW:11,StraetenPM:11,RederE:12,PuissantSM:13,MacedoGC:13,CunhaMG:14}. %\alcino{Acho que nao e boa ideia usar a designacao ``constraints'' dado que e a designacao da feature de topo tambem, mas nao me ocorre melhor. Talvez ``logic''?}
These may also be used to define inter-model consistency, assuming they are able to refer to elements from different models~\cite{FinkelsteinGHKN:94,EasterbrookN:96,StraetenMSJ:03}. The expressiveness of such constraints is typically that of first-order logic, although they may be extended with other operators like transitive closure to allow the specification of reachability properties~\cite{StraetenMSJ:03,StraetenPM:11,PuissantSM:13,MacedoC:14,CunhaMG:14}.

Approaches built over graph data structures are often based on \emph{pattern matching}, most of the times enhanced with negative application conditions (NACs)~\cite{EndersHGTT:02,LiuEM:02,HausmannHS:02,WagnerGN:03,AmelunxenLSS:07,BeckerHLW:07,MensSD:06,GieseW:09,HegedusHRBV:11}. Pattern matching is well-suited to specify structural properties but not behavioral ones, and as it is not very expressive, some approaches allow the patterns to be attached with additional attribute constraints~\cite{GieseW:09} or imperative code snippets~\cite{AmelunxenLSS:07}. %\alcino{Falta referencia.} 

Techniques with dedicated support for inter-model consistency may rely on the definition of a \emph{traceability}~\cite{Easterbrook:91,EasterbrookN:96,SpanoudakisF:97,OlssonG:02,HausmannHS:02,IvkovicK:04,KonigsS:06,BeckerHLW:07,EramoPRV:08,GieseW:09,ChechikLNDESS:09,Kortgen:10,MafaziMS:14} that connects elements from different models. Constraints~\cite{EasterbrookN:96,EramoPRV:08} or patterns~\cite{EndersHGTT:02,HausmannHS:02,KonigsS:06,BeckerHLW:07,GieseW:09,Kortgen:10} may then be defined over the traceability links that denote the notion of inter-model consistency, although some techniques assume fixed constraints over these links~\cite{MafaziMS:14}. The traceability links may either be explicitly defined by the user~\cite{Easterbrook:91,EasterbrookN:96,EndersHGTT:02,MafaziMS:14}---by manually indicating which elements correspond to each other---or be implicitly introduced either by the repair rules~\cite{HausmannHS:02,KonigsS:06,BeckerHLW:07,GieseW:09,Kortgen:10} or by calculation~\cite{SpanoudakisF:97}. The expressiveness of such techniques depends on the ability to define properties over traceability links, like their multiplicity~\cite{HausmannHS:02}
Another way to define inter-model consistency is through the definition of \emph{consistency relations}~\cite{Meertens:98,CicchettiREP:10,MacedoC:14,MacedoCP:14}, that define which sets of model instances are considered to be consistent with each other.
Finally, some frameworks assume a notion of consistency that is implicitly defined by a \emph{transformation}~\cite{HidakaHIKMN:10}. This is typical in multi-view frameworks with a reference model, from which each view is calculated through transformation.

%The shape of the constraints also of the constraint specification language establishes the classes of inconsistencies that the technique is able to detect. Typically, techniques support first-order logic constraints. Techniques may additionally support transitive closure operators that allow the specification of reachability constraints~\cite{StraetenMSJ:03,StraetenPM:11,PuissantSM:13,MacedoC:14,CunhaMG:14}. More advanced techniques may natively support temporal logic~\cite{x} \tiago{ref missing?}. %\alcino{Porque e que ``first order logic'' ou ``transitive closure'' nao aparecem como sub-features?} \nuno{Como nunca sera exaustivo, nao sei se e' boa ideia especificar...}\alcino{E o mesmo problema que referi antes. E normal deixar este tipo de features nos surveys que usaram feature diagrams?} 
%On inter-model constraints defined as consistency relations or transformations, this axis is determined by the associated specification language. On those relying on traceability, expressiveness regards the ability to define properties over traceability links, like their multiplicity~\cite{HausmannHS:02}. %\nuno{in \cite{SpanoudakisF:97} traceability is an isomorphism}

\subsection{Update} \label{sec:fm_update}
\begin{figure}[t]
    \centering
    \includegraphics[scale=0.43]{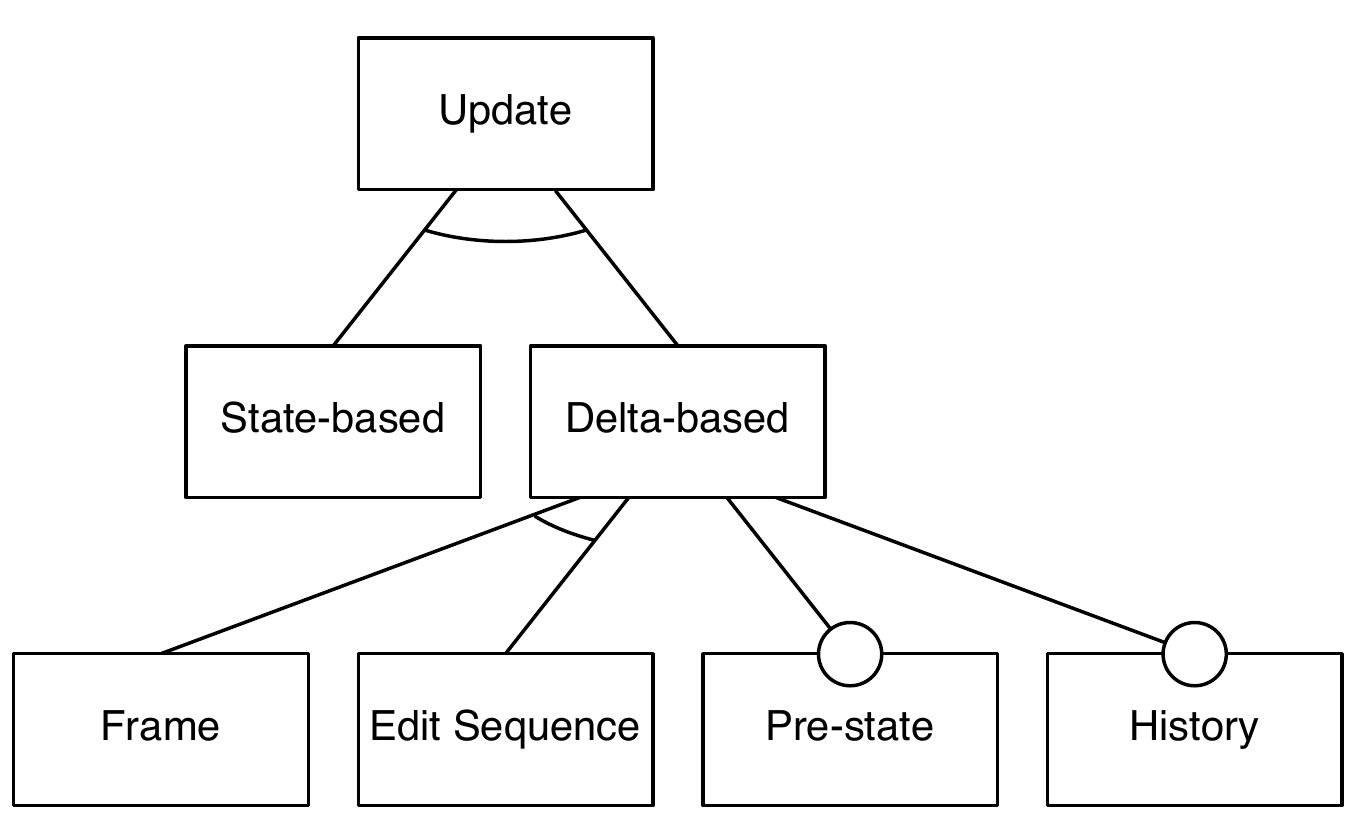}
    \caption{Update features.}
    \label{fig:fm_update}
\end{figure}

Update features define what information is available to the fixing procedures regarding the evolution of the models from the previous known state to the current one, i.e., the shape of the updates $\upd$. These are summarized in Fig.~\ref{fig:fm_update}. %\tiago{o update pode ser state based e operation based ao msm tempo?}

\subsubsection{State-based} \label{sec:fm_update_state} The simplest techniques are purely \emph{state-based}, where the repair procedure simply considers the post-state of the update (i.e., the current state of the environment), in which case updates from $\upd$ simply amount to model instances $\boldsymbol{m}$~\cite{NuseibehR:99,LiuEM:02,HausmannHS:02,NentwichEF:03,StraetenMSJ:03,MensSD:06,BeckerHLW:07,KusterR:07,AmelunxenLSS:07,KolovosPP:08,Kortgen:10,PuissantMS:10,DamW:10,DamLG:10,KleinerFA:10,HegedusHRBV:11,StraetenPM:11,DamW:11,MacedoC:14,CunhaMG:14,SchoenboeckKEKSWW:14}. Such techniques are not able to detect which user actions caused the introduction of inconsistencies. 

%Some techniques are essentially state-based but with operation-based flavor (for instance, in~\cite{PuissantSM:13} model elements are annotated with authorship and versioning information). \tiago{substituir por: for instance, in~\cite{PuissantSM:13}, a model state is not represented by a set of model elements, but rather by a sequence of edit operations} \tiago{e uma sequencia, ordem importa, um estado equivale a um historico}

\subsubsection{Delta-based} \label{sec:fm_update_operation}
In contrast, \emph{delta-based} techniques require information regarding the user actions that led to the current state. These techniques are able to more easily identify problematic portions of the model, but require the online tracking of the applied modifications, which may not be possible in heterogeneous and distributed development environments. Moreover, they may also improve the overall efficiency of the technique, as they allow the identification of which rules must be reassessed after the update.

Some techniques consider a \emph{frame} condition associated with the post-state of the environment that indicates the portion of the state that was effectively modified, allowing the procedure to diagnose inconsistencies more effectively~\cite{EgyedLF:08,GieseW:09,ChechikLNDESS:09,RederE:12}. Alternatively, techniques may require the exact \emph{sequence} of \emph{edit} operations that led to the current state of the environment~\cite{PuissantSM:13,EndersHGTT:02,WagnerGN:03,IvkovicK:04,StraetenD:06,XiongHZSTM:09,MafaziMS:14}. Such techniques may not even have access to the current state of the environment, acting only over the provided edit sequence.
Delta-based techniques may also be provided with the \emph{pre-state} of the environment (i.e., the state $\boldsymbol{m}_0$ prior to the user update)~\cite{IvkovicK:04,EramoPRV:08,PuissantSM:13} in order to more effectively determine the impact of user updates. The technique may also try to derive the frame or the edit sequence from this pre-state, but there is no guarantee that the result will exactly mirror the real user actions. These pre-states are not necessarily consistent, as we assume that inconsistencies are tolerated throughout the development process. Finally, techniques may keep the full \emph{history} of the evolution of the environment~\cite{FinkelsteinGHKN:94,EasterbrookN:96,SilvaMBB:10,PuissantSM:13}, in which case the repair procedure can access not only the most recent update, but also the complete historic. 
%Within operation-based techniques, the update may be represented in a variety of shapes. %Some techniques just require the \emph{traceability} between pre- and post-state elements~\cite{x}, \alcino{Falta referencia.} thus discriminating which elements are identical. 
 %\alcino{Acho que nao e muito claro o que e o delta.} \nuno{tambem estou com dificuldade em definir... o que eu queria era encaixar tecnicas tipo \cite{GieseW:09}, q sao apenas notificadas que certa parte do modelo foi modificada, voltando checkar se ainda esta consistente a partir dai (aparentemente, nao tendo acesso a edit sequence exata)} \alcino{Isto continua a parecer state based, mas so mostra o delta. Seria de colocar no state-based como feature ortogonal as outras duas?} 
 %\alcino{Nao devia haver aqui uns excludes. Ou seja, se e ``edit  operations'' nao deveria ser incompativel com o delta e o traceability, dado que eles sao derivados destes? Idem para o ``record'' em relacao aos outros tres.} \nuno{Ate acho que devia ser um xor.}

\subsection{Check}
\begin{figure}[t]
    \centering
    \includegraphics[scale=0.43]{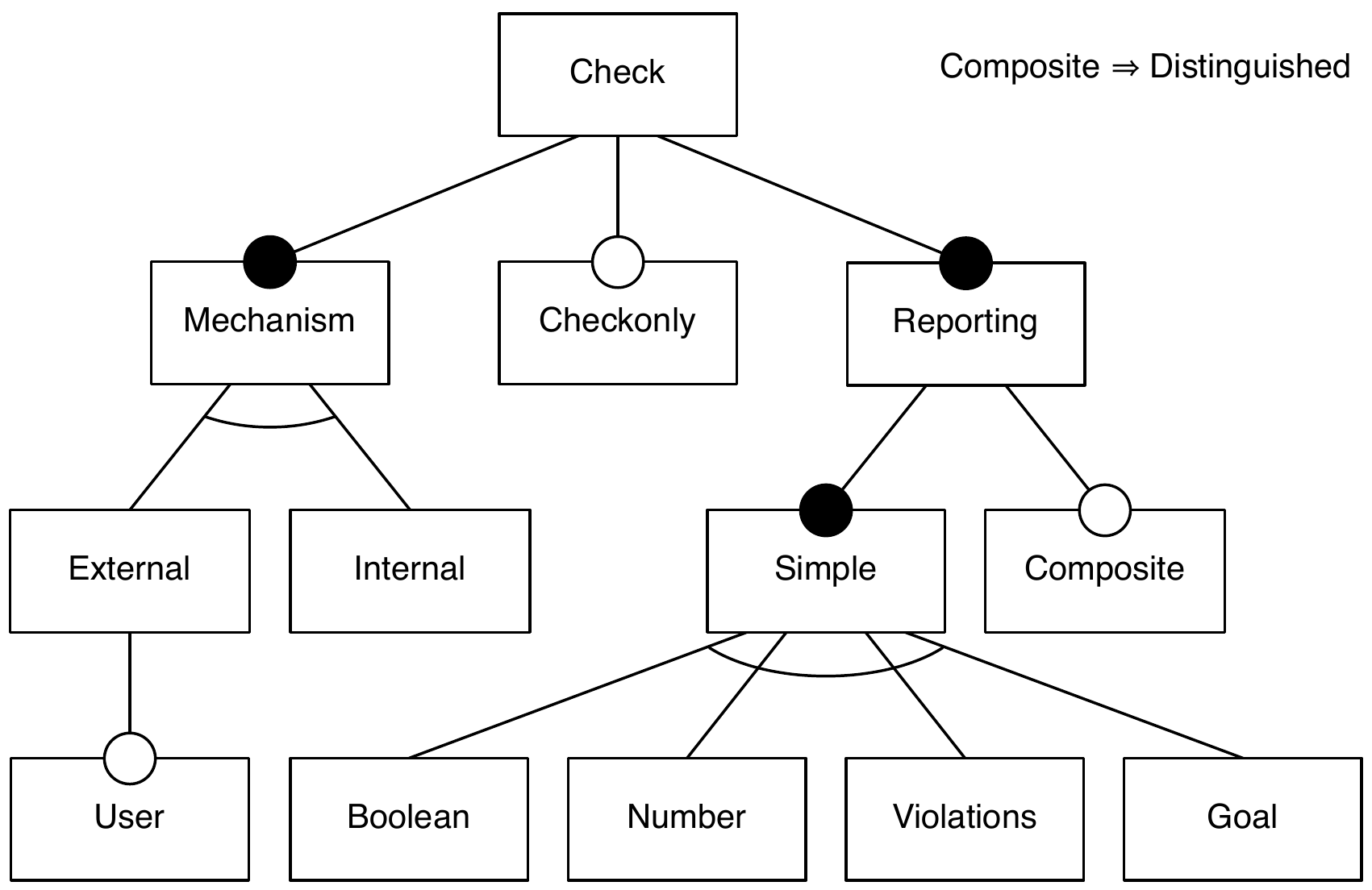}
    \caption{Check features.}
    \label{fig:fm_check}
\end{figure}

Check features regard the model repair technique's reliance on the checking procedure, whose design options are depicted in Fig.~\ref{fig:fm_check}. Since consistency checking is not the focus of this study, these features only classify the relationship between the checking and repair procedures. 

\subsubsection{Checking Mechanism} \label{sec:fm_repair_checking} 
Repair techniques may have the checking procedure as an \emph{internal} or \emph{external} mechanism. The former typically use the checking procedure as a fundamental piece in the repair procedure~\cite{EasterbrookN:96,NuseibehR:99,WagnerGN:03,NentwichEF:03,StraetenMSJ:03,StraetenD:06,MensSD:06,DamWP:06,AmelunxenLSS:07,EgyedLF:08,KolovosPP:08,RederE:12,MacedoC:14,CunhaMG:14,SchoenboeckKEKSWW:14}, while the latter may rely on external tools to detect elements that may be causing the inconsistency~\cite{StraetenPM:11,PuissantSM:13}. While the latter allow the repair procedure to be extensible by deploying state-of-the-art checking procedures, the former typically result in more efficient techniques, since the repair technique can exploit the potential of the checking procedure. Earlier techniques could also rely on the manual identification of the inconsistencies by the \emph{user}~\cite{Easterbrook:91,SpanoudakisF:97}. %\alcino{``User'' devia ser uma feature ou e apenas uma forma particular de external? No ultimo caso tambem poderia ser uma sub-feature de ``external''.}

\subsubsection{Checkonly} Typically the checking and repair procedures are distinct, and thus the user may execute the technique in \emph{checkonly} mode, so that he/she is able to simply check the model for inconsistencies before proceeding to repair it. While not exactly a functionality of model repair techniques, without this feature the user is not aware that the model is in need to be repaired. In techniques that allow violation selection (Section~\ref{sec:fm_constraint_dynamic}) such functionality is fundamental to allow the user to inspect the violations occurring in the current state.

Approaches may not have a proper checkonly mode. For instance, in rule-based approaches~\cite{StraetenMSJ:03,StraetenD:06}  (Section~\ref{sec:fm_repair_core}) the constraint may be simply defined as the pre-condition of the resolution rule\nunoHide{ and in `generative' approaches~\cite{BeckerHLW:07,GieseW:09} amounts to failing to generate the current models}. Nonetheless, rule-based techniques may provide both checking and repair rules~\cite{EasterbrookN:96,EndersHGTT:02,StraetenMSJ:03,WagnerGN:03,KonigsS:06,AmelunxenLSS:07}, some using the checking rules to flag violations whose occurrence enables the application of the repair rules~\cite{LiuEM:02,MensSD:06,DamWP:06,KusterR:07,EgyedLF:08,KolovosPP:08,SilvaMBB:10,SchoenboeckKEKSWW:14}.

\subsubsection{Reporting} \label{sec:fm_check_report}
\emph{Reporting} regards the information provided by the checking procedure about the detected inconsistencies, i.e., the shape of the inconsistency level $\inc$. Techniques may just expect a basic \emph{boolean} procedure that simply reports whether inconsistencies were found. This is typical for solver-based approaches~\cite{MacedoC:14,CunhaMG:14} (Section~\ref{sec:fm_repair_core}). Techniques may instead expect to know the \emph{number} of violations occurring in the current state~\cite{HegedusHRBV:11}. 
Most commonly, the checking procedure returns a set of \emph{violations} detected in the model instances~\cite{Easterbrook:91,LiuEM:02,WagnerGN:03,NentwichEF:03,MensSD:06,KusterR:07,EgyedLF:08,DamWP:06,HegedusHRBV:11,RederE:12,SchoenboeckKEKSWW:14}, usually containing information regarding which constraint is being broken and the model elements involved. Having information about individual violations allows the user to selectively apply repairs (Section~\ref{sec:fm_constraint_dynamic}), unlike with less expressive reports. %\alcino{Nao falam do ``number'' - o que e? Se for o ``inconsistency level'', nao era melhor chamar-lhe isso?}
Techniques may also report a \emph{goal} that must be achieved by the repair procedure. These may take the shape of a formula that is suspected to have rendered a constraint false and which the repair procedure must make true~\cite{SilvaMBB:10,PuissantSM:13} or simply information regarding elements suspect of causing the inconsistency and that must be removed~\cite{NuseibehR:99,StraetenPM:11} or missing model elements that must be created~\cite{EasterbrookN:96}.

Since the shape of the partial order $\sqsubseteq$ over inconsistency levels is dependent on the shape of these reports, this feature is tightly connected with the correctness criteria that the repair technique may be expected to follow (Section~\ref{sec:fm_repair_correctness}). In most cases, there is a single sensible partial to choose from. In boolean reports, this is simply defined as
\begin{equation*}
i \sqsubseteq i' \equiv i \Leftarrow i'
\end{equation*}
just enforcing that a consistent state does not regress into an inconsistent one, with the least element $\bot_\inc = True$. In case of the numerical report, this takes the shape
\begin{equation*}
i \sqsubseteq i' \equiv i \leq i'
\end{equation*}
where $\leq$ is the standard order over naturals, stating that the number of inconsistencies at least does not increase, with $\bot_\inc = 0$. For the list of violations, this simply takes the shape
\begin{equation*}
i \sqsubseteq i' \equiv i \subseteq i'
\end{equation*}
meaning that no new violations are introduced, with $\bot_\inc = \{\}$, the empty set of violations.

\nunoHide{what is the order in goal techniques like~\cite{StraetenPM:11,PuissantSM:13}? they receive some combination of literals from the checker, which is rendered true by the repair procedure...}

\paragraph{Composite} \label{sec:fm_check_composite}

%\alcino{Nao percebo porque e que composite nao faz parte da shape tambem?} \nuno{Da shape da constraint? Calculo que sim, o checks compostos emergiam simplesmente de constraints compostas..}
In some approaches, the checking procedure reports a composite inconsistency level. These emerge from distinguished constraints (Section~\ref{sec:fm_constraint_dynamic}), which are independently checked by the procedure. In the most typical scenario of violation selection, inconsistency levels $\inc$ take the shape $\inc_1 \times \inc_2$, a pair whose first element states whether the selected violation was removed, and the second element provides information regarding the remainder environment constraints, allowing the user to be aware of possible side-effects. Composite reports may also arise when the techniques distinguish different classes of constraints, for instance intra- and inter-model constraints~\cite{MacedoGC:13}. %\alcino{O que significa o subscrito 1?}

In these composite reports there is more than a single sensible partial order over each shape of $\inc$. If both components are deemed equally important, the partial order takes the shape of the product order:
\begin{equation*}
(i_1,i_2) \sqsubseteq (i'_1,i'_2) \equiv i_1 \sqsubseteq i'_1 \wedge i_2 \sqsubseteq i'_2
\end{equation*}
meaning that the inconsistency level is improved if either %\tiago{both?} \nuno{either: $\sqsubset$ se um ficar igual e o outro melhorar}
of the components is.
%\alcino{Nao devia ser sqsubseteq tambem no lado esquerdo?}
The least element of this partial ordered set is simply $(\bot_{\inc_1},\bot_{\inc_2})$. Many, however, prioritize the amelioration of the first component. One of the weaker partial orders in this case is the lexicographic order, under which improvements to the first component allow arbitrary updates on the second one:
\begin{equation*}
(i_1,i_2) \sqsubseteq (i'_1,i'_2) \equiv i_1 \sqsubset i'_1 \vee (i_1 = i'_1 \wedge i_2 \sqsubseteq i'_2)
\end{equation*}
In such case, the least element is still $(\bot_{\inc_1},\bot_{\inc_2})$.
Alternatively, techniques may prioritize the improvement of the first component but disallow damage to the second one. This is typical in techniques that forbid negative side-effects: the selected violation must be removed but no new ones may be introduced in the process. This order can be defined as: %\tiago{e um pouco estranho nao permitir positive side-eff quand $i_1 = i'_1$} \nuno{sim, mas foi a unica maneira de arranjar uma pre-ordem que fizesse o que queremos, pq $\sqsubset$ degenera em $(i_1,i_2) \sqsubseteq (i'_1,i'_2) \equiv i_1 \sqsubset i'_1 \vee (i_1 = i'_1 \wedge i_2 \sqsubset i'_2)$}
\begin{equation*}
(i_1,i_2) \sqsubseteq (i'_1,i'_2) \equiv (i_1 \sqsubset i'_1 \wedge i_2 \sqsubseteq i_2') \vee (i_1 = i'_1 \wedge i_2 = i'_2)
\end{equation*}
This partial order does not have a least element but several minimal elements: once the selected violation is removed, nothing can be done to improve the consistency of the environment. Since these techniques cannot be executed once the selected violation is removed, this actually entails the expected behavior. %\tiago{nao percebi isto dos minimal} \nuno{o problema desta pre-ordem e' que so esta preocupada em reparar a violacao selecionada, mas nao as restantes violacoes; por isso qualquer $i$ em que a selecionada e' corrigida tem a inconsistencia minima}
%\alcino{Talvez acrescentar ``but several minimal elements''.} 
%\alcino{No entanto, isto nao contradiz a formalizacao que dizia que tinha que ter um least element? Talvez seja melhor flexibilizar a formalizacao para dizer que o least element e minimal... } \nuno{Acho que nao formalizacao dizermos que pode nao haver least element.}

\subsection{Repair}
\begin{figure*}[t]
    \centering
    \includegraphics[scale=0.43]{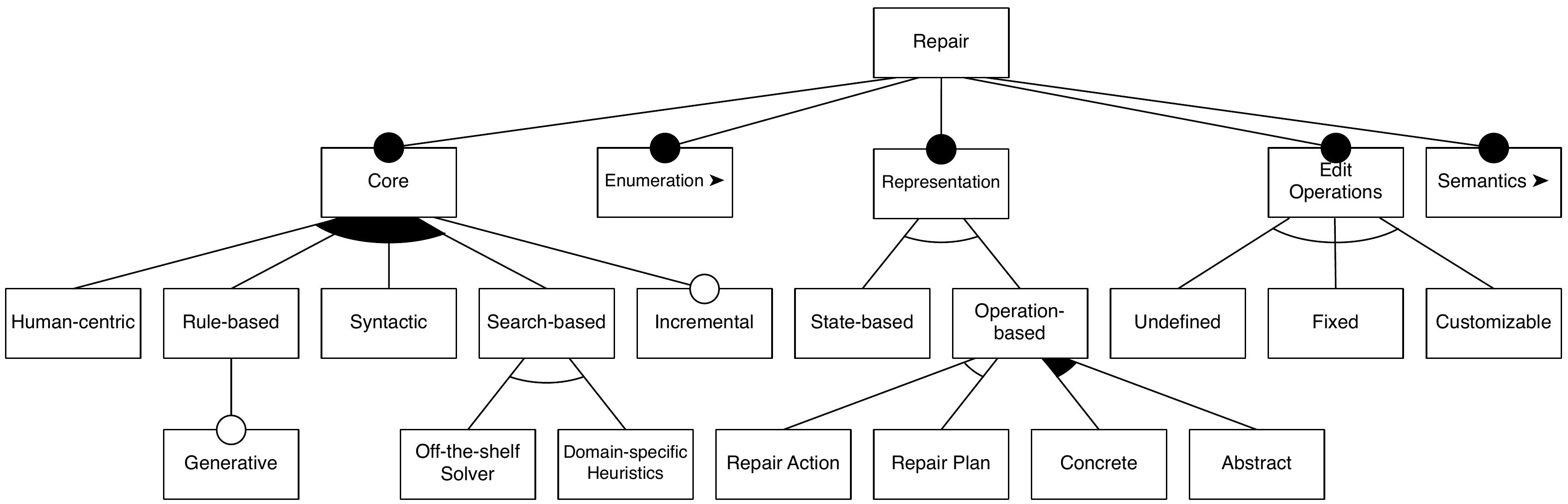}
    \caption{Repair features.}
    \label{fig:fm_repair}
\end{figure*}

These features, depicted in Fig.~\ref{fig:fm_repair}, classify the behavior of the model repair procedure, including the shape and enumeration of the generated repairs, as well as the user's ability to control it. The semantics of the generated repairs are explored in the following section.
%It also explores the possible properties that this procedure may guarantee, like correctness and completeness of the repairs. %\tiago{typo in figure: customazible}

\subsubsection{Core} \label{sec:fm_repair_core}
This feature classifies the engine underlying the repair generation procedure. \emph{Rule-based} techniques~\cite{FinkelsteinGHKN:94,EasterbrookN:96,OlssonG:02,LiuEM:02,EndersHGTT:02,HausmannHS:02,WagnerGN:03,StraetenMSJ:03,StraetenD:06,MensSD:06,DamWP:06,AmelunxenLSS:07,KusterR:07,EgyedLF:08} rely on a set of previously defined rules that are applied whenever an inconsistency is detected. While providing full control over the resolution of inconsistencies, it puts the weight on the designer that must specify how constraints are fixed. Moreover, having a fixed set of resolution rules greatly reduces the flexibility of the technique. \emph{Generative} approaches derive their transformation rules from production rules that define what is a well-formed model~\cite{KonigsS:06,BeckerHLW:07,GieseW:09,Kortgen:10,SilvaMBB:10}. The classical example of such approaches are those based on TGGs, where the rules are derived from the grammar productions. \nunoHide{everything related to these generative approaches must be revised}

In contrast, \emph{syntactic} techniques automatically derive repair plans by syntactic analysis of the constraints~\cite{NentwichEF:03,DamW:10,DamLG:10,DamW:11,RederE:12}. Typically, these repair plans are calculated at static-time and then instantiated to concrete model instances at run-time when an inconsistency is found. While these techniques may be able to generate repair alternatives without user input, the number of generated plans may become overwhelming for the user to choose from. Syntactic techniques are also not well suited to deal with multiple inconsistencies, nor inconsistencies that affect a large portion of the model.

\emph{Search-based} approaches interpret model repair as a model search problem. These are able to automatically find fully-consistent models, but suffer from scalability issues. Moreover, they are well-suited to fix inconsistencies that affect a large portion of the model, like reachability properties. Some approaches rely on \emph{off-the-shelf solvers}~\cite{EramoPRV:08,KleinerFA:10,StraetenPM:11,MacedoC:14,CunhaMG:14} to search for consistent states. These solvers are oblivious of the application domain, and may produce unpredictable solutions. In contrast, other techniques rely on \emph{domain-specific} search procedures~\cite{NuseibehR:99,PuissantMS:10,PuissantSM:13} that rely on domain-specific knowledge, like heuristics and the available edit operations, that allow a finer control on the generation of repairs. % \nuno{develop}\alcino{yep, nao percebi o que e um plan.}

Some hybrid techniques are drawn from more than one of these axes. Such is the case of rule-based approaches that rely on search-based techniques to calculate repair plans from those rules~\cite{SilvaMBB:10,SchoenboeckKEKSWW:14}. Some earlier approaches are \emph{human-centric}, relying on the user to manually flag inconsistencies and propose repairs~\cite{Easterbrook:91}, focusing on the negotiation and education between different stakeholders. %\tiago{nao esta no diagrama, kill?}

\paragraph{Incremental} \label{sec:fm_repair_incremental}
\emph{Incremental} approaches reuse data from previous checking or repair executions, improving efficiency and localization of inconsistencies. 
%\alcino{Nao esta muito clara a diferenca. Talvez ``unlike our definition, incremental techniques are all those that do not just perform batch transformations, that is generate the target model anew from the soure without taking into acount its current state.''} 
Such techniques must be typically run in an online setting so that the required information is preserved between executions. Thus, they are also typically delta-based (Section~\ref{sec:fm_update_operation}) so that this information is more easily managed. Incrementality can be essential to preserve the consistency of the environment---as in the case of TGGs~\cite{HausmannHS:02,KonigsS:06,BeckerHLW:07,GieseW:09,Kortgen:10}, which rely on implicit inter-model traceabilities calculated in previous executions---or simply a mechanism to improve efficiency---as in the technique from~\cite{EgyedLF:08,RederE:12}, that stores the instantiations of the constraints so that inconsistencies can be more efficiently checked and repaired. Frameworks that record the whole evolution of the model instances (Section~\ref{sec:fm_update_operation}) may also be seen as incremental~\cite{FinkelsteinGHKN:94,EasterbrookN:96,SilvaMBB:10} since this history may be used to guide the generation of repairs.

\subsubsection{Enumeration} \label{sec:fm_repair_enum} 
\begin{figure}[t]
    \centering
    \includegraphics[scale=0.43]{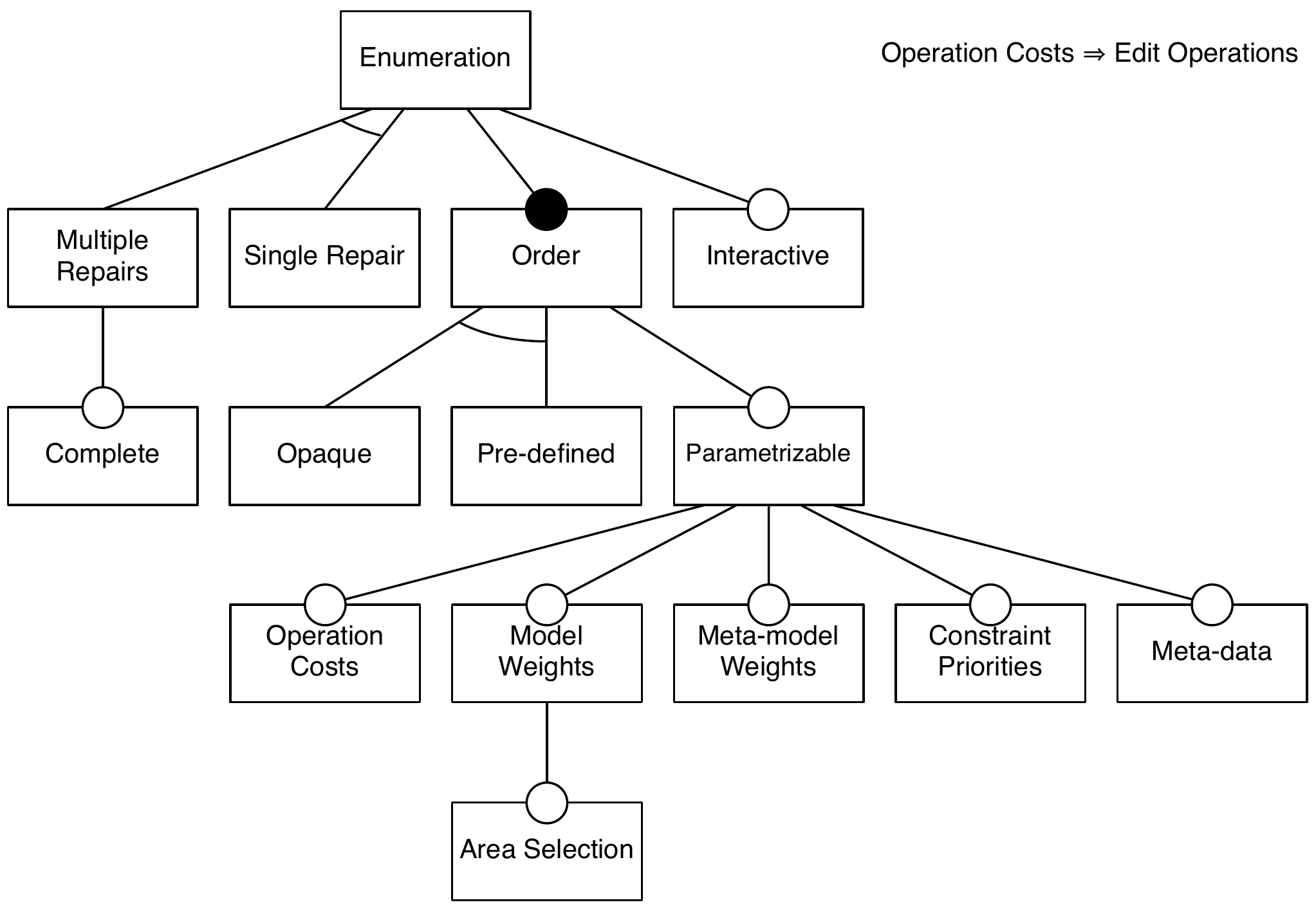}
    \caption{Repair enumeration features.}
    \label{fig:fm_enum}
\end{figure}

This feature defines the mechanism through which the calculated repairs are selected and presented to the user, whose design options are depicted in Fig.~\ref{fig:fm_enum}. 

Since the number of possible repairs may be overwhelming, to be manageable techniques usually restrict themselves to a subset of the acceptable repairs. This may still amount to \emph{multiple} repair options~\cite{EasterbrookN:96,SpanoudakisF:97,LiuEM:02,NentwichEF:03,StraetenD:06,MensSD:06,KusterR:07,AmelunxenLSS:07,EramoPRV:08,KolovosPP:08,DamLG:10,KleinerFA:10,Kortgen:10,HegedusHRBV:11,StraetenPM:11,DamW:11,RederE:12,PuissantSM:13,MacedoC:14,CunhaMG:14,SchoenboeckKEKSWW:14}, although some are able to select \emph{single repairs}~\cite{LiuEM:02,StraetenMSJ:03,IvkovicK:04,BeckerHLW:07,GieseW:09,XiongHZSTM:09,SilvaMBB:10,PuissantMS:10}. The means through which these repairs are selected may or not have been influenced by the user, as will be shown below.

\paragraph{Complete}
Techniques that return multiple repairs are said to be \emph{complete} if they return every possible repair within the parameters of the execution (i.e., the bounds of the search space, the allowed edit operations and any restriction imposed by the enforced semantic properties)~\cite{NentwichEF:03,EramoPRV:08,DamW:10,DamLG:10,DamW:11,MacedoC:14,CunhaMG:14,SchoenboeckKEKSWW:14}. Techniques that are not complete may discard interesting repair alternatives or fail to repair certain inconsistencies. %Techniques that require the user to define repair hints typically suffer from completeness problems. 

\nunoHide{typically, rule-based techniques are said to suffer from completeness issues; however, according to this definition, rule-based techniques are typically complete because their edit operations are precisely the defined rules...}

\nunoHide{\cite{KleinerFA:10,StraetenPM:11} are complete, but within very strict bounds; \cite{PuissantSM:13}?}

\paragraph{Order} \label{sec:fm_order}
This feature regards the \emph{order} through which the repair procedure selects the repairs from among those acceptable. In procedures that return a single repair, this order determines which repair will be selected; in procedures that return multiple repair alternatives, it determines the set of selected repairs as well as the order in which they are produced. This order can be embodied in a distance metric $\Delta : \upd \times \upd \rightarrow \mathbb{N}$ over updates which the procedure tries to minimize.
While related to least-change (Section~\ref{sec:fm_repair_lc}), techniques with ordered repair enumeration that are not complete are not necessarily least-change, as the minimal repair among the generated ones may not be the minimal repair overall.

Although in theory this order always exists for each procedure, it may not have been defined by the developer of the technique, and thus be \emph{opaque} to the user, rendering the procedure unpredictable. Approaches that may not be predictable include those that return a single repair with a solver-based core, that may be sensible to the translation into its language, or rule-based approaches that provide no control on how the rule application is selected (Section~\ref{sec:fm_repair_core}). For instance, in rule-based approaches it may be achieved by establishing a priority order for the rules~\cite{LiuEM:02}, which may be hidden from the user.
Some frameworks with these cores may somewhat circumvent the unpredictability problem by generating every available repair alternative~\cite{MacedoC:14}.

Other approaches have this order \emph{pre-defined}, rendering the technique more predictable and useful. Typically fixed metrics include the graph-edit distance, that counts insertions and removals of model elements, and operation-based distances, that count the number of edit steps between two models, given a set of valid edit operations (Section~\ref{sec:fm_repair_ops}). %\alcino{Estes sao bons exemplos de pre-defined ou fixed orders.}

%Such metric can be used to restrict the generated repairs (Section~\ref{sec:fm_repair_lc}) or to affect the order in which the repairs are enumerated (Section~\ref{sec:fm_repair_order}). More generally, such information could be simply provided to the user to help him choose a resolution from those calculated by the repair procedure~\cite{x}.

%\alcino{Eu acho que ``order'', ``non-deterministic'' e esta e tudo a mesma coisa e deviam estar agrupadas no ``order'' directamente debaixo do enumeration, pois também afecta o single-repair. Talvez debaixo do order ter ``undefined'' (ou ``unpredictable''), ``pre-defined'' (ou ``fixed'') e o ``parametrizable'' que esta aqui.}

%\nuno{aqui nao sei se e' realmente o nao determinismo que queremos classificar, mas em vez disso se e' previsivel para o utilizador qual o repair que vai ser selecionado, ou se esse procedimento e' opaco; nomeadamente, queria encaixar aqui as tecnicas rule-based que nao explicam como selecionam as regras de entre as aplicaveis}

%\alcino{Sim, acho que non-deterministic nao e correcto. Continuo a achar que um solver concreto e deterministico.}
%\alcino{Talvez seja melhor usar ``Predictable'' e de facto fazer o merge com a feature ``order'', pois esta ultima tambem se pode aplicar a single repair.}

%\nuno{\cite{Konigs:09} claims that the selection of the rules in \cite{GieseW:09,GreenyerK:07} is non-deterministic. actually, I can't find any proof of this, but I can't also find how the rules are scheduled either. in the perspective of the user, they might as well be non-deterministic.}

\paragraph{Parametrizable}
\label{sec:fm_domain_parametrizable}
Allowing users to \emph{parameterize} the distance function $\Delta$ enables them to control the behavior of the repair procedure. 
One such way to achieve this, under graph-edit distance, is to assign different weights to different parts of the \emph{meta-model}~\cite{PuissantSM:13,MacedoCG:15}. This allows the user to prioritize changes over certain types of model elements over others. Alternatively, the weights may me assigned directly to the \emph{model} elements, prioritizing changes over concrete parts of the model instances~\cite{PuissantSM:13}. An extreme form of this feature is in \emph{area selection}, in techniques that allow the user to freeze portions of the model instances, as in bidirectional transformation.
Instead of focusing on the models, the user may instead be allowed to control the application of the edit operations (if these are well-defined (Section~\ref{sec:fm_repair_ops})) by attaching them with \emph{costs}~\cite{DamW:10,DamW:11,DamG:14,CunhaMG:14,SchoenboeckKEKSWW:14}.  %Allowing the user to control the set of valid edit operations would affect this distance metric~\cite{MacedoC:14}. % \tiago{corrigir requires para customizable}
Users may also be able to assign different \emph{priorities} over the defined constraints. This provides the user with more information regarding the impact of each possible repair~\cite{KusterR:07} or may be used by the repair procedure to select repairs that best improve the inconsistency level. Such weights can also be used by the checking procedure to return more informative reports. Finally, the user may be able to control the procedure by relying on some additional \emph{meta-data} from the environment, like authoring and versioning information~\cite{PuissantSM:13}.
%\nuno{se o peso sobre os elementos do model estao no `Repair', nao deviam tambem as prioridades das constraints estar?}\alcino{Devia era ser coerente. Ou os pesos dos elementos estao no domain, ou este passa tambem para o repair.}

%\nuno{Shouldn't weights over constraints be here too (Section~\ref{sec:fm_constraint_weight})?}\alcino{Acho que sim.}

%\nuno{Besides action, meta-model and model costs, \cite{PuissantSM:13} also allows weights on authors and versions.}\alcino{Esquece...}

\paragraph{Interactive}
Techniques may rely on an \emph{interactive} dialog with the user to refine the set of possible repairs~\cite{Easterbrook:91,BeckerHLW:07,EramoPRV:08}. This process precedes the actual enumeration of repairs to the user, and may have as a goal the retrieval of a single resolution from the set of acceptable ones~\cite{BeckerHLW:07}. This is not the same as providing the user with abstract repair plans that he must instantiate posteriorly.

%If the technique allows the user is to somehow customize the distance function, it provides an additional mechanism through which he can control the generation procedure~\cite{DamW:10,DamW:11,MacedoC:14,MacedoCG:15,DamG:14}.

\subsubsection{Representation} \label{sec:fm_repair_represent}
This feature regards the actual shape of the artifacts $\rps$ returned by the repair procedure. Techniques may be \emph{state-based}, and simply return the newly generated consistent model~\cite{StraetenMSJ:03,IvkovicK:04,KleinerFA:10,StraetenPM:11,MacedoC:14,CunhaMG:14,SchoenboeckKEKSWW:14}. In such cases, a repair $r \in \rps$ simply amounts to a new state $\boldsymbol{m} \in \dom$.  More elaborate procedures may be \emph{operation-based}, returning instead a set of edit operations to be performed by the user to restore consistency.

Note that the shape of repairs $r \in \rps$ is not necessarily the same as the one of updates $u \in \upd$ (Section~\ref{sec:fm_update}). For instance, it is common for rule-based approaches to consider state-based updates but produce operation-based repairs.

\paragraph{Operation-based} \label{sec:fm_repair_ob}
In operation-based approaches, a repair proposed to the user may take the shape of a \emph{repair action}~\cite{EasterbrookN:96,SpanoudakisF:97,NuseibehR:99,LiuEM:02,EndersHGTT:02,WagnerGN:03,NentwichEF:03,MensSD:06,DamWP:06,StraetenD:06,KolovosPP:08,XiongHZSTM:09,Kortgen:10,MafaziMS:14}, consisting of an atomic edit operation (as defined in Section~\ref{sec:fm_repair_ops}), or of a \emph{repair plan}~\cite{BeckerHLW:07,EramoPRV:08,GieseW:09,SilvaMBB:10,PuissantMS:10,DamW:10,DamLG:10,DamW:11,HegedusHRBV:11,RederE:12,PuissantSM:13}, built from the sequential composition of valid edit operations. Note that this feature does not regard the enumeration of multiple repair alternatives (Section~\ref{sec:fm_repair_enum}) but rather the shape of each particular alternative.

Moreover, these repairs may be \emph{concrete}~\cite{MensSD:06,KolovosPP:08,GieseW:09,XiongHZSTM:09,Kortgen:10,HegedusHRBV:11}, which can be directly applied to the model, or \emph{abstract}, requiring input from the user to be instantiated. The latter typically occur when some repair edit requires a parameter that the fixing procedure is not able (or was not designed) to provide, relying instead on the user to define it~\cite{EasterbrookN:96,SpanoudakisF:97,LiuEM:02,WagnerGN:03,NentwichEF:03,StraetenD:06,DamWP:06,AmelunxenLSS:07,EramoPRV:08,DamW:11,RederE:12,PuissantSM:13}. This makes it prone for the user to provide a value that does not fix the inconsistency, or introduces new ones (which may become common as the complexity of the modeling environment increases).

\subsubsection{Edit Operations}
\label{sec:fm_repair_ops}  
This feature defines the set of edit operations available to the repair procedure to calculate the repair alternatives. 
For state-based techniques, typically those solver-based (Section~\ref{sec:fm_repair_core}), this set may be \emph{undefined}, since the repair procedure simply searches for consistent states.

In rule-based approaches~\cite{EasterbrookN:96,OlssonG:02,HausmannHS:02,LiuEM:02,EndersHGTT:02,WagnerGN:03,StraetenMSJ:03,StraetenD:06,MensSD:06,AmelunxenLSS:07,SilvaMBB:10}, this set amounts to the rules defined in the framework. %\nuno{isto faz sentido?} \alcino{Acho que sim, mas entao sao sempre customizable e devia haver um implicacao entre as duas, certo?} \nuno{nao necessariamente, ha rule-based approaches em que as regras estao fixas}
In contrast, in syntactic and other search-based approaches, this amounts to the set of operations available to the procedure to form repair plans (Section~\ref{sec:fm_repair_ob}). These usually amount simply to creation, modification and deletion operations~\cite{NentwichEF:03}, although some do not allow the creation of elements~\cite{NuseibehR:99}.

\nunoHide{in TGGs~\cite{BeckerHLW:07,KonigsS:06,GieseW:09,Kortgen:10} grammar productions may not delete, so deletes are achieved artificially, meaning that they are not available when defining rules} \nunoHide{in \cite{HegedusHRBV:11} they are hard-coded by are not standard CMD, but domain specific operations}

%\alcino{Eu acho que faltam sub-features - assim ou e customizable ou nada. Para alem do customizable, incluir pelo menos as sub-features que ja sao mencionadas impliciatamente acima: pre-defined (ou fixed) e undefined (tudo num xor, acho eu).}

%\nuno{warning: does this definition render all rule-based techniques complete?}

Although in many techniques this set of valid edit operations is \emph{fixed}~\cite{PuissantSM:13,RederE:12}, the user may also be allowed to \emph{customize} it, either by allowing him to define the set of valid edits~\cite{IvkovicK:04,MacedoC:14} or disable some of those predefined. Such is the case in rule-based approaches~\cite{LiuEM:02,WagnerGN:03} and some syntactic approaches that generate the repair plans for each constraint at static-time~\cite{NentwichEF:03,DamW:10,DamW:11}. 
(This can also be achieved through external means by assigning high costs to edit operations (Section~\ref{sec:fm_domain_parametrizable})).

While techniques may use this set of edit operations to return operation-based repairs (Section~\ref{sec:fm_repair_ob}) to the user~\cite{NentwichEF:03,EgyedLF:08,DamW:10,DamLG:10}, this is not necessarily the case. For instance, techniques may internally use operations to calculate the repaired model but still present state-based repairs~\cite{SchoenboeckKEKSWW:14,MacedoC:14}.

\subsection{Semantics} %\nuno{the final formalization of these properties is pending on the final formalization from Section~\ref{sec:formal}}
\begin{figure}[t]
    \centering
    \includegraphics[scale=0.43]{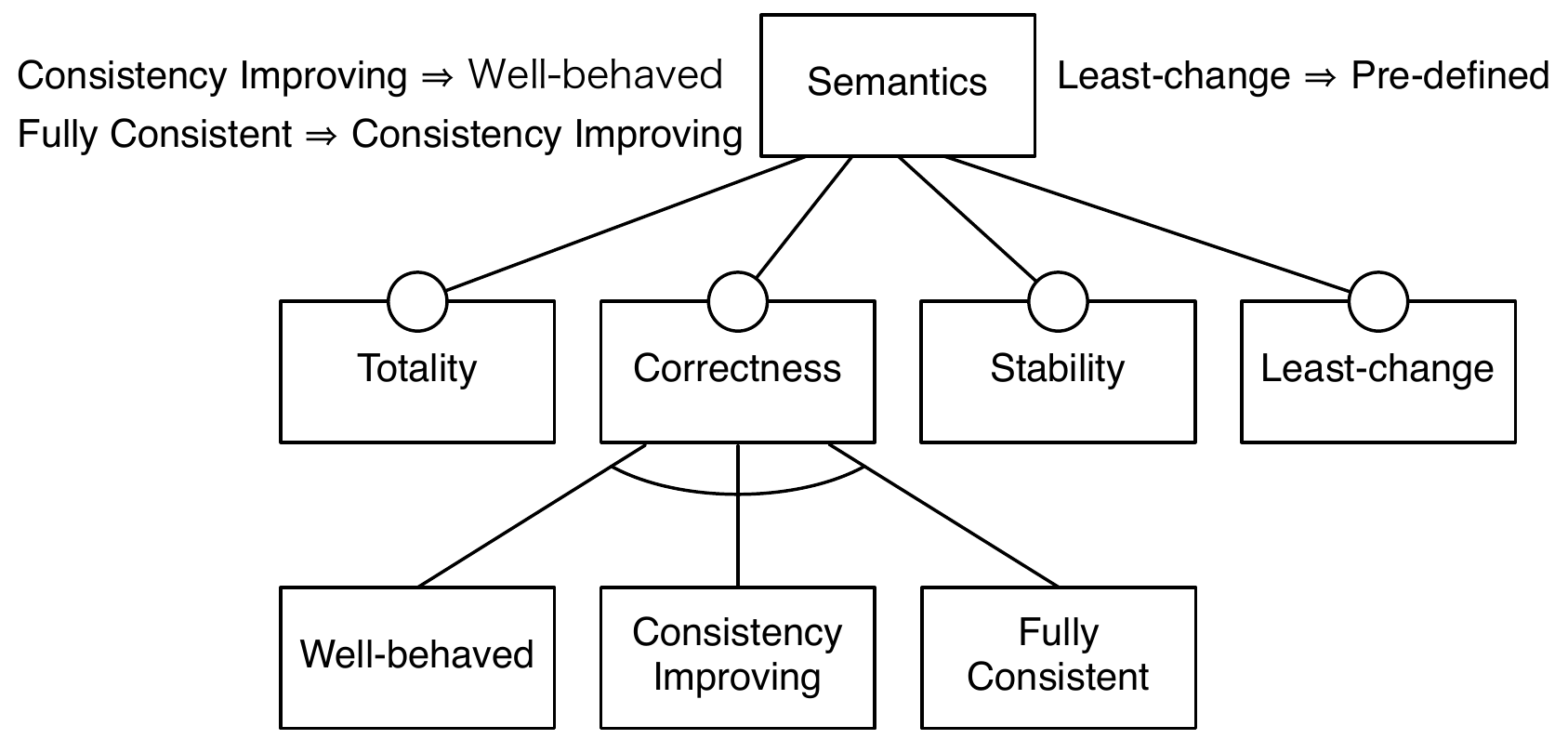}
    \caption{Repair semantics features.}
    \label{fig:fm_semantic}
\end{figure}

This axis explores the \emph{semantic} properties that the repair procedure is guaranteed to follow, which are depicted in Fig.~\ref{fig:fm_semantic}. These properties may be difficult to assess, especially if dependent on user input, like in some rule-based approaches. Thus, we follow a conservative approach and only assume properties explicitly referred by the authors of the technique. %\tiago{kill update preservation}

\subsubsection{Totality} A technique is said to be \emph{total} if for every user update that results in an inconsistent state, it is able to produce a repair (if there is one such repair over the current state). This property can be formalized, for a set of constraints $\boldsymbol{c}$ and an update $u$, in the following manner:
\begin{align*}
    &(\exists u' \in \upd \cdot \mathsf{pre}(u') = \mathsf{post}(u) \wedge \chk_{\boldsymbol{c}} \, u' \sqsubset \chk_{\boldsymbol{c}} \, u) \Rightarrow \\
    &\quad (\exists r \in \rep_{\boldsymbol{c}} \, u)
\end{align*}
meaning that, if there is an update $u'$ from the current state that reduces the level of inconsistency, then the repair procedure will always return a repair alternative. %\tiago{mas n pode ser + inconsistente certo? ou esta a assumir se que pertencendo a REPAIR implica ser well behav plo -?, se sim nota?} \nuno{sao propriedades ortogonais; a totalidade apenas diz que retorna alguma csa (se for possivel); a inconsistencia e' avaliada pelas outras propriedades atras}
We assume that if the updates do not preserve the information regarding the pre-state, then $\mathsf{pre}(u') = \mathsf{post}(u)$ always holds.

The most simple instantiation of this rule occurs in purely state-based approaches with a boolean checking procedure. For a model $m$, it takes the shape:
\begin{align*}
    &(\exists m' \in \upd \cdot \chk_{\boldsymbol{c}} \, m' = True) \Rightarrow \\
    &\quad (\exists r \in \rep_{\boldsymbol{c}} \, m)
\end{align*}
meaning that, if there exists a model that is consistent under $\boldsymbol{c}$, the repair procedure will return a model.
%\alcino{Nao sei se vale a pena estar a dar mais um exemplo de instanciacao. Ja se percebe bem o que e totalidade.}
%If instead we are dealing with sequences of edit operations $(m,s)$ and checking procedures that report a set of violations, this rule takes the following shape:
%\begin{align*}
%    &(\exists (s(m),s') \in \upd \cdot \chk_{\boldsymbol{c}} \, (s(m),s') \subset \chk_{\boldsymbol{c}} \, (m,s)) \Rightarrow \\ 
%    &\quad (\exists r \in \rep_{\boldsymbol{c}} \, (m,s))
%\end{align*}
%Meaning that, for any update $(m,s)$, if there exists an update from state $s(m)$ that reduces the level of inconsistency, then the repair procedure will generate a solution.

Solver-based techniques are naturally total, as they simply search for consistent states~\cite{MacedoC:14,CunhaMG:14}. Rule-based techniques are also total in general, since the detection of the inconsistencies is connected to the resolution rule. Syntactic techniques that focus in single violations at a time are typically total~\cite{NentwichEF:03,RederE:12}, while those that consider every inconsistency at once may encounter conflicts and fail to produce a repair. Approaches with need for repair hints may fail if the user-defined resolutions do not restore consistency~\cite{IvkovicK:04}.

%\nuno{tricky: \cite{CicchettiREP:10} is BX, and (allegedly) only returns inconsistent solutions if there is no consistent solution available; this only happens because only one the models is updated at a time; however, according to this definition, it would already be total, since it needs only to return a solution if there is a valid update}

\nunoHide{\cite{DamW:10,DamW:11} parece total...}

\subsubsection{Correctness} \label{sec:fm_repair_correctness}
Since the goal of repair procedures is to remove inconsistencies from the environment's state, they must provide some correctness guarantees. In fact, we have already defined model repair (Def.~\ref{def:rep}) under the assumption this notion can be formalized by a partial order $\sqsubseteq$ over inconsistency levels $\inc$. As seen in Section~\ref{sec:fm_check_report}, most of the times the shape of $\inc$ entails the partial order. The correctness of a repair procedure is defined from its behavior in relation to this partial order.

%\alcino{Antes ja tinha sido definido o que e uma tecnica ``well-behaved'', mas agora esse conceito nem sequer e mencionado? Qual a relacao com a correctness? Nao era melhor voltar a aparecer aqui uma feature designada ``well-behaved'' com a mesma lei de antes? Ou retirar isso da formalizacao e colocar apenas aqui. Presumo que nem todas as tecnicas sejam sequer well-behaved, pelo que ter isso como feature e muito relevante.}

\paragraph{Well-behaved}
A model repair procedure is said to be \emph{well-behaved} if the inconsistency level at least does not increase whenever one of these repairs is applied, i.e., 
\begin{equation*}
    \forall r \in \rep_{\boldsymbol{c}} \, u \cdot \neg (\chk_{\boldsymbol{c}} \, u \sqsubset \chk_{\boldsymbol{c}} \, r(u))
\end{equation*}
%\nuno{$\neg \chk_{\boldsymbol{c}} \, u \sqsubset \chk_{\boldsymbol{c}} \, r(u)$ or $\chk_{\boldsymbol{c}} \, r(u) \subseteq \chk_{\boldsymbol{c}} \, u$? I chose the (looser) former since it allows incomparable updates (for instance, violations $\{v_1,v_2\}$ may turn into $\{v_3\}$.) }
This is the minimal correctness behavior expected from a repair procedure. For instance, in boolean procedures, this means not turning completely consistent environments into inconsistent ones.

\paragraph{Consistency Improving}
%\alcino{Nao gosto dos nomes ``partial correctness'' e ``full correctness'', talvez ``consistency improving'' e ``fully consistent'', ou qq coisa do genero...} 
\emph{Consistency improving} procedures guarantee that the state of the environment is effectively ameliorated, reducing its inconsistency level (unless it is already at a minimum inconsistency level).  For a set of constraints $\boldsymbol{c}$ and update $u$, this property can be specified as:
\begin{align*}
    &\forall r \in \rep_{\boldsymbol{c}} \, u \cdot \\
    &\quad \chk_{\boldsymbol{c}} \, r(u) \sqsubset \chk_{\boldsymbol{c}} \, u \vee \neg \exists i \in \inc \cdot i \sqsubset \chk_{\boldsymbol{c}} \, r(u)
\end{align*}
Consistency improving procedures are always well-behaved. % \tiago{indicar no diagrama?}. 
If there is a single minimal inconsistency level $\bot_{\inc}$, then it can be simplified as:
\begin{align*}
    &\forall r \in \rep_{\boldsymbol{c}} \, u \cdot \\
    &\quad \chk_{\boldsymbol{c}} \, r(u) \sqsubset \chk_{\boldsymbol{c}} \, u \vee \chk_{\boldsymbol{c}} \, r(u) = \bot_{\inc}
\end{align*}
 %\nuno{aqui tbm tem q ser adaptado para quando nao ha $\bot$}

Under boolean checking procedures this property degenerates into fully consistent procedures, defined below. Under more expressive checking procedures, like those reporting a set of violations, this behavior may occur in techniques that attempt to fix violations until a certain threshold is reached~\cite{SilvaMBB:10}. Under composite inconsistency levels (Section~\ref{sec:fm_check_composite}) this is common in techniques that are only concerned with a certain class of constraints (e.g., techniques dedicated to handle inter-model constraints may disregard intra-model constraints), or that support violation selection but do not enforce the fixing of the remainder environment constraints. %\tiago{ref missing?}

%The easiest way to envision this property is in frameworks that are able to detect a set of inconsistencies. In these cases, the partial order over inconsistency levels can be defined as:
%\begin{equation*} i_1 \leq i_2 \Rightarrow i_2 \not \subseteq i_1 \end{equation*}
%That is, a set of inconsistencies $i_1$ is considered ``less inconsistent'' than another set $i_2$ if there is at least one inconsistency in $i_1$ no loner present in $i_2$. Note that this says nothing regarding the introduction of other inconsistencies.

\paragraph{Fully Consistent}

%\alcino{E preciso compatibilizar isto com o facto de por vezes nao haver least element, mas um conjunto de minimais. Pelo menos dizer que a lei aqui so se aplica quando ha um least element. Tambem nao e dificil adaptar para dizer que o resultado e minimal (e consequentemente o least element se so houver um minimal).}
Procedures are said to be \emph{fully consistent} if they guarantee that the inconsistency level is always reduced to a minimum, i.e., for every update $u$ and set of constraints $\boldsymbol{c}$:
\begin{equation*}
    \forall r \in \rep_{\boldsymbol{c}} \, u \cdot \neg \exists i \in \inc \cdot i \sqsubset \chk_{\boldsymbol{c}} \, r(u)
\end{equation*}
Fully consistent procedures are always consistency improving. % \tiago{indicar no diagrama?}. 
In case there is a least element $\bot_{\inc}$ in the inconsistency level, the law degenerates into
\begin{equation*}
    \forall r \in \rep_{\boldsymbol{c}} \, u \cdot \chk_{\boldsymbol{c}} \, r(u) = \bot_{\inc}
\end{equation*}

The impact of this property depends on the minimal elements of the partially ordered set $\inc$. For instance, under boolean checking procedures, this amounts to setting the result to true, while under procedures that return a set of violations, this amount to fixing every violation (including possible negative side-effects).
This is the typical behavior of search-based approaches, that resolve all consistencies at the same time~\cite{PuissantMS:10,KleinerFA:10,StraetenPM:11,MacedoC:14,CunhaMG:14,SchoenboeckKEKSWW:14}. 
Note that the definition of correctness is orthogonal to totality. 
%\alcino{Queres dizer ``orthogonal to totality''?}
This means that procedures that fail to produce repairs are still deemed correct. In fact, some techniques enforce correctness by simply failing if consistency is not recovered after the repair procedure is executed~\cite{IvkovicK:04,XiongHZSTM:09}. 

Fully consistent procedures are not necessarily desirable, as the model may need to undergo inconsistent states before fully recovering consistency~\cite{EasterbrookN:96}.

\nunoHide{Can `generative' approaches~\cite{GieseW:09,Kortgen:10} be correct? I think this would require the set of rules to handle every possible pattern... However, this overlaps with totality...!}

\subsubsection{Stability} A technique is said to be \emph{stable} if for every update that does not result in an inconsistent state, it returns null repairs~\cite{XiongHZSTM:09,PuissantSM:13,MacedoC:14,CunhaMG:14,RederE:12}. For an user update $u$ and constraints $\boldsymbol{c}$, this property can be formulated as:
\begin{align*}
    &\chk_{\boldsymbol{c}} \, u = \bot_\inc \Rightarrow \\
    &\quad  \forall r \in \rep_{\boldsymbol{c}} \, u \cdot \mathsf{post}(r(u)) = \mathsf{post}(u)
\end{align*}

In purely state-based approaches with boolean checking procedures, this degenerates into the following property, for a model $m$:
\begin{align*}
    &\chk_{\boldsymbol{c}} \, m = True \Rightarrow \\
    &\quad \forall r \in \rep_{\boldsymbol{c}} \, m \cdot r(m) = m
\end{align*}
\tiagoHide{tinha m' em vez de r, u passa a m mas REPAIR da sempre r certo?, mudei... (e nos outros locais tb)} \nunoHide{nao, a ideia e' que em state-based os repairs sao apenas modelos, por isso e' que dava para simplificar!}\tiagoHide{mas achei confuso, penso que ficara melhor com r}
%\alcino{Idem.}
%In operation-based approaches that return a set of violations, for an update $(m,s)$ it is instead defined as:
%\begin{align*}
%    &\chk_{\boldsymbol{c}} \, (m,s) = \emptyset \Rightarrow \\
%    &\quad \forall r \in \rep_{\boldsymbol{c}} \, (m,s) \cdot r(s(m)) = s(m)
%\end{align*}
%\nuno{ha aqui um abuso de notacao: pela definicao, $r(s,m) \in \upd$, mas estamos a assumir que como $s(m) \in \dom$, $r(s(m)) \in \dom$.}

Rule-based techniques are naturally stable, as the resolution rules are not applied unless inconsistencies are detected. Techniques are not stable if they apply update procedures regardless of the models being consistent~\cite{IvkovicK:04}.

\subsubsection{Least-change}
\label{sec:fm_repair_lc}
The principle of \textit{least-change} requires repaired models to be as close as possible to the original, according to some defined model metric $\Delta : \upd \times \upd \rightarrow \mathbb{N}$ (meaning that it must not be opaque, Section~\ref{sec:fm_order}), possibly customized by the user (Section~\ref{sec:fm_domain_parametrizable}). This renders the approach more predictable to the designer since the set of selected repairs is well-defined~\cite{CunhaMG:14}. However, while most approaches informally and loosely approximate this intuition using \emph{ad hoc} or heuristic mechanisms, providing least-change guarantees is a complex task~\cite{DamW:10,DamLG:10,DamW:11,MacedoC:14,CunhaMG:14,DamG:14}. In general, this technique is formalized as follows, for an update $u$ and constraints $\boldsymbol{c}$:
\begin{align*}
    &\forall r \in \rep_{\boldsymbol{c}} \, u \cdot \\
    &\quad \forall r' \in \rps \cdot \chk_{\boldsymbol{c}} \, r'(u) = \chk_{\boldsymbol{c}} \, r(u) \Rightarrow \\
    &\quad \quad \Delta(r(u),u) \leq \Delta(r'(u),u)
\end{align*}
Meaning that, compared with the repairs that are equally consistent, the returned repairs are closer to the current state of the environment. %\nuno{sera que so devia comparar com os $\chk_{\boldsymbol{c}} \, r'(u) = \chk_{\boldsymbol{c}} \, r(u)$?}\alcino{Pois, estava mesmo a pensar nisso... diria que sim.}
In purely state-based approaches, this degenerates into the following property, for a model $m$ and constraints $\boldsymbol{c}$:
\begin{align*}
    &\forall r \in \rep_{\boldsymbol{c}} \, m \cdot \\
    &\quad \forall r' \in \rps \cdot \chk_{\boldsymbol{c}} \, r(m) \Rightarrow \chk_{\boldsymbol{c}} \, r'(m) \Rightarrow  \\
    &\quad \quad \Delta(r(m),m) \leq \Delta(r'(m),m)
\end{align*}
%\alcino{Idem.}
%For operation-based updates $(s,m)$, this takes the shape:
%\begin{align*}
%    &\forall r \in \rep_{\boldsymbol{c}} \, (m,s) \cdot \\
%    &\quad \forall r' \in \rps \cdot \chk_{\boldsymbol{c}} \, r'(s(m)) \subseteq \chk_{\boldsymbol{c}} \, r(s(m)) \Rightarrow  \\
%    &\quad \quad \Delta(r(s(m)),(m,s)) \leq \Delta(r'(s(m)),(m,s))
%\end{align*}

\nunoHide{~\cite{SilvaMBB:10} is least-change in the following way: the procedure is guaranteed to remove the smaller number of problematic (recent) elements; this means that it may be possible to achieve consistency by removing a smaller number of elements, but these would be older than the selected ones.}

If the identity of indiscernibles holds for the distance function ($\Delta(m,m') = 0 \equiv m = m'$), then least-change entails stability. Otherwise there are minimal updates other than the null update.

\nunoHide{\cite{XiongHZSTM:09}}

%%%%%%%%%%%%%%%%%%%%%%%%%%%%%%%%%%%%%%%%%%%%%%%%%%%%%%%%%%%%%%%

\section{Classifying techniques}
\label{sec:classes}

Throughout the paper, the proposed features were supported with references to various papers, not only as a way of providing a rich collection of related work, but also to ensure that only literature supported features were included in the taxonomy.
In this section we attempt to validate the proposed features by classifying three recent and very distinct model repair approaches in terms of our feature model. The major purpose here is to demonstrate that classifying techniques through our feature model helps in obtaining structured and complete descriptions which allow a better understanding and clear comparison of different approaches. Although some features may be difficult to assess for a particular technique (mainly due to lack of information or ambiguity), we end up with quite complete profiles with good taxonomy coverage and, most importantly, drawn from a common view point, making similarities and differences more obvious. Notice that, for the sake of brevity, in most cases we do not mention optional features which do not apply to the approach addressed.

To better illustrate the differences between the techniques, as well as the impact of the design decisions, we define a simple running example on which they are applied. Since we could not access the implementations of all these
approaches, in order to define our profiles and infer actual repairs, we resorted to the explicit information available in the literature and to our understanding of the techniques after a thorough study.

\begin{figure}[t]
    \centering
    \subfloat[][UML class diagram.]{\label{fig:cd_example}\includegraphics[scale=0.5]{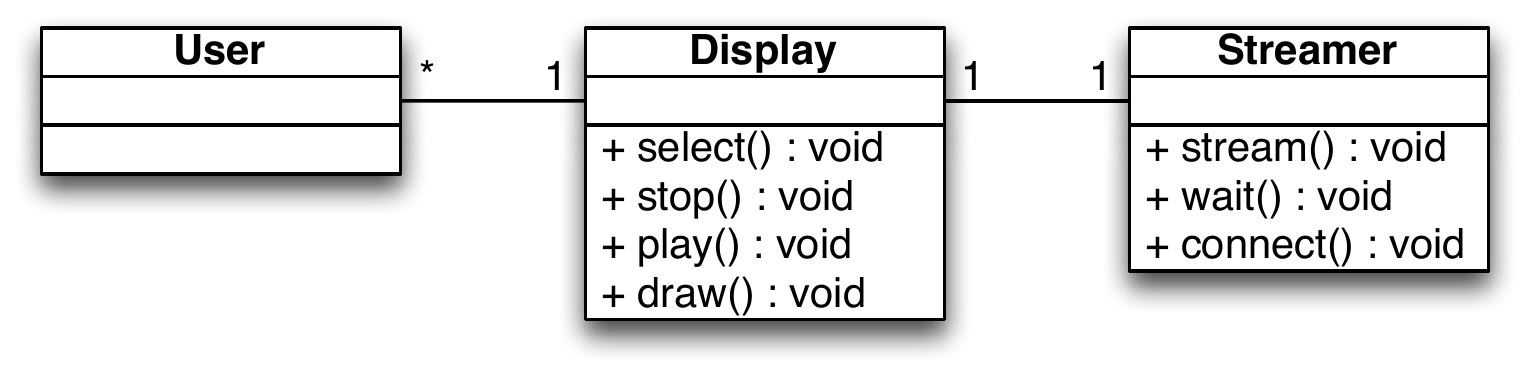}} \qquad
    \subfloat[][UML sequence diagram.]{\label{fig:sm_example}\includegraphics[scale=0.5]{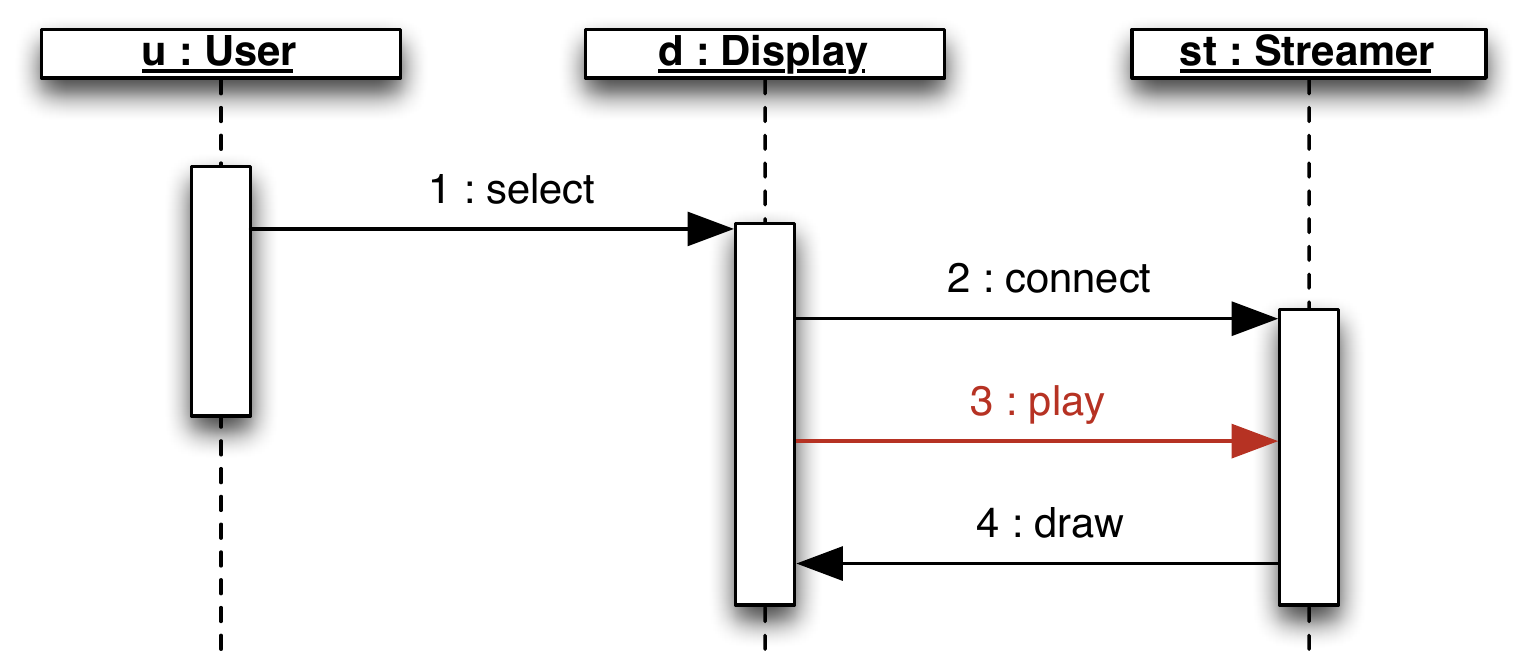}} 
    \caption{Simple VOD system.}
\end{figure}

The example (borrowed from~\cite{PuissantSM:13}) represents a more developed version of the VOD system and is composed by the class and sequence diagrams shown in Figs.~\ref{fig:cd_example} and~\ref{fig:sm_example}, respectively. The class diagram captures the structure of the system, while the sequence diagram describes the steps required in the process of playing a movie.
Recalling constraint \texttt{message\_operation} from Section~\ref{sec:overview}, in order for this model to be consistent, for every message in the sequence diagram, there must exist an operation in the class of the receiver lifeline, whose name equals that of the message. Since there is no operation \texttt{play} in class \texttt{Streamer}, and display \texttt{d} sends message \texttt{play} to streamer \texttt{st}, this model is inconsistent. 
We only consider this single constraint in isolation, the repair space not being subject to any other restrictions, for instance related to some state diagram or to the associations between classes.

\subsection{The {\Badger} Approach} \label{sec:almeida}

{\Badger}~\cite{PuissantSM:13} is a regression planner, implemented in Prolog, that generates repair plans for resolving design model inconsistencies by applying the artificial intelligence technique of automated planning. This technique aims to generate sequences of actions that lead from an initial state to a state meeting a specific predefined goal. Requiring as input a model and a set of inconsistencies, {\Badger} performs a regression planning by starting from the negation of these inconsistencies as the goal state, and searching backwards to find a sequence of actions that reach the initial state.\\

\paragraph{Domain}
{\Badger} is based on a \emph{logical formalism}, as models and meta-models are represented by logic facts, specified in a Prolog embedded Domain Specific Language (eDSL). The technique provides rules for defining meta-model elements, their properties and relationships, thus being \emph{meta-model independent}. However, having the Prolog eDSL as its \emph{technical space} (\emph{other}), it does not provide any automated mechanism to allow the embedding of models nor meta-models persisted in standard languages.

\paragraph{Constraint}
Similarly, constraints are \emph{user definable} in the eDSL as \emph{intra-model} \emph{logical} constraints expressed in first-order logic with transitive closure. Since these constraints are defined in the same technical space as the models and meta-models, rather than being attached to the meta-model, they may refer to concrete model elements. 
  
\paragraph{Update}
In {\Badger}, models and updates are indistinguishable since they are not represented by the elements they contain, but rather by \emph{sequences} of \emph{edit} operations. The entire \emph{history record} is kept (thus each \emph{pre-state} also), with authorship and versioning information attached to each edit step.

\paragraph{Check}
For detecting inconsistencies, {\Badger} relies on the \emph{external} checking procedure proposed in~\cite{BlancMMM:08}, which returns model-level predicates corresponding to existing inconsistencies. These predicates are then required by the repair procedure for specifying a \emph{goal} state which negates them.

\paragraph{Repair}
Prolog's built-in backtracking mechanism allows {\Badger} to generate \emph{multiple} \emph{repair plans}, each one consisting of a set of repair actions that renders the goal true. The tool is a \emph{domain-specific} planner based on a recursive best-first search (RBFS) algorithm.
Although this is an improvement of the well-known A* algorithm, which is known to be complete, it is not clear in the paper whether {\Badger} provides a complete enumeration of plans or not. {\Badger} has a \emph{fixed} set of \emph{edit operations} for creating and deleting objects, as well as for creating, modifying and deleting properties or references on those objects.
The repair procedure enumerates the repair plans under a \emph{parametrizable order}, which the user can control by tweaking the cost function used by the planner algorithm. For instance, the metric can be parameterized by assigning \emph{costs} to edit \emph{operations}, or \emph{weights} to \emph{meta-model} and \emph{model} elements. \emph{Area selection} and operation disabling can be achieved by assigning infinite costs. Since it keeps the whole history record, costs over \emph{meta-data} such as authors and versions can also be assigned.
In order to avoid the multiplication of resolution plans, for modifying references only (other operations are \emph{concrete}), {\Badger} resorts to temporary (\emph{abstract}) elements which the user must replace by concrete ones when effectively applying the repair.
%Semantics
Concerning semantics, {\Badger} applies a \emph{consistency improving} procedure as it generates plans transforming the erroneous model into a model which does not have the detected inconsistencies (negated in the desired goal). It is not, however, fully consistent, since other violations may arise (negative side-effects).
%If the planner is in fact complete, by definition it will also be total and, because it generates ordered plans, least-change 
Finally, the solution function used by {\Badger}, which verifies whether there are no more unsatisfied literals in the desired goal, should ensure \emph{stability}.\\	

% example
By default, the generated plans are ordered in terms of the number of actions they contain. 
For the example, {\Badger} generates the following eight plans to resolve the violation~\cite{PuissantSM:13}:
\begin{enumerate}
\item modify reference $target$ of message \texttt{play}
\item set property $name$ of message \texttt{play} to
\texttt{stream}
\item set property $name$ of operation \texttt{stream}
to \texttt{play}
\item set property $name$ of message \texttt{play} to
\texttt{wait}
\item set property $name$ of operation \texttt{wait} to
\texttt{play}
\item set property $name$ of message \texttt{play} to
\texttt{connect}
\item set property $name$ of operation \texttt{connect} to \texttt{play}
\item delete message \texttt{play} and its references $source$ and $target$
\end{enumerate}
 
Alternative cost functions change the order in which resolution plans are generated (disabling some if infinite costs are assigned). For instance, if one were to set a higher priority to the sequence diagram by assigning smaller costs to actions that create, modify or delete an element belonging to it, the order in which these plans would be generated becomes 1, 2, 4, 6, 8, 3, 5 and 7.

The first generated plan, which suggests modifying reference $target$ of message \texttt{play}, is an example of an abstract repair. It avoids enumerating every lifeline, requiring the user to choose one when applying the repair. 

Despite resolving previously detected inconsistencies, Badger is not free from negative side-effects. This is demonstrated with plan 7, where it repairs message \texttt{play} but introduces another violation of the same type for message \texttt{connect}.
Considering the abstract syntax presented in its paper for the class and sequence diagrams, as well as all the types of repair actions supported by {\Badger}, the list of plans does not appear complete. For instance, adding the missing operation to class \texttt{Streamer} or modifying reference $class$ of \texttt{st}, would also be valid repairs.

\subsection{The {\ModelAnalyzer} Approach} \label{se::reder}

The {\ModelAnalyzer}~\cite{RederE:10, RederE:12} is a tool which follows an incremental approach to model repair, mainly focusing on efficiency. Using the syntactic structure of constraints, it determines which specific parts of a model must be checked and repaired. To achieve this, a form of profiling is used to dynamically observe constraint instances \footnote{A constraint instance corresponds to the evaluation of a constraint for a particular element of its context. For instance, in the example one would have one constraint instance per message.} during evaluation in order to identify what model elements they must assess. Building upon this tracking mechanism, once a constraint instance is evaluated, the tool is able to generate a corresponding tree of repair actions.\\

\paragraph{Domain}
{\ModelAnalyzer} is built over an \emph{object-oriented} formalism, and even though the underlying repair technique is in theory applicable to any kind of models, the tool is implemented for UML (\emph{MDA}) diagrams only, not providing any meta-modelling language.

\paragraph{Constraint}
In the shape of \emph{intra-model} \emph{logical} rules, constraints are \emph{user definable} by means of a generic language, called abstract rule language (ARL), to which it is possible to map arbitrary constraint languages, such as OCL. Once evaluated, the user is supposed to \emph{select} a specific \emph{violation} to be fixed, instead of resolving all inconsistencies at once.

\paragraph{Update}
%Model elements are constantly tracked in a online manner through the leafs of trees which follow the syntactic structure of the constraints. The set of leafs of each tree (its scope) changes depending on what elements are necessary to evaluate for a given model state. When an element changes due to a modification in the model, every constraint instance having that element in its evaluation scope is checked.
For each instance of each constraint, a consistency tree following its syntactic structure is kept in memory and dynamically evaluated in response to identified model updates (\emph{delta-based}). When an element changes due to a modification in the model, every constraint instance having that element in its evaluation scope is notified. This works as a \emph{frame} condition indicating the portion of the state that was effectively changed.

\paragraph{Check}
This is an \emph{internal} checking procedure tightly coupled to the repair mechanism. In fact, it is the core of this technique, while the repair procedure is built over it and thus can be naturally run in \emph{checkonly} mode.
%Each quantifier operator (universal or existential) generates a distinct tree for each element in their current range. For instance, for each message $m$ an additional subtree is kept for each operation in $m.target.class.operations$. Since model elements hang at the leaves of the trees, the delta-based updates allow the checking procedure to only re-evaluate portions of the tree whose valuation may have changed.
A \emph{violation} is reported for each constraint instance that evaluates to false, an evaluated tree being returned. Since the involved model elements are localized through their leaves, one is able to understand where and why they failed.

\paragraph{Repair} 
The repair procedure is based on the comparison of the expected truth value of each consistency tree node, derived from its parent (ultimately, that of the root being true), with its actual observed valuation. Wherever these values differ, a corresponding repair node is generated accordingly to the type of consistency node (logical operator) and observed valuations. Since there may be more than a way to modify the valuation of a logical operator, alternative \emph{repair plans} are returned for each violation, consisting of sequences of \emph{abstract} and \emph{fixed} edit operations (element creation, deletion, and modification). This results in repair trees which also follow the \emph{syntactic} structure of the design constraint and represent enumerations of \emph{multiple} repair plans. 
The approach is \emph{incremental} because once an update is performed, only those trees (and tree branches in particular) are evaluated which are affected by that particular change.
Regarding semantics, the repair procedure is \emph{consistency improving} because it fixes those inconsistencies/trees selected by the user. Yet, similarly to {\Badger}, it is not fully consistent because other trees may be negatively affected. Besides easily ensuring \emph{totality}, this approach is also \emph{stable}, as the repair generation only occurs if the truth value of the consistency tree is false.\\

For the defined example, {\ModelAnalyzer} is expected to produce seven alternative repair plans, each consisting of a single repair action. Here we present the repair tree flattened into a set of alternative repairs \footnote{This is done for conciseness, and possible because the tree would only include disjunction nodes.}. Notice that, unlike the list of repairs generated by {\Badger}, here the alternative plans are not ordered:
\begin{itemize}
  \item modify reference $target$ of message \texttt{play}
  \item modify reference $target.class$ of message \texttt{play}
  \item add operation to $target.class.operations$ of message \texttt{play}
  \item modify property $name$ of message \texttt{play}
  \item modify property $name$ of operation \texttt{stream}
  \item modify property $name$ of operation \texttt{wait}
  \item modify property $name$ of operation \texttt{connect}
\end{itemize}

Repair plans are generated either to fix the ranges of the quantifiers, or their predicates. In the former case, a repair action is suggested for each property referenced on the range's expression, while in the latter case, a repair subtree is calculated for each element contained in that range. For message \texttt{play} in particular, the top three plans fix the range of the existential quantifier, while the other four repair its predicate. Notice that modifying the class of the receiving lifeline, as well as adding an operation to its current class (respectively the second and third plan) are two particular fixes missing in {\Badger}'s repair list. However, compared with that previous technique, this approach is instead missing the possibility of deleting message \texttt{play} itself. In fact, we did not find any information about how {\ModelAnalyzer} handles additions and removals of context elements, so the repair enumeration might not be complete.

Unlike {\Badger}, where a plan may suggest a concrete value to be assigned to some property, here all repair actions are abstract. For instance, the action suggesting to add an operation does not state whether this should be created anew or should come from another class, nor any suggestion to modify a name reveals what value should be used.

As a given tree is seen in isolation, one repair may render (once instantiated) another tree inconsistent (negative side-effect). For instance, as {\Badger} also suggests, modifying property $name$ of operation \texttt{connect} (last plan) can only make message \texttt{play} consistent, if it also makes message \texttt{connect} inconsistent. 
Nevertheless, the authors stress that such potential side-effects are detectable by checking whether a repair action of a repair tree references a model element belonging to the validation scope of other trees.

\subsection{The {\Echo} Approach} \label{sec:macedo}

{\Echo} is a tool for consistency management based on the relational model finder Alloy, developed on top of the popular EMF. While initially built as a bidirectional model transformation framework~\cite{MacedoC:13,MacedoC:14}, it eventually evolved to also handle intra-model consistency~\cite{MacedoGC:13} and multidirectional transformation~\cite{MacedoCP:14}. Thus, {\Echo} is able to check and repair both inter- and intra-model consistency.\\

\paragraph{Domain}
Since {\Echo}'s kernel is the Alloy model finder, it is based on a \emph{relational} formalism. Both models---following the standard structured language XMI---and meta-models---defined in \emph{EMF's} Ecore meta-modeling language---are processed into this formalism, rendering the technique \emph{meta-model independent}. Moreover, {\Echo} has support for \emph{multi-model} environments, so multiple Ecore meta-models may be provided. Although its core engine is \emph{bounded}, the repair procedure, presented below, guarantees that this characteristic is hidden from the user.

\paragraph{Constraint}
Constraints are \emph{user-definable}, either through the embedding of OCL \emph{intra-model} \emph{logical} constraints as meta-model annotations, or by following QVT-R, a declarative language designed to specify \emph{inter-model} \emph{consistency relations} between related models. Both these types of constraints are expressed in first-order logic with transitive closure, and are also embedded into the Alloy core. % \tiago{estes graus de express sempre vao para o diagrama?}.

\paragraph{Update}
{\Echo} is \emph{state-based} since it simply considers the post-state resulting from a user update. While this allows the technique to be run offline---since it does not need to record the user's edit operations---this will require the procedure to check the consistency of the whole model at every execution. %\tiago{single-model? fica confuso o que se quer dizer ao certo com ``only able to handle updates on a single model"} \nuno{nao, pode se fazer update aos dois}

\paragraph{Check}
The checking procedure is \emph{internal} to {\Echo} and can be run in \emph{checkonly} mode: once models, meta-models and the constraints are embedded into Alloy, its model checking capabilities are used to check the consistency of the environment. Thus, the checking procedure is essentially \emph{boolean}. However, intra- and inter-model constraints are distinguished, with {\Echo} testing them independently, resulting in a \emph{composite} checking report. %\nuno{agora que penso nisso a ordem parcial no Echo e' manhosa: corrigir intra-model pode causar inconsistencias inter-model; corrigir inter-model garante que nao ha inconsistencias intra-model.} \alcino{Mas porque e que nao corrige tudo?} \nuno{nenhum motivo em particular, simplesmente foi implementado assim}

\paragraph{Repair}
The core of the repair procedure is similar to that of the checking, but relying instead on Alloy's model finding capabilities. Thus, it automated through the use of a \emph{solver} that calculates new model \emph{states} that satisfy the constraints. Being built over model finding, the procedure is naturally \emph{complete}, enumerating \emph{multiple} model states. Despite being state-based, the user is able to \emph{customize} the set of allowed edit operations that give rise to the calculated states, thus controlling their generation. %(\emph{customizable}) and navigate \tiago{non-deterministic? ou nao previsivel?} through a \emph{complete} enumeration of \emph{multiple} \emph{state-based} repairs. 
%In multi-model environments, {\Echo} supports limited \emph{area selection} by allowing the user to repair a single model at a time.
%\tiago{afinal qual e a ordem? parece me pre-defined... se sim, corrigir a referencia la em cima...} \nuno{pois, e' tricky; de facto as metricas estao pre-definidas, mas e' possivel customiza-las atraves das edit operations; se calhar o parametrizable nao devia estar no XOR, mas ser opcional; alem disso, customizable edit operations devia implicar parametrizable operation costs?}
%Semantics
The tool follows the principle of \emph{least-change}, which is achieved by instructing the model finder to iteratively search for models at an increasing distance. Two \emph{pre-defined} %\tiago{ah ok, ainda assim a referencia parece n estar bem, e la em baixo parece implicar que e parameterizable} \nuno{ver acima}
metrics are supported by {\Echo}: graph-edit distance, that counts insertions and removals of atomic model entities, or an operation-based distance that counts the number of user-defined operations applied. The latter is \emph{parametrizable} through the definition of the valid edit operations. Finally, due to its core based on model finding, this technique is naturally \emph{total}, \emph{fully consistent} and \emph{stable}. The trade-off is performance, since this procedure does not scale for large models.\\ %\nuno{fazer intra-model repair pode causar inter-model inconsistencies, por isso nao sei se e' bem fully correct...} \alcino{Acho que quando estao ambas presentes devia corrigir tudo. Se calhar e apenas um ``bug'' da implementacao a ser corrigido.} \nuno{sim} 

\begin{figure*}[t]
    \centering
    \subfloat[][\smash{Message \emph{3 : play} renamed to \emph{stream}.}]{\label{fig:ex_echo_2}\includegraphics[scale=0.4]{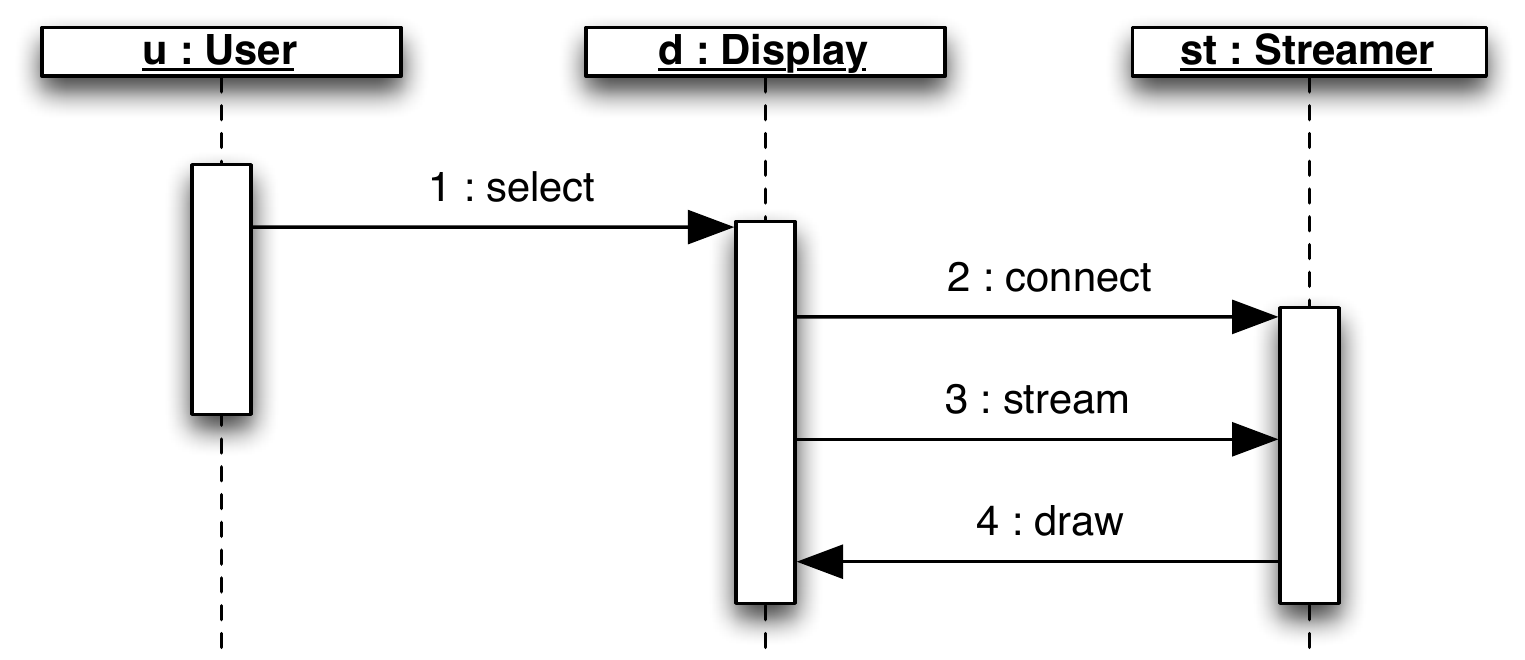}} 
    \subfloat[][\smash{Message \emph{3 : play} renamed to \emph{wait}.}]{\label{fig:ex_echo_3}\includegraphics[scale=0.4]{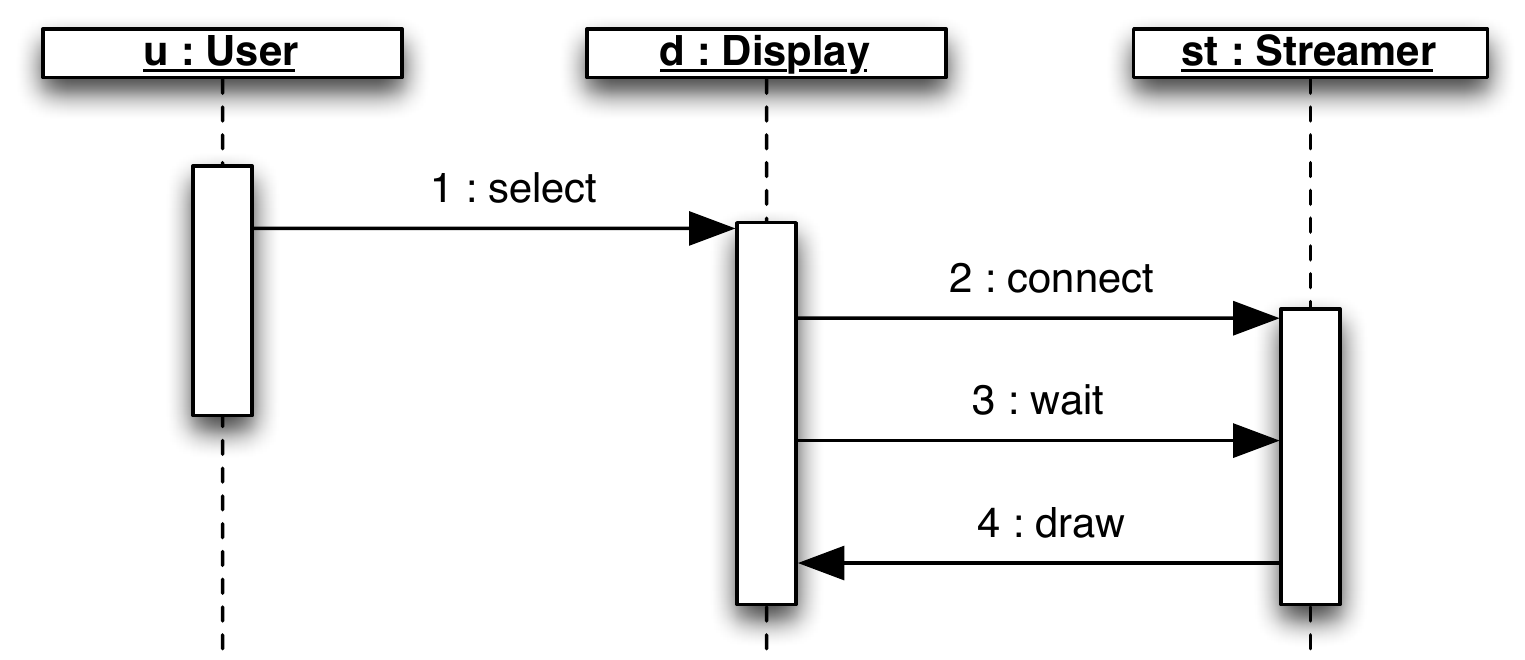}}  
    \subfloat[][\smash{Message \emph{3 : play} renamed to \emph{connect}.}]{\label{fig:ex_echo_4}\includegraphics[scale=0.4]{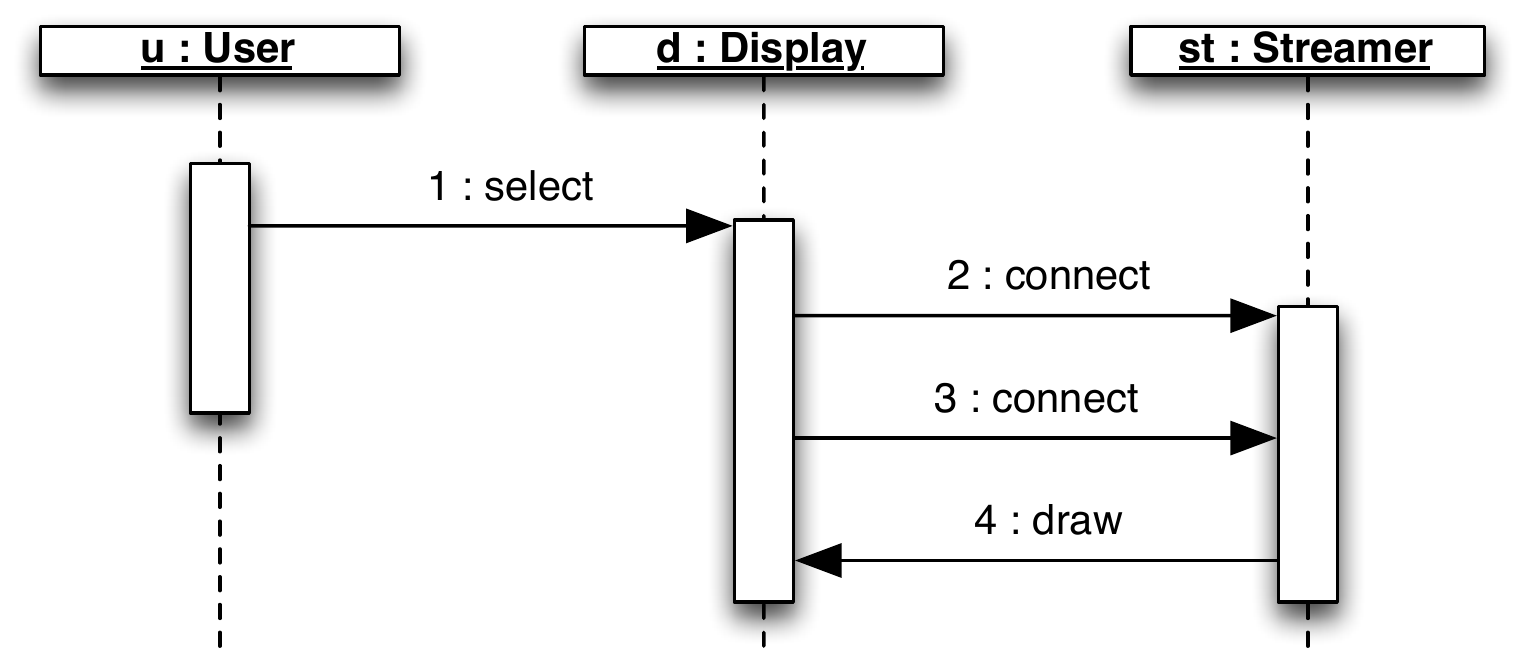}} \\
    \subfloat[][\smash{Target of message \emph{3 : play} moved do \emph{d}.}]{\label{fig:ex_echo_1}\includegraphics[scale=0.4]{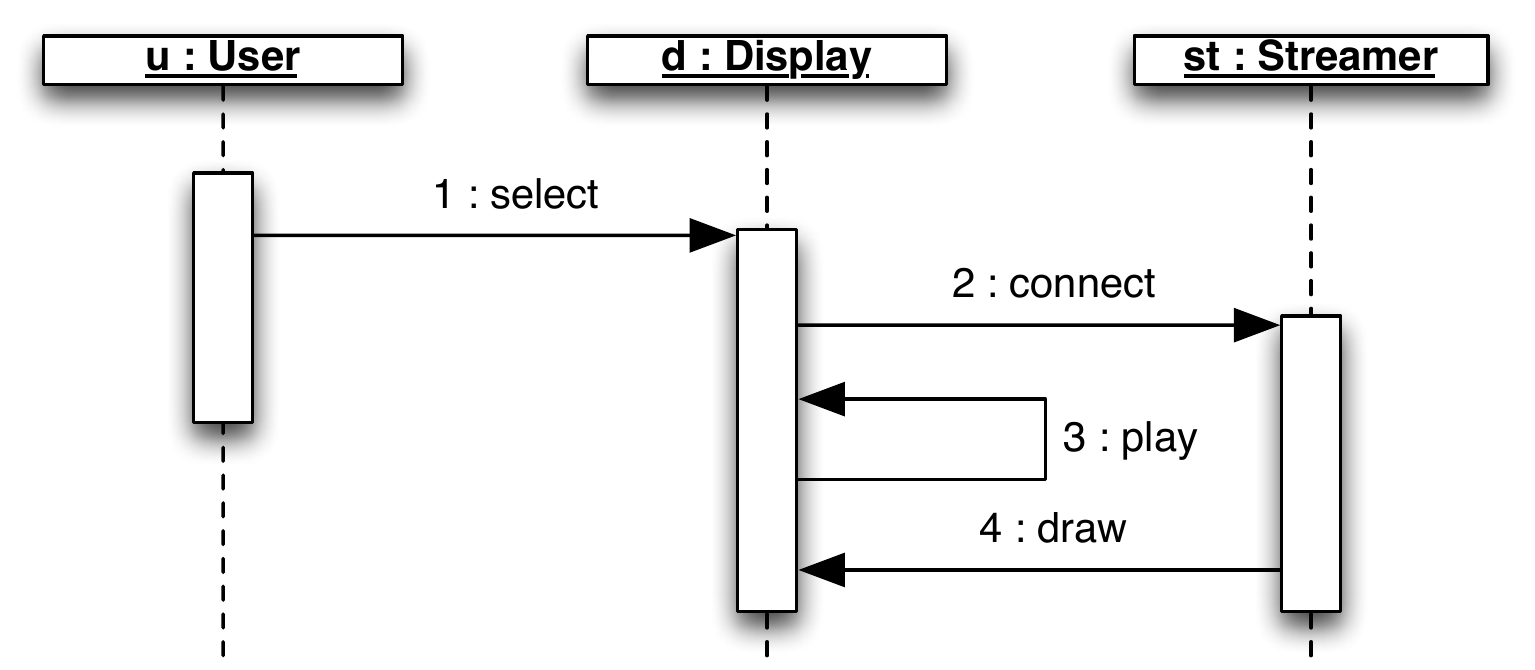}} 
    \subfloat[][\smash{\emph{Streamer} operation \emph{wait} renamed to \emph{play}.}]{\label{fig:ex_echo_5}\includegraphics[scale=0.4]{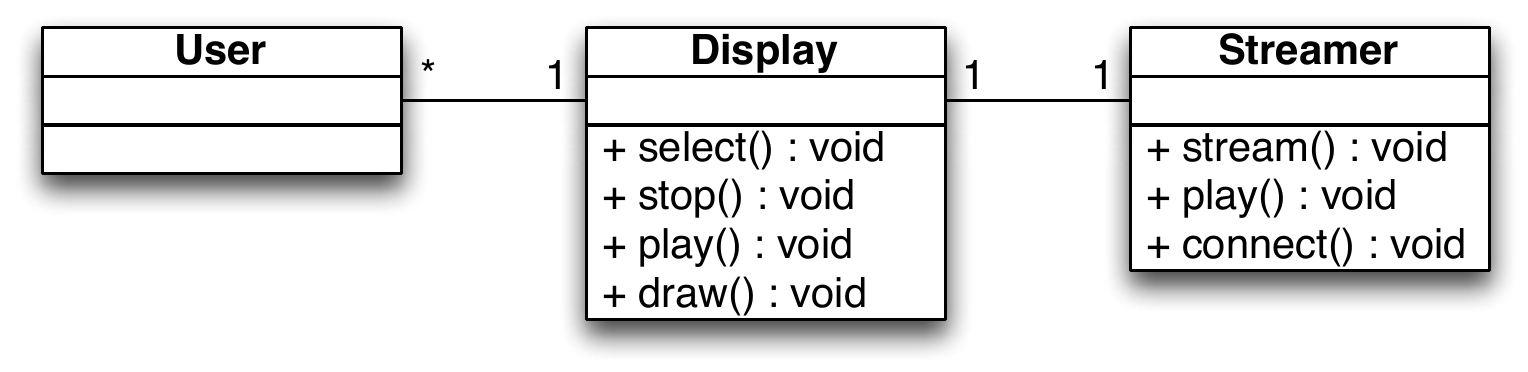}} 
    \subfloat[][\smash{\emph{Streamer} operation \emph{stream} renamed to \emph{play}.}]{\label{fig:ex_echo_6}\includegraphics[scale=0.4]{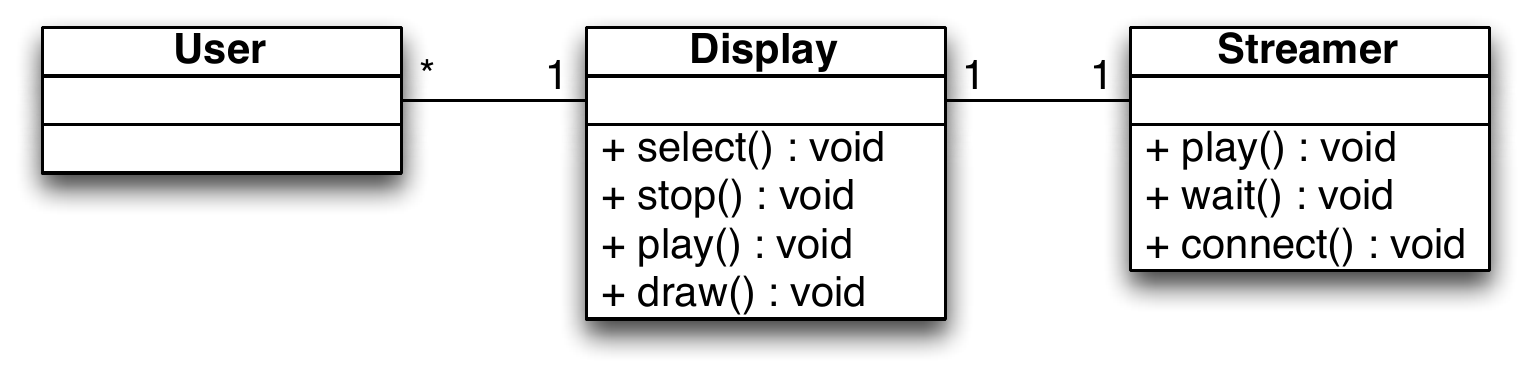}} 
    \caption{Model repair alternatives generated by {\Echo}.}
    \label{fig:ex_echo}
\end{figure*}

% \tiago{no diagrama devera estar um xor com partial. e side-effect aware passaria para baixo de partial certo?}
Regarding our example, it was encoded in {\Echo} %\tiago{referir que temos acesso a ferramenta? ate pk abaixo damos mto detalhe...?} \nuno{nao sei, assim parece q estamos a usar inside information sem fazer todos os esforcos para experimentar as outras}
as an inter-model consistency problem to be compared with the other approaches. Since {\Echo}'s repairs are state-based, it returns new model states rather than repair plans. One of the consequences is that the user is not directly aware of the performed updates. Figure~\ref{fig:ex_echo} shows the repair alternatives for this problem that are closest to the original model under regular graph-edit distance. %\tiago{a) mudar 3. para play} \nuno{nao percebi} 
For models at the same distance from the inconsistent model, the order in which they are returned is arbitrary. %\alcino{Confuso. Talvez referir que estas sao as mais perto e, para as que estao a mesma distancia e que e arbitrario.} 
Note that only fully consistent models are returned (e.g., no alternative renames operation \texttt{connect}, as it is being referred by another message).
The creation of a new operation and the deletion of the message are not among this initial set of alternatives, because they are not at minimal distance from the initial model. Nevertheless, once the minimal ones are enumerated, {\Echo} starts producing the next closest ones, which would include those repair alternatives.  %\alcino{Mas o echo nao permite tambem ver as que estao mais longe? Assim parece que e incompleto.} \nuno{sim, permite} 
Note that renaming the \texttt{connect} operation to \texttt{play} and changing the class of the \texttt{st} lifeline, although embodying minimal updates, are not produced by {\Echo}, since they create negative side-effects and would render the environment inconsistent. %\tiago{tb podes referir que, comparando ao egyed, nao considerou alterar o tipo da lifeline (que, presumo, estaria proximo) pk como e concreto e correto nao encontrou repair proximo valido (quanto muito mudava para tipo Display (diplay invocava outro display) mas gerava side effect negativo relativo a message connect)}

The order of the returned solutions could be modified by defining the valid edit operations (through OCL pre- and post-conditions) and enforcing the operation-based distance. %\tiago{hmm assim parece parameterizable...} \nuno{ver acima}.
For instance, defining only operations to rename or delete messages, {\Echo} would only return the models from Figs.~\ref{fig:ex_echo_2},~\ref{fig:ex_echo_3} and~\ref{fig:ex_echo_4}, and one where the message is deleted.

\section{Related Work}
\label{sec:related}
Although several taxonomies have been proposed to a variety of MDE activities, no systematic classification has been proposed for model repair specifically. To date, the most comprehensive study on consistency management, including inconsistency fixing, is still~\cite{SpanoudakisZ:01}, which, based on previous definitions from~\cite{NuseibehER:00} and~\cite{FinkelsteiinST:96}, surveyed and classified the existing approaches. % topic \tiago{parece que se esta a referir a model repair...} 
Robust feature-based classifications have been proposed for model transformation~\cite{CzarneckiH:06}, model synchronization~\cite{AntkiewiczC:07} and bidirectional transformation~\cite{HidakaTCH:15}. While some features of model repair approaches overlap with features from those areas, there are many topics that are specific to this domain.

A feature-based classification of model repair techniques was previously presented in~\cite{Puissant:12}, addressing the flexibility, usability and extensibility of the approaches. Our classification largely subsumes that proposal, not only with a thorough classification of the behavior of the repair procedure and on the user's ability to control it, but also by addressing the relationship of such procedures with the remainder artifacts of the MDE environment. Our validation through literature review is also more exhaustive.

Although our formalization of the semantic properties of the model repair procedure is novel, they are inspired by those proposed for constraint maintainers in the context of bidirectional transformation~\cite{Meertens:98}.

%focusing on flexibility (coverage: all, some or single inconsistencies; monotonicity: side-effects; resolution space: single or multiple repairs; language independence: modeling language), 

%usability (selection: subset of inconsistencies, subset of rules, sub-part of the model; 

%tool integration: integrated with CASE tools; user intervention: guide generation of resolutions) 

%extensibility features (solver: support multiple solvers; resolution: add more resolution rules; automation: adding new rules does not require specifying resolutions). 

%\alcino{Ao listarem tudo o que outro cobre ate da a impressao que o nosso nao e assim tao mais exaustivo... Omitir esta lista exaustiva? Tentar justificar melhor porque e que no nosso e melhor (mais exaustifivo, diferente)?} 

%Our classifications also addresses these topics but is much more exhaustive, as is our literature review.

%In this survey we focused on repair approaches for design models. While other applications, like data repair~\cite{x}, could eventually be mapped to this taxonomy, as their concerns vary from MDE some features may not be exactly reflected.

\section{Conclusion}
\label{sec:conclusion}

Inconsistency fixing methods are vital to any development process within the increasingly adopted MDE. Resulting from an exhaustive and systematic analysis of their diverse landscape, in this paper we propose a novel feature-based classification system for such model repair techniques, with the intent to aid the MDE practitioner in choosing the approach most suitable for his/her particular needs, but also the developer interested in developing new techniques. Supported by an underlying formalization of the problem of model repair, our feature-based taxonomy comprises five major classification axis, organized as hierarchical models defining valid configurations of features. With a deep focus on the behavior of the repair procedure, the shape of all other relevant MDE artifacts was also addressed, as well as the role of the user in specifying and customize them.

We classify some modern approaches to model repair under our classification system, obtaining normative profiles which assist in understanding the techniques, and, since drawn from a common view point, make similarities and differences more obvious. A comparison is done, and the impact of some design decisions demonstrated, by applying these techniques to a simple example.

Although approaches to model repair are rather heterogeneous, our experiments show that the proposed classification is sufficiently flexible to classify most existing approaches. Nonetheless, we plan to further refine our taxonomy by encompassing new methodologies and rigorously reviewing it as needed, thus ensuring that it remains applicable, complete and understandable.

\bibliographystyle{abbrv}
\bibliography{surv}

\begin{thebibliography}{10}

\bibitem{AmelunxenLSS:07}
C.~Amelunxen, E.~Legros, A.~Sch{\"{u}}rr, and I.~St{\"{u}}rmer.
\newblock Checking and enforcement of modeling guidelines with graph
  transformations.
\newblock In {\em AGTIVE 2007}, volume 5088 of {\em LNCS}, pages 313--328.
  Springer, 2007.

\bibitem{AntkiewiczC:07}
M.~Antkiewicz and K.~Czarnecki.
\newblock Design space of heterogeneous synchronization.
\newblock In {\em GTTSE}, volume 5235 of {\em LNCS}, pages 3--46. Springer,
  2007.

\bibitem{Balzer:91}
R.~Balzer.
\newblock Tolerating inconsistency.
\newblock In {\em ICSE 1991}, pages 158--165. IEEE / ACM, 1991.

\bibitem{BeckerHLW:07}
S.~M. Becker, S.~Herold, S.~Lohmann, and B.~Westfechtel.
\newblock A graph-based algorithm for consistency maintenance in incremental
  and interactive integration tools.
\newblock {\em Software \& Systems Modeling}, 6(3):287--315, 2007.

\bibitem{BenavidesSC:10}
D.~Benavides, S.~Segura, and A.~R. Cort{\'{e}}s.
\newblock Automated analysis of feature models 20 years later: {A} literature
  review.
\newblock {\em Inf. Syst.}, 35(6):615--636, 2010.

\bibitem{BlancMMM:08}
X.~Blanc, I.~Mounier, A.~Mougenot, and T.~Mens.
\newblock Detecting model inconsistency through operation-based model
  construction.
\newblock In {\em ICSE 2008}, pages 511--520. IEEE, 2008.

\bibitem{ChechikLNDESS:09}
M.~Chechik, W.~Lai, S.~Nejati, J.~Cabot, Z.~Diskin, S.~Easterbrook,
  M.~Sabetzadeh, and R.~Salay.
\newblock Relationship-based change propagation: A case study.
\newblock In {\em MISE 2009}, pages 7--12. IEEE, 2009.

\bibitem{CicchettiREP:10}
A.~Cicchetti, D.~D. Ruscio, R.~Eramo, and A.~Pierantonio.
\newblock {JTL}: a bidirectional and change propagating transformation
  language.
\newblock In {\em SLE'10}, volume 6563 of {\em LNCS}, pages 183--202. Springer,
  2010.

\bibitem{MacedoCG:15}
A.~Cunha, N.~Macedo, and T.~Guimar{\~a}es.
\newblock Exploring scenario exploration.
\newblock Submitted.

\bibitem{CunhaMG:14}
A.~Cunha, N.~Macedo, and T.~Guimar{\~a}es.
\newblock Target oriented relational model finding.
\newblock In {\em FASE 2014}, volume 8411 of {\em LNCS}, pages 17--31.
  Springer, 2014.

\bibitem{CzarneckiH:06}
K.~Czarnecki and S.~Helsen.
\newblock Feature-based survey of model transformation approaches.
\newblock {\em {IBM} Systems Journal}, 45(3):621--646, 2006.

\bibitem{DamG:14}
H.~K. Dam and A.~Ghose.
\newblock Towards rational and minimal change propagation in model evolution.
\newblock {\em CoRR}, abs/1402.6046, 2014.

\bibitem{DamLG:10}
H.~K. Dam, L.~L{\^{e}}, and A.~K. Ghose.
\newblock Supporting change propagation in the evolution of enterprise
  architectures.
\newblock In {\em EDOC 2010}, pages 24--33. {IEEE} Computer Society, 2010.

\bibitem{DamRE:14}
H.~K. Dam, A.~Reder, and A.~Egyed.
\newblock Inconsistency resolution in merging versions of architectural models.
\newblock In {\em WICSA 2014}, pages 153--162. {IEEE} Computer Society, 2014.

\bibitem{DamW:10}
H.~K. Dam and M.~Winikoff.
\newblock Supporting change propagation in {UML} models.
\newblock In {\em ICSM 2010)}, pages 1--10. {IEEE} Computer Society, 2010.

\bibitem{DamW:11}
H.~K. Dam and M.~Winikoff.
\newblock An agent-oriented approach to change propagation in software
  maintenance.
\newblock {\em Autonomous Agents and Multi-Agent Systems}, 23(3):384--452,
  2011.

\bibitem{DamWP:06}
K.~H. Dam, M.~Winikoff, and L.~Padgham.
\newblock An agent-oriented approach to change propagation in software
  evolution.
\newblock In {\em ASWEC 2006}, pages 309--318. {IEEE} Computer Society, 2006.

\bibitem{DemuthLE:13}
A.~Demuth, R.~E. Lopez{-}Herrejon, and A.~Egyed.
\newblock Supporting the co-evolution of metamodels and constraints through
  incremental constraint management.
\newblock In {\em MODELS 2013}, volume 8107 of {\em LNCS}, pages 287--303.
  Springer, 2013.

\bibitem{Easterbrook:91}
S.~M. Easterbrook.
\newblock Handling conflict between domain descriptions with computer-supported
  negotiation.
\newblock {\em Knowledge Acquisition}, 3(3):255--–289, 1991.

\bibitem{EasterbrookN:96}
S.~M. Easterbrook and B.~Nuseibeh.
\newblock Using {ViewPoints} for inconsistency management.
\newblock {\em Software Engineering Journal}, 11(1):31--43, 1996.

\bibitem{EgyedLF:08}
A.~Egyed, E.~Letier, and A.~Finkelstein.
\newblock Generating and evaluating choices for fixing inconsistencies in {UML}
  design models.
\newblock In {\em ASE 2008}, pages 99--108. IEEE, 2008.

\bibitem{EndersHGTT:02}
B.~Enders, T.~Heverhagen, M.~Goedicke, P.~Tr{\"{o}}pfner, and R.~Tracht.
\newblock Towards an integration of different specification methods by using
  the viewpoint framework.
\newblock {\em Transactions of the {SDPS}}, 6(2):1--23, 2002.

\bibitem{EramoPRV:08}
R.~Eramo, A.~Pierantonio, J.~R. Romero, and A.~Vallecillo.
\newblock Change management in multi-viewpoint system using {ASP}.
\newblock In {\em ECOCW 2008}, pages 433--440. {IEEE} Computer Society, 2008.

\bibitem{EtzlstorferKKLRSSW:13}
J.~Etzlstorfer, A.~Kusel, E.~Kapsammer, P.~Langer, W.~Retschitzegger,
  J.~Schoenboeck, W.~Schwinger, and M.~Wimmer.
\newblock A survey on incremental model transformation approaches.
\newblock In {\em MoDELS 2013 Workshops}, volume 1090 of {\em {CEUR} Workshop
  Proceedings}, pages 4--13. CEUR-WS.org, 2013.

\bibitem{FinkelsteiinST:96}
A.~Finkelsteiin, G.~Spanoudakis, and D.~Till.
\newblock Managing interference.
\newblock In {\em Joint proceedings of the second international software
  architecture workshop (ISAW-2) and international workshop on multiple
  perspectives in software development (Viewpoints' 96) on SIGSOFT'96
  workshops}, pages 172--174. ACM, 1996.

\bibitem{FinkelsteinGHKN:94}
A.~Finkelstein, D.~M. Gabbay, A.~Hunter, J.~Kramer, and B.~Nuseibeh.
\newblock Inconsistency handling in multperspective specifications.
\newblock {\em {IEEE} Trans. Software Eng.}, 20(8):569--578, 1994.

\bibitem{GieseW:09}
H.~Giese and R.~Wagner.
\newblock From model transformation to incremental bidirectional model
  synchronization.
\newblock {\em Software \& Systems Modeling}, 8(1):21--43, 2009.

\bibitem{GrundyHM:98}
J.~C. Grundy, J.~G. Hosking, and W.~B. Mugridge.
\newblock Inconsistency management for multiple-view software development
  environments.
\newblock {\em {IEEE} Trans. Software Eng.}, 24(11):960--981, 1998.

\bibitem{HausmannHS:02}
J.~H. Hausmann, R.~Heckel, and S.~Sauer.
\newblock Extended model relations with graphical consistency conditions.
\newblock In {\em UML 2002 Workshop on Consistency Problems in {UML}-based
  Software Development}, pages 61--74, 2002.

\bibitem{HegedusHRBV:11}
A.~Hegedus, A.~Horv{\'a}th, I.~R{\'a}th, M.~C. Branco, and D.~Varr{\'o}.
\newblock Quick fix generation for {DSMLs}.
\newblock In {\em Visual Languages and Human-Centric Computing (VL/HCC), 2011
  IEEE Symposium on}, pages 17--24. IEEE, 2011.

\bibitem{HidakaHIKMN:10}
S.~Hidaka, Z.~Hu, K.~Inaba, H.~Kato, K.~Matsuda, and K.~Nakano.
\newblock Bidirectionalizing graph transformations.
\newblock In {\em ICFP 2010}, pages 205--216. ACM, 2010.

\bibitem{HidakaTCH:15}
S.~Hidaka, M.~Tisi, J.~Cabot, and Z.~Hu.
\newblock Feature-based classification of bidirectional transformation
  approaches.
\newblock {\em Software \& Systems Modeling}, 2015.

\bibitem{IvkovicK:04}
I.~Ivkovic and K.~Kontogiannis.
\newblock Tracing evolution changes of software artifacts through model
  synchronization.
\newblock In {\em ICSM 2004}, pages 252--261. {IEEE} Computer Society, 2004.

\bibitem{KangCHNP:90}
K.~C. Kang, S.~G. Cohen, J.~A. Hess, W.~E. Novak, and A.~S. Peterson.
\newblock Feature-oriented domain analysis ({FODA}) feasibility study.
\newblock Technical Report CMU/SEI-90-TR-21, Software Engineering Institute,
  Carnegie Mellon University, 1990.

\bibitem{KleinerFA:10}
M.~Kleiner, M.~D.~D. Fabro, and P.~Albert.
\newblock Model search: Formalizing and automating constraint solving in {MDE}
  platforms.
\newblock In {\em Proceedings of the 6th European Conference on Modelling
  Foundations and Applications (ECMFA 2010)}, volume 6138 of {\em LNCS}, pages
  173--188. Springer, 2010.

\bibitem{KolovosPP:08}
D.~S. Kolovos, R.~F. Paige, and F.~Polack.
\newblock Detecting and repairing inconsistencies across heterogeneous models.
\newblock In {\em ICST 2008}, pages 356--364. {IEEE} Computer Society, 2008.

\bibitem{KonigsS:06}
A.~K{\"{o}}nigs and A.~Sch{\"{u}}rr.
\newblock {MDI:} a rule-based multi-document and tool integration approach.
\newblock {\em Software and System Modeling}, 5(4):349--368, 2006.

\bibitem{Kortgen:10}
A.-T. K{\"o}rtgen.
\newblock New strategies to resolve inconsistencies between models of decoupled
  tools.
\newblock In {\em LWI 2010}, volume 661, pages 21--31. CEUR, 2010.

\bibitem{KusterR:07}
J.~M. K{\"{u}}ster and K.~Ryndina.
\newblock Improving inconsistency resolution with side-effect evaluation and
  costs.
\newblock In {\em MoDELS 2007}, volume 4735 of {\em LNCS}, pages 136--150.
  Springer, 2007.

\bibitem{LiuEM:02}
W.~Liu, S.~Easterbrook, and J.~Mylopoulos.
\newblock Rule-based detection of inconsistency in {UML} models.
\newblock In {\em Workshop on Consistency Problems in {UML}-Based Software
  Development}, volume~5, 2002.

\bibitem{Macedo:14}
N.~Macedo.
\newblock {\em A Relational Approach to Bidirectional Transformation}.
\newblock PhD thesis, Universidade do Minho, 2014.
\newblock Submitted.

\bibitem{MacedoC:13}
N.~Macedo and A.~Cunha.
\newblock Implementing {QVT-R} bidirectional model transformations using
  {A}lloy.
\newblock In {\em FASE 2013}, volume 7793 of {\em LNCS}, pages 297--311.
  Springer, 2013.

\bibitem{MacedoC:14}
N.~Macedo and A.~Cunha.
\newblock Least-change bidirectional model transformation with {QVT-R} and
  {ATL}.
\newblock {\em Software \& Systems Modeling}, 2014.

\bibitem{MacedoCP:14}
N.~Macedo, A.~Cunha, and H.~Pacheco.
\newblock Towards a framework for multi-directional model transformations.
\newblock In {\em Workshops of EDBT/ICDT 2014}, volume 1133 of {\em CEUR
  Workshop Proceedings}, pages 71--74. CEUR-WS, 2014.

\bibitem{MacedoGC:13}
N.~Macedo, T.~Guimar{\~a}es, and A.~Cunha.
\newblock Model repair and transformation with {Echo}.
\newblock In {\em ASE 2013}, pages 694--697. IEEE, 2013.

\bibitem{MafaziMS:14}
S.~Mafazi, W.~Mayer, and M.~Stumptner.
\newblock Conflict resolution for on-the-fly change propagation in business
  processes.
\newblock In {\em Proceedings of the Asia-Pacific Conference on Conceptual
  Modelling (APCCM)}, Conferences in Research and Practice in Information
  Technology (CRPIT). Australian Computer Society, 2014.

\bibitem{Meertens:98}
L.~Meertens.
\newblock Designing constraint maintainers for user interaction.
\newblock available at \url{http://www.kestrel.edu/home/people/meertens}, 1998.

\bibitem{MensSD:06}
T.~Mens, R.~Van Der~Straeten, and M.~D’Hondt.
\newblock Detecting and resolving model inconsistencies using transformation
  dependency analysis.
\newblock In {\em Model driven engineering languages and systems}, pages
  200--214. Springer, 2006.

\bibitem{NentwichEF:03}
C.~Nentwich, W.~Emmerich, and A.~Finkelstein.
\newblock Consistency management with repair actions.
\newblock In {\em ICSE 2003}, pages 455--464. IEEE, 2003.

\bibitem{NuseibehER:00}
B.~Nuseibeh, S.~M. Easterbrook, and A.~Russo.
\newblock Leveraging inconsistency in software development.
\newblock {\em {IEEE} Computer}, 33(4):24--29, 2000.

\bibitem{NuseibehR:99}
B.~Nuseibeh and A.~Russo.
\newblock Using abduction to evolve inconsistent requirements specification.
\newblock {\em Australasian J. of Inf. Systems}, 6(2), 1999.

\bibitem{OlssonG:02}
T.~Olsson and J.~Grundy.
\newblock Supporting traceability and inconsistency management between software
  artifacts.
\newblock In {\em 2002 IASTED International Conference on Software Engineering
  and Applications}. IASTED Press, 2002.

\bibitem{QVT:11}
OMG.
\newblock {\em {MOF} 2.0 {Query/View/Transformation} Specification ({QVT}),
  Version 1.1}, January 2011.
\newblock Available at \url{http://www.omg.org/spec/QVT/1.1/}.

\bibitem{OCL:12}
OMG.
\newblock {\em {OMG} Object Constraint Language {(OCL)}, Version 2.3.1},
  January 2012.
\newblock Available at \url{http://www.omg.org/spec/OCL/2.3.1/}.

\bibitem{MOF:14}
OMG.
\newblock {\em {OMG} {Meta Object Facility} ({MOF}) Core Specification, Version
  2.4.2}, June 2014.
\newblock Available at \url{http://www.omg.org/spec/MOF/2.4.2/}.

\bibitem{Puissant:12}
J.~P. Puissant.
\newblock {\em Resolving Inconsistencies in Model-Driven Engineering using
  Automated Planning}.
\newblock PhD thesis, Université de Mons, 2012.

\bibitem{PuissantMS:10}
J.~P. Puissant, T.~Mens, R.~Van, and D.~Straeten.
\newblock Resolving model inconsistencies with automated planning.
\newblock In {\em In Proceedings of the 3rd Workshop on Living with
  Inconsistencies in Software Development}, pages 8--14, 2010.

\bibitem{PuissantSM:13}
J.~P. Puissant, R.~Van Der~Straeten, and T.~Mens.
\newblock Resolving model inconsistencies using automated regression planning.
\newblock {\em Software \& Systems Modeling}, pages 1--21, 2013.

\bibitem{RederE:10}
A.~Reder and A.~Egyed.
\newblock Model/analyzer: a tool for detecting, visualizing and fixing design
  errors in {UML}.
\newblock In {\em ASE 2010}, pages 347--348. {ACM}, 2010.

\bibitem{RederE:12}
A.~Reder and A.~Egyed.
\newblock Computing repair trees for resolving inconsistencies in design
  models.
\newblock In {\em ASE 2012}, pages 220--229. IEEE, 2012.

\bibitem{SchoenboeckKEKSWW:14}
J.~Schoenboeck, A.~Kusel, E.~J., E.~Kapsammer, W.~Schwinger, M.~Wimmer, and
  M.~Wischenbart.
\newblock {CARE} -- a constraint-based approach for re-establishing
  conformance-relationships.
\newblock In {\em APCCM 2014}, volume 154 of {\em CRPIT}, pages 19--28. ACS,
  2014.

\bibitem{SilvaMBB:10}
M.~A.~A. Silva, A.~Mougenot, X.~Blanc, and R.~Bendraou.
\newblock Towards automated inconsistency handling in design models.
\newblock In {\em Advanced Information Systems Engineering}, pages 348--362.
  Springer, 2010.

\bibitem{SpanoudakisF:97}
G.~Spanoudakis and A.~Finkelstein.
\newblock Reconciling requirements: A method for managing interference,
  inconsistency and conflict.
\newblock {\em Ann. Software Eng.}, 3:433--457, 1997.

\bibitem{SpanoudakisZ:01}
G.~Spanoudakis and A.~Zisman.
\newblock Inconsistency management in software engineering: Survey and open
  research issues.
\newblock {\em Handbook of software engineering and knowledge engineering},
  1:329--380, 2001.

\bibitem{Stevens:07}
P.~Stevens.
\newblock A landscape of bidirectional model transformations.
\newblock In {\em GTTSE 2007}, volume 5235 of {\em LNCS}, pages 408--424.
  Springer, 2007.

\bibitem{Stevens:14}
P.~Stevens.
\newblock Bidirectionally tolerating inconsistency: Partial transformations.
\newblock In {\em FASE 2014}, volume 8411 of {\em LNCS}, pages 32--46.
  Springer, 2014.

\bibitem{StraetenD:06}
R.~V.~D. Straeten and M.~D'Hondt.
\newblock Model refactorings through rule-based inconsistency resolution.
\newblock In {\em SAC 2006}, pages 1210--1217. {ACM}, 2006.

\bibitem{StraetenMSJ:03}
R.~V.~D. Straeten, T.~Mens, J.~Simmonds, and V.~Jonckers.
\newblock Using description logic to maintain consistency between {UML} models.
\newblock In {\em UML 2003}, volume 2863 of {\em LNCS}, pages 326--340.
  Springer, 2003.

\bibitem{StraetenPM:11}
R.~Van Der~Straeten, J.~P. Puissant, and T.~Mens.
\newblock Assessing the {Kodkod} model finder for resolving model
  inconsistencies.
\newblock In {\em Modelling Foundations and Applications}, pages 69--84.
  Springer, 2011.

\bibitem{WagnerGN:03}
R.~Wagner, H.~Giese, and U.~Nickel.
\newblock Plug-in for flexible and incremental consistency management.
\newblock In {\em Workshop on Consistency Problems in {UML} based Software
  Development}, 2003.

\bibitem{XiongHZSTM:09}
Y.~Xiong, Z.~Hu, H.~Zhao, H.~Song, M.~Takeichi, and H.~Mei.
\newblock Supporting automatic model inconsistency fixing.
\newblock In {\em Proceedings of the the 7th joint meeting of the European
  software engineering conference and the ACM SIGSOFT symposium on The
  foundations of software engineering}, pages 315--324. ACM, 2009.

\end{thebibliography}
%\putbib[IEEEabrv,surv]
%\end{bibunit}

\end{document}